\documentclass[10pt,twoside,onecolumn,letterpaper,journal,draftcls]{IEEEtran}

\usepackage[utf8]{inputenc}
\usepackage[T1]{fontenc}
\usepackage{ifthen}

\usepackage{hyphenat}
\usepackage{fancyhdr}
\usepackage{graphicx}
\usepackage{color}
\usepackage{float}
\usepackage[nottoc,notlot,notlof]{tocbibind}
\usepackage[linesnumbered,ruled,vlined]{algorithm2e}
\SetAlgorithmName{Protocol}{List of protocols}

\usepackage{algpseudocode}

\usepackage{longtable}
\usepackage{tabularx}
\usepackage{booktabs}

\newcolumntype{L}[1]{>{\raggedright\arraybackslash}p{#1}}
\newcolumntype{C}[1]{>{\centering\arraybackslash}p{#1}}
\newcolumntype{R}[1]{>{\raggedleft\arraybackslash}p{#1}}

\usepackage{enumerate}

\usepackage{dsfont}
\usepackage{bbm}

\usepackage{amsmath}   % From the American Mathematical Society
\usepackage{mathtools}
\usepackage{amssymb}
\usepackage{amsbsy}
\usepackage{amstext}
\usepackage{amscd}
\usepackage{amsopn}
\usepackage{amsthm}
\usepackage{amsfonts}
\usepackage{bm}
\usepackage[artemisia]{textgreek}
\usepackage{xfrac}
\usepackage{mathabx}
\usepackage{framed}
\usepackage{adjustbox}

\usepackage{setspace}

\usepackage{tikz}
\usetikzlibrary{backgrounds,fit,decorations.pathreplacing,calc}
\usetikzlibrary{shapes}
\usetikzlibrary{circuits.logic.US}
\usetikzlibrary{patterns}

\usetikzlibrary{mindmap,backgrounds}

\usetikzlibrary{arrows,snakes,backgrounds}
\usetikzlibrary{positioning,matrix,through,calc,arrows,fit,shapes,decorations.pathreplacing,decorations.markings}

\usepackage[numbers,sort&compress]{natbib}
\usepackage{bibentry}

\newcommand{\RR}{\mathds{R}}
\newcommand{\NN}{\mathds{N}}

\newcommand{\pr}[1]{\mathrm{Pr}\left\{#1\right\}} %Probability
\newcommand{\ev}[1]{\mathds{E}\left\{#1\right\}}  %Expected value
\newcommand{\myvar}[1]{\mathds{V}\!\mathrm{ar}\left\{#1\right\}}  %Variance
\newcommand{\sd}{\mathbf{SD}}

\newcommand{\myop}[1]{\mathrm{#1}}

\newcommand{\eps}{\epsilon}
\renewcommand{\varepsilon}{\epsilon}

\newcommand{\mc}[1]{\mathcal{#1}}
\newcommand{\supp}{\mathrm{supp}}

\newcommand{\floor}[1]{\lfloor #1 \rfloor}
\providecommand{\eqref}[1]{(\ref{#1})}
\newcommand{\vect}[1]{\boldsymbol{\mathrm{#1}}}

\newcommand{\mymtrx}[1]{\mathrm{#1}}

\newcommand{\remove}[1]{}

\makeatletter
\def\th@plain{%
	\thm@notefont{}% same as heading font
	\itshape % body font
}
\def\th@definition{%
	\thm@notefont{}% same as heading font
	\normalfont % body font
}
\makeatother

\newtheorem{theorem}{Theorem}
\newtheorem{proposition}[theorem]{Proposition}
\newtheorem{lemma}[theorem]{Lemma}

\newtheorem{corollary}{Corollary}[theorem]

\theoremstyle{definition}
\newtheorem{definition}{Definition}
\newtheorem{remark}{Remark}
\theoremstyle{definition}
\newtheorem{example}{Example}

\usepackage{chngcntr}

\newlength{\eqboxstorage} \newcommand{\eqbox}[1]{ \setlength{\eqboxstorage}{\fboxsep} \setlength{\fboxsep}{6pt} \boxed{#1} \setlength{\fboxsep}{\eqboxstorage} }

\usepackage[colorlinks=true, allcolors=red]{hyperref}
\hypersetup{allcolors=red!80!blue}

\begin{document}
	%\title{Secret Key Agreement in Wiretapped Tree-PIN}
	\title{Wiretap Secret Key Capacity of Tree-PIN}
	
	\author{Alireza~Poostindouz,~\IEEEmembership{Student~Member,~IEEE,} 
		and~Reihaneh~Safavi-Naini,~\IEEEmembership{Member,~IEEE}%
		\thanks{The  authors  are  with  the  Department  of  Computer  Science,  University  of Calgary,  Calgary,  AB  T2N  1N4,  Canada  (e-mail: alireza.poostindouz@ucalgary.ca; rei@ucalgary.ca)}%
		\thanks{Parts of the results in this work were presented in ISIT, Paris, July 2019.}%
	}%

	\markboth{TO BE SUBMITTED TO IEEE TRANSACTIONS ON INFORMATION THEORY,~Vol.~XX, No.~X, August~20XX}%
	{POOSTINDOUZ AND SAFAVI-NAINI: Wiretap Secret Key Capacity of Tree-PIN}
	%{POOSTINDOUZ AND SAFAVI-NAINI: SECRET KEY AGREEMENT IN WIRETAPPED TREE-PIN}
	%
	%
	%
	%
	%
	%
	%

	\maketitle
	
	\begin{abstract}
		
		We consider 
		the problem of multiterminal secret key agreement (SKA) 
		in wiretapped source model where
		terminals have access to samples of correlated random variables from a 
		publicly known joint probability distribution.
		The adversary has access to a 
		side information variable, 
		that is correlated with terminals' variables. 
		We focus on a special type of terminal variables in this model,  known as
		Tree-PIN,  where the relation between variables  of the terminals can be represented by a tree. 
		The study of Tree-PIN source model is of practical importance 
		as it can be realized in wireless network environments. 
		We derive the wiretap secret key capacity of Tree-PIN with independent leakage,  and give
		lower and upper bounds on the  maximum achievable secret key length in finite-length regime. 
		We then prove  
		an upper bound and a lower bound for 
		the wiretap secret key capacity of a wiretapped PIN and  
		give two conditions for which these bounds are tight. 
		We also extend our main result to two other  
		related models 
		and prove their corresponding capacities. At the end, we argue how our 
		analysis suggests that public interaction is 
		required for achieving the multiterminal WSK capacity.

	\end{abstract}

	\section{Introduction}
	\label{sec:intro}
	
	In a multiterminal secret key agreement (SKA) problem, a designated group of users (terminals) collaborate to obtain a shared secret  key (SK) such that users outside the group do not have any information about the key.
	We study the problem of SKA in  
	source model \cite{Csiszar2004a},  
	where there is a set of 
	$m$ terminals $\mc{M}={\{1,\ldots,m\}}$  and
	the goal is to  establish 
	a shared secret key among  
	a  subset   ${\mc{A} \subseteq \mc{M}}$ of terminals. 
	Terminals have access to samples of correlated random variables where
	random variable $X_j$ is observed by the $j$th terminal, 
	and $X_{\mc M} = (X_1, \ldots, X_m)$ denotes the  set %
	variables of  all terminals. 
	To obtain a shared key, 
	terminals use a public channel to exchange messages that are visible by the eavesdropper, Eve. %
	All terminals,  including the helper terminals %
	in  ${\mc{A}^c=\mc{M}\setminus \mc{A}}$,  
	cooperate to establish a %
	shared secret key. 
	Eve, will %
	see and record public messages, denoted by $\vect F$, 
	and has access to the side information $Z$ that is correlated with $X_{\mc{M}}$.

	For a %
	key agreement protocol  
	that establishes a %
	key of length $\ell$,  
	the {\em key rate}  
	is defined  for the case that the terminals'  random variables consist of {a vector of} $n$ independent and identically distributed (IID) samples of  the source distribution $P_{ZX_{\mc M}}$,  and is given  by $\ell/n$.  The  {\em key capacity} of a protocol for a given source distribution is the highest achievable  key rate associated with that distribution,   
	and {for this general case of variable $Z$,} is referred to as %
	\emph{wiretap secret key (WSK) capacity}. 
	For the special case  
	where $Z=\text{constant}$ and there is no wiretapper, the model is called {\em non-wiretapped} and the key capacity is called \emph{secret key (SK) capacity}.  
	An 
	important special case is when 
	the adversary ``wiretaps'' and their side information 
	{is obtained from a set ${\mc D\subseteq \mc A^c}$ of compromised }helper terminals.
	It is assumed that the compromised terminals of $\mc D$  
	make their RV's public, $X_{\mc D} = (X_j|~j\in\mc D)=Z$,  and remain cooperative  throughout the SKA protocol. 
	The key capacity of such %
	a source model is called \emph{private key (PK) capacity}. 
	A summary of these adversarial models %
	and their corresponding key capacities %
	are given in Table~\ref{table:key_cap_names}. 
	Single-letter expressions for SK and PK capacities of  
	multiterminal source model are known \cite{Csiszar2004a}.   
	Single-letter characterization of WSK capacity however, remains an open question %
	in general, 
	even for the case of two-party SKA (that is when ${|\mc A|=|\mc M|=2}$)  
	\cite{Csiszar2011,ElGamal2011,Narayan2016}. 
	WSK capacity of a few %
	special cases are known \cite{Csiszar2008,Poostindouz2019,Vippathalla2021}. 
	In this work, we prove the WSK capacity of  
	another special subclass of %
	multiterminal model, referred to as %
	the wiretapped Tree-PIN model with independent leakage. 
	In the following, we first give a brief overview of relevant %
	related works,
	and then outline  %
	our contributions.

	\begin{table}[!t]
		\caption[Different Types of Key Capacities Based on Different Adversarial Assumptions.]{Different Types of Key Capacities Based on The Assumption About The Adversary.}
		\label{table:key_cap_names}
		\centering
		\begin{tabular}{ccc}
			&                                   &              \\ 
			Source Model   & Eve's Side Information            & Key Capacity \\
			\hline 
			\hline
			Wiretapped     & $Z$ not known publicly            & WSK          \\
			Compromised    & $Z=X_{\mc D}$, and known publicly & PK           \\
			Non-wiretapped & $Z=\text{constant}$               & SK           \\
			\hline
		\end{tabular}
	\end{table}

	\subsection{Related Works}
	\paragraph*{\textbf{Capacity results}}
	The SKA problem for two terminals was first considered, independently, in~\cite{Ahlswede1993} and \cite{Maurer1993}. 
	The SK capacity was proved to be   $I(X_1;X_2)$ \cite{Ahlswede1993,Maurer1993}.  
	It was also %
	proved that $I(X_1;X_2|Z)$ is an achievable key rate if the terminals know 
	Eve's side information $Z$. 
	Therefore the conditional mutual information $I(X_1;X_2|Z)$  is an 
	upper bound for the WSK capacity,  %
	and {it was shown \cite{Ahlswede1993} that } it is tight if the Markov Relation $X_1-X_2-Z$ (or $X_2 - X_1 - Z$) holds. 
	\citeauthor{Csiszar2004a} extended the two-party source model of \cite{Ahlswede1993,Maurer1993} 
	to the multiterminal model and 
	proved 		
	single-letter expressions for SK and PK capacities 
	of %
	multiterminal source models %
	\cite{Csiszar2004a}.
	Similar to the two-party scenario,  it was showed that 
	multiterminal PK capacity provides %
	an upper bound on the WSK capacity. 
	The PK (and SK) capacity achieving {protocol} of \cite{Csiszar2004a} has two steps: 
	in the first step, terminals 
	communicate over the public channel to obtain omniscience,  
	that is %
	terminals in %
	$\mc D^c$ learn $X_{\mc M}$, and in
	the second step, %
	terminals in {$\mc A$} extract their copy of the key 
	from the common shared randomness $X^n_{\mc M}$.  
	While WSK capacity remains unknown in general, the characterization of WSK and also
	alternative formulations of SK  and PK  
	capacities for %
	special cases of 
	multiterminal models have been %
	studied, extending 
	the general results of \cite{Csiszar2004a}. 
	We briefly  review two of these special case models 
	that are related to our %
	work.

	The %
	Markov Tree model is a special case of the 
	general multiterminal source model that was introduced and studied in \cite{Csiszar2004a,Csiszar2008}. 
	In a non-wiretapped Markov Tree, the correlation between source variables is given by an 
	undirected tree  $G=(\mc M,\mc E)$ in which each terminal is represented by a node in $G$, 
	and for 
	any path from terminal $i_1$ to $i_f$, 
	denoted by $\mathrm{Path}(i_1\rightarrow i_f) = (e_{i_1 i_2},e_{i_2 i_3},\cdots, e_{i_{f-1} i_f})$, 
	the Markov chain $X_{i_1} - X_{i_2} - X_{i_3} - \cdots - X_{i_{f-1}} - X_{i_f}$ holds. 
	The source model is called wiretapped Markov Tree, if the source variables form a Markov Tree, and 
	the variable associated with each terminal
	is independently and partially leaked to Eve -- i.e., with respect to each $X_j$ there exists a $Z_j$ component 
	available to Eve, where $Z_j$ is {a noisy version} of $X_j$. 
	In a wiretapped Markov Tree,  
	corresponding to a %
	path from terminal $i_1$ to $i_f$  
	as above, 
	the Markov chain $Z_{i_1} - X_{i_1} - X_{i_2} - X_{i_3} - \cdots - X_{i_{f-1}} - X_{i_f} - Z_{i_f}$ holds. 
	The SK and PK capacities of 
	the 
	Markov Tree source model where derived in \cite[Example 7]{Csiszar2004a}. 
	The WSK capacity of  wiretapped  Markov Tree however remains an open problem 
	even for the case of two-party SKA 
	(i.e., when $m=2$ and $Z_1 - X_1 - X_2 - Z_2$). 
	For the case that  the variable associated with only \emph{one} of the leaf terminals 
	is leaked (i.e., $Z_i = \text{constant}, \forall i\neq j$ where $j\in \mc M$ is a leaf node of $G$),
	the WSK capacity of  the wiretapped  Markov Tree is proved in \cite[Theorem 5.1]{Csiszar2008}.

	A second %
	special case %
	of the multiterminal model %
	is the Pairwise Independent Network (PIN) model 
	\cite{Ye2007}, %
	inspired by a wireless setting where each pair of terminals can obtain %
	correlated variables from the channel connecting the two.
	Source variables in PIN are defined by an undirected graph $G=(\mc M,\mc E)$ with node (vertex) set $\mc M$ and edge set $\mc E$, where for %
	an edge $e_{ij}=e_{ji}\in\mc E$ between $i$ and $j$ ($i\neq j\in\mc M$), there exists a variable $V_{ij}$ accessible to terminal $i$, and a second variable $V_{ji}$ (correlated with $V_{ij}$) accessible to terminal $j$. The set of all ``reciprocal correlated pairs'' of variables (i.e., $\{(V_{ij},V_{ji})|~e_{ij}\in\mc E\}$) are assumed mutually independent\footnote{This means that $P_{X_{\mc M}} = \prod_{e_{ij}} P_{V_{ij}V_{ji}}$.}.
	An upper bound on the SK capacity of PIN is given in~\cite{Ye2007}, and a capacity achieving  SKA protocol  %
	when $\mc A=\mc M$, or when $|\mc A|=2$, was proposed %
	in~\cite{Nitinawarat2010c}.
	The PIN model has been well studied %
	\cite{Chan2011, Nitinawarat2010a,Kashyap2013,Xu2015}, 
	and has inspired other %
	multiterminal models %
	\cite{ChungChan2010,Courtade2016,Chan2018,Chan2018c}.
	An important subclass of the PIN model is defined when the defining graph $G$ is 
	an undirected tree. This model is called %
	Tree-PIN \cite{Poostindouz2019}. 
	{In this work, we  focus  on  %
		wiretapped Tree-PIN model.} %
	We observe %
	that a %
	non-wiretapped Tree-PIN is a non-wiretapped Markov Tree, but the converse is not true. 
	Similarly, we will show %
	that, every wiretapped Tree-PIN with independent leakage is a 
	wiretapped Markov Tree, but the converse does not necessarily hold.

	\paragraph*{\textbf{Finite-length performance}}
	The finite-length analysis of %
	coding schemes %
	has found much attention in recent years %
	\cite{Hayashi2008,Hayashi2009,Polyanskiy2010,Hayashi2013a,Tyagi2014,Hayashi2016,Tyagi2017,Hayashi2019}. Such analysis is important theoretically, and also 
	in practice.  
	While SKA %
	key capacities capture the best asymptotic efficiency of a source model, in practice one needs to obtain bounds on the achievable key length when a  finite number ($n$) of source samples is available. %
	For wiretapped multiterminal source model, a single-shot ($n=1$) upper bound on the key length is given in \cite{Tyagi2014}.
	Finite-length upper and lower bounds for two-party SKA, when $X_1 - X_2 - Z$ holds, have been obtained in \cite{Hayashi2016}.
	For multiterminal key agreement when Eve has no side information, a finite-length lower bound (of the form $nC_{SK} - \mc O(\sqrt{n\log n})$) is given in \cite{Tyagi2017}.

	\paragraph*{\textbf{Communication and computation costs}} 
	The %
	key rate measures efficiency   of SKA protocol in %
	using the initial correlated randomness, 
	it is also important in practice to measure communication and computation costs. 

	The {\em computational efficiency}  %
	of an SKA protocol is in terms of the computational complexity of  
	terminals' operations. 
	An SKA protocol is considered computationally efficient if its computational complexity is quasi-linear in $n$, and is of the form $\mc O(n \log n)$.  
	The known computationally efficient capacity achieving SKA protocols are given in %
	\cite{Renes2013,Chou2015a,Holenstein2005,Holenstein2006a,Poostindouz2021}. In most cases the protocols have not been analysed for 
	finite-length performance. 
	
	{\em Communication efficiency} of an SKA protocol is measured using (i) the  %
	public communication, that for asymptotic case can be measured in terms of asymptotic rate $r_{PC}$, and for finite-length case, in terms of the total number of  bits, of the public communication, %
	and  (ii) the total number of rounds $N_{PC}$ of public discussion.    
	We define these measures in Section~\ref{sec:Multiterminal_SKA}. %
	Informally, the asymptotic rate of public communication measures %
	the number of  bits of public communication that is used per each observation bit.
	{In a round of public discussion the messages of the terminals %
		only depend on the private samples  of the corresponding  terminals, %
		and the public messages of the previous rounds. }
	The SKA protocols in \cite{Ahlswede1993,Csiszar2004a,Poostindouz2019} are noninteractive: %
	they have %
	one round of public communication, $N_{PC}=1$. 
	Interactive SKA's have two or more rounds of public communication; e.g., the SKA protocol of \cite{Nitinawarat2010a} has $N_{PC}=2$ and the two-party SKA protocol of \cite{Hayashi2016} has $N_{PC}\in\mc O(n)$. 
	{For source models, %
		the minimum asymptotic  rate of public communication, and the minimum number of public discussion rounds that are required 
		for achieving the key capacity, are important parameters of the system.} 
	{For SK and PK capacity, the result of  \cite{Csiszar2004a} 
		implies that the minimum asymptotic  rate of public communication for omniscience, is 
		an %
		upper bound for the minimum asymptotic rate of public communication 
		that is required for achieving the 
		corresponding capacity.} %
	The minimum asymptotic  rate of public communication for SKA for various source models were studied in 
	\cite{Tyagi2013,Mukherjee2015,Mukherjee2016,Chan2017}.

	\subsection{Our Contributions}
	In this work, %
	we introduce and study 
	\emph{wiretapped PIN model} and \emph{wiretapped Tree-PIN model}. 
	The wiretapped PIN model with independent leakage %
	is defined 
	as a PIN with an underlying %
	undirected graph $G=(\mc M,\mc E)$ where legitimate  terminals are represented 
	by vertices (nodes) of the graph. %
	An undirected edge between the nodes $i$ and $j$ 
	is represented by $e_{ij} \in \mc E$.   
	Corresponding to each %
	edge $e_{ij} \in\mc E$, there exists a variable $V_{ij}$ 
	accessible to terminal $i$, 
	and a second variable $V_{ji}$ accessible to terminal $j$. 
	Also, with respect to each edge $e_{ij}\in\mc E$ Eve 
	has access to a component variable $Z_{ij}$, and %
	the set of all triplets of variables 
	$\{(V_{ij},V_{ji},Z_{ij})|~e_{ij}\in\mc E\}$ 
	are assumed mutually independent,
	and for each  $e_{ij}\in\mc E$ 
	either $V_{ij}-V_{ji}-Z_{ij}$ or $V_{ji}-V_{ij}-Z_{ij}$ hold\footnote{Only one wiretapped component $Z_{ij}$ is accessible to Eve for each connection $e_{ij}\in\mc E$-- e.g., $Z_{ij}-V_{ij}-V_{ji}-Z_{ji}$ is not allowed.}. 
	Since $G$ is undirected, we have %
	$Z_{ij}=Z_{ji}$,  
	and %
	denote the adversary's side information by $Z=(Z_{ij}|i<j)$. 
	A wiretapped Tree-PIN is a special case of wiretapped PIN for which the corresponding  %
	undirected graph $G$ is a tree. 
	A simple example of such wiretapped Tree-PIN is depicted in Figure~\ref{fig:ex1}.

	\paragraph*{\textbf{Main results}} 
	We derive the WSK capacity of wiretapped Tree-PIN with independent leakage as described above, and %
	present an SKA protocol that achieves this capacity. 
	Our SKA protocol has two rounds of public communication ($N_{PC}=2$) and 
	as shown in Remark~\ref{remark:R_SK-vs-Capacity-Tree},
	has a lower asymptotic public communication rate than other SKA
	protocols (including %
	the protocol in %
	\cite{Csiszar2004a}) that have the two steps of achieving omniscience followed by privacy amplification.  
	We note that the adversary in our model is more powerful 
	than the adversary in the wiretapped Markov Tree model of \cite[Theorem 5.1]{Csiszar2008}, 
	as in the capacity result of \cite[Theorem 5.1]{Csiszar2008} Eve only 
	wiretaps one terminal's variable, while in our model of wiretapped Tree-PIN 
	Eve wiretaps all terminals' variables 
	by wiretapping all pairs of correlated variables $(V_{ij}, V_{ji})$. 
	For the case of two-party SKA, our capacity result also reduces to the %
	result in \cite{Ahlswede1993} when   
	$X_1 - X_2 - Z$ (or when $Z - X_1 - X_2$) holds. %
	A simplified version of the wiretapped Tree-PIN model where 
	it is assumed that $V_{ij} = V_{ji}$, was studied and its capacity was derived 
	in our previous work presented in \cite{Poostindouz2019}.

	\begin{figure}[t]
		\centering
		
		\begin{adjustbox}{width=0.8\linewidth}
			\begin{tikzpicture}[font=\bfseries\Huge]
				\tikzstyle{place}=[circle,draw=black!90,fill=blue!10,very thick,inner sep=3pt, minimum size=42pt,line width=2pt,font=\bfseries\Huge]
				\tikzstyle{eve}=[circle,draw=black!90,fill=red!10,very thick,inner sep=3pt, minimum size=52pt,line width=2pt]
				{\color{black}
					\node (n1) at ( -6,0) [place] {1};
					\node (n2) at ( 0,0)  [place] {2};
					\node (n3) at ( 6,0)   [place] {3};
					\node (n4) at ( 12,0)  [place] {4};

					\node (eve) at (3, -6) [eve] {Eve};

					\node (v1) at (16,0)  [right] {$X_1 = (V_{12})$};
					\node (v2) at (16,-1.1) [right] {$X_2 = (V_{21}, V_{23})$};
					\node (v3) at (16,-2.2) [right] {$X_3 = (V_{32}, V_{34})$};
					\node (v4) at (16,-3.3) [right] {$X_4 = (V_{43})$};
					\node (e)  at (16,-6) [right] {$Z = (Z_{12}, Z_{23}, Z_{34})$};

					\draw[-,black!60!green,line width=6pt,round cap-round cap,shorten <=1mm,shorten >=1mm] (n1) -- (n2) node [pos=0.01, above right] {$V_{12}$}  node [pos=0.99, above left] {$V_{21}$};
					\draw[snake=snake, line before snake=10mm, line after snake=10mm,segment aspect=5, segment amplitude=5pt,->,black!60!green,line width=3pt,round cap-round cap,shorten <=2mm,shorten >=1mm] (n2.south) -- (eve.north west) node [pos=0.8, below left] {$Z_{12}$};
					
					\draw[-,black!60!blue,line width=6pt,round cap-round cap,shorten <=1mm,shorten >=1mm] (n3) -- (n2) node [pos=0.01, above left] {$V_{32}$}  node [pos=0.99, above right] {$V_{23}$};
					\draw[snake=snake, line before snake=10mm, line after snake=10mm,segment aspect=5, segment amplitude=5pt,->,black!60!blue,line width=3pt,round cap-round cap,shorten <=2mm,shorten >=1mm] (n2.south east) -- (eve.north) node [pos=0.8, above right] {$Z_{23}$};

					\draw[-,red!70!yellow,line width=6pt,round cap-round cap,shorten <=1mm,shorten >=1mm] (n3) -- (n4) node [pos=0.01, above right] {$V_{34}$}  node [pos=0.99, above left] {$V_{43}$};
					\draw[snake=snake, line before snake=10mm, line after snake=10mm,segment aspect=5, segment amplitude=5pt,->,red!70!yellow,line width=3pt,round cap-round cap,shorten <=2mm,shorten >=1mm] (n4.south) -- (eve.north east) node [pos=0.8, below right] {$Z_{34}$};

				}%
				
			\end{tikzpicture}
		\end{adjustbox}
		\caption[An example of wiretapped Tree-PIN with independent leakages.]{An example of wiretapped Tree-PIN with independent leakages defined over $\mc M =\{1,2,3,4\}$ and $\mc E = \{e_{12},e_{23},e_{34}\}$. The solid lines (edges) show the independent connections between terminals, and the curly lines (with the same color) show the corresponding independent wiretapping RV's of Eve. The RV associated with each terminal $i\in\mc M$ is of the form $X_i = (V_{ij}|~e_{ij}\in\mc E)$. 
			In this example, the following Markov relations hold $V_{12}-V_{21}-Z_{12}$, $V_{32}-V_{23}-Z_{23}$, $V_{34}-V_{43}-Z_{34}$. Eve's RV is a collection of independent wiretapped components, i.e., $Z=(Z_{12}, Z_{23}, Z_{34})$ 
		}
		\label{fig:ex1} %
	\end{figure}

	In Section~\ref{sec:FLA-notsym},  we give  a  
	finite-length upper bound and three finite-length lower bounds 
	for the maximum achievable secret key length of a wiretapped Tree-PIN, 
	where each lower bound is due to a different concrete construction of our SKA protocol. 
	We will 
	discuss and compare the  three construction approaches in terms of their corresponding 
	lower bounds, their computational complexity,  and their communication costs.   
	Our SKA protocol is capacity achieving; however, 
	its achieved  key length for $n$ source samples 
	(finite-length analysis) does not match 
	the 
	finite-length upper bound, and the
	construction of a capacity achieving protocol 
	that achieves the finite-length upper bound 
	of wiretapped Tree-PIN remains open.
	\paragraph*{\textbf{Related models}}
	Tree-PIN model has attracted attention over the past years as it can be 
	extended and used to study a number of other related practically important models. %
	In Section~\ref{sec:disc}, 
	we extend our main capacity result for wiretapped Tree-PIN to the following more general scenarios.  
	For \emph{wiretapped PIN models}  where $G$ %
	can have loops, we show that, a SKA protocol based  on Steiner Tree Packing can achieve the WSK capacity when  $\mc A=\mc M$ or $|\mc A|=2$. 
	This is similar to the results obtained in \cite{Poostindouz2019,Nitinawarat2010c}, 
	for SKA in non-wiretapped PIN.   
	Next, we note that an important open problem in SKA is 
	finding the WSK capacity of the two-party model when Markov Relation 
	$Z_1 - X_1 - X_2 - Z_2$ 
	holds, where $Z=(Z_1,Z_2)$ is Eve's wiretapped side information \cite{Csiszar2008}. 
	We 
	extend our Tree-PIN to the case where %
	corresponding to each $e_{ij}\in\mc E$, 
	we have $V_{ij}^a - V_{ji}^a - Z_{ij}^a$ and $Z_{ij}^b - V_{ij}^b - V_{ji}^b$, 
	which implies $Z_{ij}^b - X_i - X_j - Z_{ij}^a$.  
	For $|\mc M|=2$, this extended model is an special case of the 
	open problem where Markov relation $Z_1 - X_1 - X_2 - Z_2$ 
	holds. 
	{We prove the WSK capacity of this extended model which is (naturally) higher than 
		the WSK capacity of a simple Tree-PIN -- as terminals have access to more correlated sources.} 
	Lastly, we also prove the key capacity of a PIN model in which not only source variables are wiretapped but also 
	one of the terminals is compromised and is not cooperating. 
	{In this case we show that the WSK capacity reduces to the WSK capacity of the 
		associated model where the compromised terminal and terminals' variables associated with the 
		the compromised terminal are removed (ignored.)}

	\paragraph*{\textbf{Need for interaction}}
	\citeauthor{Csiszar2004a} proved that %
	SK and PK capacities can be achieved
	noninteractively \cite{Csiszar2004a}.  	
	For some special cases also WSK capacity can be achieved noninteractively \cite{Ahlswede1993,Poostindouz2019}. 
	Our proposed capacity achieving SKA protocol %
	for wiretapped Tree-PIN 
	is \emph{interactive}. 
	In Section~\ref{sec:interaction}, we discuss the number of public communication rounds %
	that is required for achieving the WSK capacity.  
	{We analyze known models and constructions \cite{Ahlswede1993, Gohari2010, Hayashi2016} and study a number of examples that suggest that in general achieving the WSK capacity requires interaction.}
	Proving this result however remains %
	an interesting open question for future research.

	\subsection{Organization}
	The rest of this paper is organized as follows. 
	We review %
	security basic notions and definitions in~Section~\ref{sec:Multiterminal_SKA}, and %
	present our main result  in~Section~\ref{sec:main-res}. %
	Section~\ref{sec:FLA-notsym} gives finite-length analysis of wiretapped Tree-PIN, and 
	Section~\ref{sec:disc} is on extensions of our main result including for 
	the wiretap secret key capacity of PIN. 
	Section~\ref{sec:interaction} discusses the problem of whether interaction is necessary 
	to attain the WSK capacity, and Section~\ref{sec:concl_treepin} concludes the paper.

	\section{Multiterminal Source Model for SKA}
	\label{sec:Multiterminal_SKA}

	{\color{black}

		In the general multiterminal source model \cite{Csiszar2004a}, %
		we have  %
		a set of $m$ terminals  %
		denoted by $\mc{M} =[m]=\{1,\ldots, m \}$, and %
		each terminal $j\in[m]$ has access to a random variable $X_j$. %
		We denote the collection of $m$ correlated random variables $X_1, \ldots, X_m$ 
		by $X_{\mc{M}}=(X_1, \ldots, X_m)$. %
		Terminals collaborate by %
		public discussion  %
		over a public channel that is %
		reliable and %
		authenticated.
		A  message that is sent by a terminal $j$ is %
		a function of  the terminal's observations of  $X_j$, and the previous public messages. 
		Public discussion happens over a finite number of rounds, denoted by, $N_{PC}$. 
		We denote by $\vect{F}$ the set of all messages sent over the public channel.

		Eve %
		has  access to  the %
		side information $Z$  which is correlated with %
		$X_\mc{M}$,  and has full read  access  to %
		public messages $\vect{F}$. Eve is a passive adversary, which means they will not %
		change, %
		or block public messages communicated messages.
		The joint distribution $P_{X_\mc{M}Z}$ is publicly known. %
		We denote the multiterminal source model by $P_{X_\mc{M}Z}$ or the 
		discrete multiple memoryless source (DMMS) notation $(X_{\mc M}, Z)$.

		Let $\mc A\subseteq \mc M$ be %
		the set of terminals  who want to establish %
		a shared secret key  %
		$K$. The key %
		need not %
		be concealed %
		from the helper terminals in $\mc{A}^c$. 
		The secret key $K$ is secure against Eve  if it satisfies the reliability and secrecy conditions.

		\begin{definition}\label{def:SK-Key}
			Consider %
			a source model $(X_\mc{M},Z)$  with adversary's side information, $Z$,  and
			$\mc A\subseteq \mc M$ denoting the  set of terminals that will share a %
			key  $K \in \mc{K}$. %
			The key  is %
			an $(\epsilon,\sigma)$-Secret Key (in short $(\epsilon,\sigma)$-SK) for $\mc A$, %
			if there exists a protocol with public communication $\vect F$, and output RVs $\{K_j\}_{j\in\mc{A}}$ such that %
			\begin{align}
				\text{(reliability)}\quad & \pr{K_j=K}\geq 1-\epsilon, \quad \forall j\in\mc A,             \\
				\text{(secrecy)}\quad     & \sd\left((K,{\vect{F}} ,Z),(U,{\vect{F}},Z)\right) \leq \sigma, 
			\end{align}
			where $\sd$ denotes the statistical distance and $U$ is the uniform probability distribution over alphabet $\mc{K}$. The length of a key $K$ is given by $\log |\mc K|$. 
		\end{definition}

		\begin{definition}\label{def:SK-length}
			For a %
			source model $(X_\mc{M},Z)$  with adversary's side information, $Z$,  and
			$\mc A\subseteq \mc M$ denoting the  set of terminals that want to share a secret key, 
			let 
			$S_{\epsilon,\sigma}(X_\mc{A}|Z)$ denote the maximum length $\log |\mc K|$ of all the $(\epsilon,\sigma)$-SKs 
			that can be established for %
			$\mc A\subseteq \mc M$. 
		\end{definition}

		\paragraph*{\textbf{SKA for IID variables}}
		Consider a source model $(X_{\mc M},Z)$ described by $P_{X_{\mc M}Z}$, where all
		terminals  cooperate for %
		to establish a shared secret key for terminals in $\mc A$. 
		{To increase the key length, %
			terminal $j\in\mc M$ use %
			a vector, $X_j^n$, of $n$ independent and identically distributed ($n-$IID) samples of $X_j$.}    
		Let $\Pi$ be an SKA protocol family that, for any $n$, establishes a secret key $K^{(n)}$ for $\mc A\subseteq \mc M$. 
		The public communication of $\Pi$, denoted by $\vect F =  \vect F(\Pi)$, can be interactive %
		and %
		be comprised of $N_{PC}(\Pi)\geq 1$ rounds where in each round $t\in[ N_{PC}(\Pi) ]$ each terminal $j$ %
		sends up to one public message $F_{tj}$.
		A message %
		is a function of $X_j^n$ and all public messages of the previous rounds that is denoted by $F^{t-1}$, and %
		so $F_{tj} = F_{tj}(X_j^n , F^{t-1})$.   
		We denote all messages of round $t$  by $F_t=(F_{t1},\ldots,F_{tm})$.
		{The public messages of terminals in each round do not depend on other messages of that round, and can be sent in any order.
			The maximum number of the rounds of public communication, $N_{PC}(\Pi)$, may in general %
			be a function of $n$. 
			The SKA protocol $\Pi$ with %
			public communication $\vect F$ is %
			called \emph{noninteractive} if $N_{PC}(\Pi)=1$, meaning that in one round each terminal %
			sends up to a single public message, and $\vect F= (F_1,\ldots,F_m)$, where $F_{j} = F_{j}(X_j^n)$.
			
			The \emph{asymptotic public communication rate} of $\Pi$ is  defined by 
			$$r_{PC}(\Pi) = \limsup_{n \rightarrow \infty} \frac{1}{n} \log (\supp(\vect F(\Pi))),$$ 
			where $\vect F(\Pi)$ is the public communication of $\Pi$. 
			Public communication cost of %
			$\Pi$ can be quantified by $r_{PC}(\Pi)$ and $N_{PC}(\Pi)$. 
		}

		Suppose SKA protocol $\Pi$ establishes an $(\eps_n,\sigma_n)-$SK $K^{(n)}$ for a subset ${\mc A\subseteq \mc M}$, and  
		let $\ell_{\Pi}(n) = \log |\mc K^{(n)}|$ denote the length of $K^{(n)}$.  
		The key rate of $\Pi$ for $n-$IID observations is given by  $\sfrac{1}{n} \ell_{\Pi}(n)$, and $r_K(\Pi) = \liminf_{n \rightarrow \infty} \sfrac{1}{n} \ell_{\Pi}(n)$ is called the \emph{asymptotic key rate} of $\Pi$. 
		The {asymptotic key rate} %
		$r_K(\Pi)$  is %
		\emph{achievable} if 
		$\lim_{n\to\infty} \eps_n = \lim_{n\to\infty} \sigma_n = 0$.    
		The {\em key capacity of a source model} is %
		the maximum of all  
		achievable asymptotic key rates of SKA protocols for the model. See Definition~\ref{def:SK-Capacity}.   
		For an integer $n\in\NN$, and %
		$\eps,\sigma\in [0,1)$, 
		define  $S_{\epsilon,\sigma}(X_\mc{A}^n|Z^n)$ to be the maximum length of all $(\epsilon,\sigma)$-SK protocols 
		for establishing a secret key for %
		${\mc A\subseteq \mc M}$.

		\begin{definition}[Key Capacity -- Definition 17.16 of \cite{Csiszar2011}]\label{def:SK-Capacity}
			Consider multiterminal SKA for a subset  $\mc A\subseteq \mc M$ in  
			a the %
			source model $(X_\mc{M},Z)$ for %
			the joint distribution $P_{X_{\mc M}Z}$,  
			where $Z$ denotes %
			Eve's  side information  %
			about $X_{\mc M}$.
			A real number $R\geq 0$ is an achievable SK rate if there exists an SKA protocol that for a given %
			$n$ establishes an $(\eps_n,\sigma_n)-$SK $K\in\mc K$ where $\lim_{n\to\infty} \eps_n = 0$, $\lim_{n\to\infty} \sigma_n = 0$, and $\liminf_{n\to\infty} \frac{1}{n}\log|\mc K| = R$. The maximum of all achievable SK rates
			is called the key capacity of the %
			model.   
		\end{definition}

		\paragraph*{\textbf{SK, PK, and WSK Capacities}}
		When $Z=\text{constant}$ (i.e., independent of $X_{\mc M}$),  
		the capacity is called SK capacity and is denoted  by $C_{SK}^{\mc A}(P_{X_\mc{M}})$. %
		When $Z=X_{\mc{D}}=(X_j~|~j\in\mc{D})$ with $\mc D$ being
		the set of (known)  compromised %
		terminals, %
		the capacity  is called PK capacity and is denoted by $C_{PK}^{\mc A|\mc D}(P_{X_\mc{M}})$. 
		In this case it is assumed that $Z$ is known publicly. 
		{In the general case when
			the side information $Z$ %
			is correlated with $X_{\mc M}$ and is not known by the terminals,} 
		the  
		key capacity  is called {WSK capacity} and is denoted by $C_{WSK}^{\mc A}(P_{X_\mc{M} Z})$. 
		An SKA protocol $\Pi$ is capacity achieving  for a source model %
		if $r_K(\Pi)$ is equal to the key capacity of the source. %

		For a source model $(X_{\mc M},Z)$  with the %
		joint probability distribution $P_{X_{\mc M}Z}$, 
		let
		$R_{SK}(X_{\mc M})$, $R_{PK}(X_{\mc M}|X_{\mc{D}})$, and $R_{WSK}(X_{\mc M}|Z)$  
		denote the minimum public communication rate %
		to achieve 
		the SK, PK, and WSK capacities,
		respectively.  
		These quantities give the minimum public communication cost of the SKA, %
		and
		are often referred to as \emph{communication complexity} of $(X_{\mc M},Z)$ \cite{Mukherjee2015,Mukherjee2016,Chan2017}.  
		Characterizations of $R_{SK}(X_{\mc M})$ for two-party SKA, and for a special case of  PIN models, 
		are given in \cite{Tyagi2013} and \cite{Chan2017}, receptively. 
		An SKA protocol $\Pi$ that achieves the WSK capacity of a source model $(X_{\mc M},Z)$  implies 
		$R_{WSK}(X_{\mc M}|Z) \leq r_{PC}(\Pi)$.
		A similar statement %
		holds for the case of SK and PK.

		The single-letter characterization of SK and PK capacities of the general multiterminal source model was derived in \cite{Csiszar2004a}. Next Theorem states this result.

		\begin{theorem}[PK Capacity \cite{Csiszar2004a}]\label{thm:PK-Cap-CN04}
			In a given source model ${X_\mc{M}}$ for sharing a secret key among terminals in $\mc{A}\subsetneq \mc{M}$, with compromised terminals $\mc{D}\subseteq \mc{A}^c$, the PK capacity is
			\begin{IEEEeqnarray}{rCl}\label{eq:PK}
				C_{PK}^{\mc A|\mc D}(P_{X_\mc{M}}) &=& H(X_{\mc{M}}|X_{\mc{D}}) - R_{CO}(X_\mc{A}|X_\mc{D}),\IEEEeqnarraynumspace
			\end{IEEEeqnarray} where ${R_{CO}(X_\mc{A}|X_\mc{D})= \min\limits_{R_{\mc{D}^c}\in\mc{R}_{CO}}\myop{sum}(R_{\mc{D}^c})}$ %
			and %
			\begin{IEEEeqnarray*}{l}
				\mc{R}_{CO} =
				\left\{ R_{\mc{D}^c} | \myop{sum}(R_\mc{B}) \geq  H(X_{\mc{B}}|X_{\mc{B}^c}),~\forall \mc{B}\subset \mc{D}^c, \mc{A}\nsubseteq \mc{B}  \right\}.
			\end{IEEEeqnarray*}
		\end{theorem}

		\begin{remark}
			{Equation~\eqref{eq:PK} also leads 
				to the SK capacity when ${\mc D = \emptyset}$.}  
			The achievability result is based on a protocol  in which first, %
			the compromised terminals (that are %
			cooperative) {publicly reveal} their observed random variables (as it is the assumption for the PK capacity,) and then the rest of the terminals in $\mc D^c$ communicate over the public channel to obtain omniscience (i.e., the state that terminals in $\mc D^c$ learn each other's initial observations). Finally, %
			terminals in $\mc A$ extract the key from the common shared randomness $X^n_{\mc M}$.  
			It was noted that this SKA protocol is noninteractive; meaning that, $N_{PC}=1$, and $\vect F= (F_1,\ldots , F_m)$, where $F_j = X_j^n$ for all $j\in\mc D$ and $F_j = F_j (X_j^n)$ for all $j\in\mc D^c$. See the achievablity part of the proof of Theorem 2, in Section IV of \cite{Csiszar2004a}. The asymptotic public communication rate of this SKA protocol is given by $r_{PC}=R_{CO}(X_\mc{A}|X_\mc{D})$, which 
			implies that $R_{PK}(X_\mc{M}|X_\mc{D})\leq R_{CO}(X_\mc{M}|X_\mc{D})$ (and $R_{SK}(X_\mc{M}) \leq R_{CO}(X_\mc{M})$). 
		\end{remark}

	}%

	Unfortunately, the WSK capacity of the general source model as defined previously, remains an open problem even for the special case of two terminals (${|\mc M|=2}$) \cite{ElGamal2011}. For the case of two-party SKA, the source model WSK capacity is upper bounded by  $I(X_1;X_2|Z)$, which is proved to be a tight bound under the additional assumption that the Markov Chain $X_1 - X_2 - Z$ holds \cite{Ahlswede1993,Maurer1993}. 
	As was mentioned before, the multiterminal WSK capacity is only known for a few limited special cases \cite{Csiszar2008,Poostindouz2019}. 
	However, PK capacity (see Theorem \ref{thm:PK-Cap-CN04}) gives a general upper bound to the WSK capacity. We show in the next section that this upper bound is tight for the case wiretapped Tree-PIN.

	\begin{lemma}[Lemma 5.1 of \cite{Csiszar2008}]\label{thm:PKisUpp}
		For a given  wiretapped source model $(X_{\mc M},Z)$, %
		let $C_{WSK}^{\mc A}(P_{X_\mc{M} Z})$ denote the WSK capacity of the wiretapped model. Let $C_{PK}^{\mc A|\{m+1\}}(P_{X_\mc{M} Z})$ be the PK capacity of an auxiliary model with $m+1$ terminals such that $X_j=X_j$ for all $j\leq m$, and $X_{m+1} = Z$, where terminal $m+1$ is compromised (i.e., $\mc D = \{m+1\}$). For any given wiretapped model such auxiliary model can be defined. By definition of the PK capacity we have $C_{WSK}^{\mc A}(P_{X_\mc{M} Z}) \leq C_{PK}^{\mc A|\{m+1\}}(P_{X_\mc{M} Z}).$
	\end{lemma}

	\section{WSK Capacity of Tree-PIN}\label{sec:main-res}
	
	Here, we first define the wiretapped PIN (Pairwise Independent Network) and wiretapped Tree-PIN models. The non-wiretapped PIN model was first defined in \cite{Ye2007} and its SK capacity was later studied in \cite{Nitinawarat2010c}. Let ${G} =(\mc M, \mc E)$ be an undirected graph.
	We denote the edge that connects the nodes $i$ and $j$ by $e_{ij}$, and
	{\em assume $e_{ij}=e_{ji}$. } In a graph ${G} =(\mc M, \mc E)$, we denote %
	the neighbours %
	of a node $j\in \mc M$ by %
	$\Gamma(j)=\{i~|~i\in\mc M, %
	e_{ij}\in\mc E\}$.

	\begin{definition}[Wiretapped PIN \& Wiretapped Tree-PIN]\label{def:Tree-PIN}
		A set of $m$ terminals form a PIN if there exists a tree $G=(\mc M,\mc E)$ with $\mc M=[m]$ such that the RV of any terminal $j\in \mc M$ can be represented by $X_j=(V_{ji}|~i \in\Gamma(j))$, %
		where all pairs of RVs in $\{(V_{ij},V_{ji}) |~ i<j \text{~and~} e_{ij}\in\mc E \}$ are mutually independent. Note that $V_{ij}\neq V_{ji}$. 
		A PIN model is called wiretapped if Eve has access to side information $Z$ which is correlated with all terminals' variables. That is, the correlation between $Z$ and all $V_{ij}$ variables can be in any general form. 
		A wiretapped PIN is called with \emph{independent leakage} if Eve's variable is of the form $Z=(Z_{ij}|~i<j)$, such that
		the set of all triplets of variables $\{(V_{ij},V_{ji},Z_{ij})|~e_{ij}\in\mc E\}$ are mutually independent
		and for each  $e_{ij}\in\mc E$ 
		either $V_{ij}-V_{ji}-Z_{ij}$ or $V_{ji}-V_{ij}-Z_{ij}$ hold. 
		A   Tree-PIN is a PIN model for which $G$ is an undirected tree. 
		A (Tree-)PIN model is called non-wiretapped if $Z=\text{constant}$. 
	\end{definition}

	In our model of wiretapped (Tree-)PIN with independent leakage, Eve has wiretapped side information correlated with every component variable of every terminal, and thus our wiretap model not only strongly resembles the case of general wiretapped PIN model it is also a special case of the wiretapped Markov Tree model for which the WSK capacity is still unknown \cite{Csiszar2008}.  
	The main results of this work are giving the WSK capacity of wiretapped Tree-PIN with independent leakage for any $\mc A$ (Theorem~\ref{thm:Tree-PIN}), and wiretapped PIN with independent leakage for $\mc A=\mc M$ or $|\mc A| =2$ (Corollary~\ref{thm:tightness-AeqM}). 
	These results are more general than previous results on wiretapped multiterminal models. 
	We will compare our results with the aforementioned past results in Section \ref{sec:disc}.  
	In this section, we focus on wiretapped Tree-PIN. 
	The WSK capacity of  wiretapped Tree-PIN is given by the following theorem\footnote{The proof for a special case of Theorem~\ref{thm:Tree-PIN} when $V_{ij} = V_{ji}$ was presented in ISIT 2019 \cite{Poostindouz2019}. An extension of this special model to the case of finite linear sources \cite{Chan2011} with a linear wiretapper was studied in \cite{Vippathalla2021}.}.

	\begin{theorem}\label{thm:Tree-PIN}%
		WSK capacity of a given wiretapped Tree-PIN ${(X_{\mc M},Z)}$ with independent leakage, defined as in Definition~\ref{def:Tree-PIN},
		for %
		any subset $\mc A\subseteq\mc M$ is given by %
		\begin{equation}
			C_{WSK}^{\mc A}(P_{X_\mc{M} Z}) = \min_{\substack{i,j\in\mc M \\ \mathrm{~s.t.~}e_{ij}\in\mc E_{\mc A}}} I(V_{ij};V_{ji}|Z_{ij}),
		\end{equation} 
		where  
		$G_{\mc A} = (\mc M_{\mc A}, \mc E_{\mc A})$ is the subgraph of $G$ with the smallest number of 
		edges  connecting all nodes of $\mc A$.
	\end{theorem}

	We emphasis that the WSK capacity of a more general wiretapped PIN model in which $Z=(Z_{ij}|~i<j)$ and for any $i$ and $j$ the Markov relation $V_{ij}- V_{ji} - Z_{ij}$ does not necessarily hold remains an open problem, even for the case of two-party SKA, $m=2$.

	\begin{IEEEproof}[Proof of Theorem~\ref{thm:Tree-PIN}] 
		The proof is in two parts: (i)~the converse, and (ii)~the achievability. In the converse part of the proof we prove an upper bound on WSK capacity, that is given by Lemma \ref{lemma:Upper-bound-Tree}.
		
		\begin{lemma}[The~converse]\label{lemma:Upper-bound-Tree}
			For a Tree-PIN ${(X_{\mc M},Z)}$ %
			defined as in Definition~\ref{def:Tree-PIN}, %
			we have
			\begin{equation*}%
				C_{WSK}^{\mc A}(P_{X_{\mc{M}} Z}) \leq 	C_{PK}^{\mc A|\{m+1\}}(P_{X_\mc{M} Z})  = \min_{\substack{i,j\in\mc M \\ \mathrm{~s.t.~}e_{ij}\in\mc E_{\mc A}}} I(V_{ij};V_{ji}|Z_{ij}),
			\end{equation*}
			where $G_{\mc A} = (\mc M_{\mc A}, \mc E_{\mc A})$ is the subtree of $G$ with the least number of edges that connects all nodes of $\mc A$ and dummy terminal $m+1$ represents the adversary.
		\end{lemma}

		In the achievability (direct) part we prove that the above upper bound is indeed achievable. That is given by Lemma \ref{lemma:Ach-Tree}. 
		
		\begin{lemma}[The~achievability]\label{lemma:Ach-Tree}
			For a wiretapped Tree-PIN ${(X_{\mc M},Z)}$ defined by $G=(\mc M,\mc E)$,  and $P_{ZX_{\mc M}}$, 
			and for any subset $\mc A\subseteq \mc M$, 
			the largest asymptotically achievable key rate of 
			SKA protocol~\ref{prot:SKA-Tree-PIN} 
			is given by
			\begin{align*}
				r_{K}(\vect{\Pi_{TP}}) =  \min_{\substack{i,j\in\mc M \\ \mathrm{~s.t.~}e_{ij}\in\mc E_{\mc A}}}  I(V_{ij};V_{ji}|Z_{ij}).
			\end{align*}
		\end{lemma}

		The proof of Theorem~\ref{thm:Tree-PIN} is immediately   complete by Lemmas \ref{lemma:Upper-bound-Tree} and \ref{lemma:Ach-Tree}. 
	\end{IEEEproof}

	\subsection{Proof Sketch of the Converse and Achievability} 
	
	In the following, we give an outline of the proof of the converse, and explain how protocol~\ref{prot:SKA-Tree-PIN}  of Lemma   \ref{lemma:Ach-Tree} achieves the key capacity. The full proofs of 
	Lemmas \ref{lemma:Upper-bound-Tree} and \ref{lemma:Ach-Tree} are given in Appendix \ref{sec:app:Tree-PIN-AnotM} and Appendix \ref{sec:app:Ach}, respectively.

	\paragraph*{\textbf{The Converse}}
	For simplicity, assume $\mc A=\mc M$. Also, recall that by Lemma~\ref{thm:PKisUpp}, we have  $C_{WSK}^{\mc M}(P_{X_\mc{M} Z}) \leq C_{PK}^{\mc M|\{m+1\}}(P_{X_{\mc M} Z})$, and due to Theorem~\ref{thm:PK-Cap-CN04} we know that $C_{PK}^{\mc A|\{m+1\}}(P_{X_\mc{M} Z}) = H(X_{\mc{M}}|Z) -  R_{CO}(X_\mc{M}|Z)$. Here, $R_{CO}(X_\mc{M}|Z)$
	denotes the solution to the real-valued Linear Programming (LP) %
	problem represented in Figure~\ref{fig:LP-Tree}.

	\begin{figure}[h]
		\centering
		\eqbox{
			\begin{IEEEeqnarraybox}{s.t.s}
				\text{Minimize:}  &&  $\sum\limits_{j\in\mc M} R_j$\\
				\text{Subject to:}&&  $\sum\limits_{j\in\mc B} R_j \geq H(X_{\mc B}|X_{\mc B^c},Z),~\quad\forall \mc B\subsetneq \mc M,$ \\%\text{~and}\\
				&& $R_j \in\RR^+, ~\quad\forall j\in\mc M.$
			\end{IEEEeqnarraybox}
		}
		\caption{The LP problem of finding $R_{CO}(X_\mc{M}|Z)$.}
		\label{fig:LP-Tree}
	\end{figure}
	
	We prove that
	\begin{align}\label{eq:RC-Eq}
		R_{CO}(X_\mc{M}|Z) = H(X_{\mc{M}}|Z) - \min_{i,j} I(V_{ij};V_{ji}|Z_{ij}). 
	\end{align}

	First, consider an arbitrary edge $e_{i'j'}\in\mc E$. By cutting this edge, the set of terminals will be partitioned into two parts $\mc B$ and $\mc B^c$ ($\mc B \cap \mc B^c = \emptyset$ and $\mc B \cup \mc B^c = \mc M$). Let $R_j$ be the rate of public communication of terminal $j$. Rewriting the inequalities %
	of LP of Figure~\ref{fig:LP-Tree} for these two sets of terminals, and considering the facts that 
	$\{(V_{ij},V_{ji}, Z_{ij})\}$'s are mutually %
	independent,
	we get $H(X_{\mc M}|Z)=\sum_{i,j} H(V_{ij},V_{ji}|Z_{ij})$ and thus, for any $e_{i'j'}\in\mc E$ we have
	\begin{align*}
		\sum\limits_{j\in\mc B} R_j   & \geq \sum\limits_{\substack{ i\in\mc B   \\ j\in\mc B }} H(V_{ij},V_{ji}|Z_{i'j'})+H(V_{i'j'}|V_{j'i'},Z_{i'j'}), \\
		\sum\limits_{j\in\mc B^c} R_j & \geq \sum\limits_{\substack{ i\in\mc B^c \\ j\in\mc B^c  }} H(V_{ij},V_{ji}|Z_{ji})+H(V_{j'i'}|V_{i'j'},Z_{i'j'}).
	\end{align*} 
	By adding these two inequalities, we arrive at
	\begin{align*}
		\sum\limits_{j\in\mc M} R_j 
		& \geq 
		H(X_{\mc M}|Z) -  \big(H(V_{i'j'},V_{j'i'}|Z_{i'j'}) - H(V_{i'j'}|V_{j'i'},Z_{i'j'}) - H(V_{j'i'}|V_{i'j'},Z_{ji}) \big), \\
		& =    
		H(X_{\mc M}|Z) - I(V_{i'j'};V_{j'i'}|Z_{i'j'}).
	\end{align*}
	This holds for any arbitrary $e_{i'j'}\in\mc E$, and thus, we have proved that $R_{CO}(X_\mc{M}|Z) \geq H(X_{\mc{M}}|Z) - \min_{i,j} I(V_{ij};V_{ji}|Z_{ij}).$ 
	See Appendix~\ref{sec:app:Tree-PIN-AnotM}  for the full proof when $\mc A \neq \mc M$.  
	This lower bound on $R_{CO}$ implies that $C_{PK}^{\mc A|\{m+1\}}(P_{X_\mc{M} Z})  \leq  \min_{i,j} I(V_{ij};V_{ji}|Z_{ij}),$ 
	which is essentially sufficient to prove the converse. 
	However, we further prove that this bound is tight and the equality in \eqref{eq:RC-Eq} holds. 
	To do so, we show that there exist a heuristic rate assignment for $R_1$ to $R_m$ such that $\sum_{j\in\mc M} R_j $ is always equal to the right hand side of \eqref{eq:RC-Eq}.  
	The proof is in Appendix~\ref{sec:app:Tree-PIN-AnotM}. 
	This exact formulation of $R_{CO}(X_\mc{M}|Z)$ will be used later 
	in Remark~\ref{remark:R_SK-vs-Capacity-Tree} 
	for arguing the public communication efficiency of SKA protocol~\ref{prot:SKA-Tree-PIN}.

	\begin{algorithm}[!t]
		\caption{SKA for Tree-PIN ($\vect{\Pi_{TP}}$)}
		\label{prot:SKA-Tree-PIN}
		\DontPrintSemicolon
		\SetKwInput{PorKnw}{Known}
		\SetKwInput{PorAsm}{Assumption}
		\SetKwInput{PorPar}{Final Key Length}
		\PorKnw{Undirected tree $G=(\mc M,\mc E)$ with $\mc M=[m]$, and joint distribution $P_{Z X_{\mc M}}$}
		\PorAsm{Node $2$ is the only neighbor of node $1$, i.e., $\Gamma(1)=\{2\}$ }
		\KwIn{Descriptions of $m-1$ two-party SKA protocols $\{ {\pi}_{ij}|~ i<j \text{~and~} e_{ij}\in\mc E \}$}%
		\KwIn{$n-$IID samples $(X_1^n , X_2^n, \ldots , X_m^n)$ }%
		\PorPar{$\ell$}
		\KwOut{Terminals' copies of the final key $(K_1,\ldots, K_m)$, each with length $\ell$}
		\SetKwFor{MyFor}{for}{}{end~for}
		\SetKwIF{If}{ElseIf}{Else}{if}{then}{else if}{else}{end~if}
		
		\BlankLine\BlankLine%

		\tcp*[h]{Establishing Pairwise Secret Keys}	\\
		\MyFor{$i\in\mc M$}{\MyFor{$j>i$}{\If(\tcp*[f]{Nodes (terminals) $i$ and $j$ are adjacent}){$j\in\Gamma(i)$}{Terminals $i$ and $j$ \textbf{do} reconcile on  $V_{ij}^n$ using public communication  $Q_{ij}$\\
					Terminals $i$ and $j$ \textbf{do} extract pairwise keys $S_{ij}=S_{ji}=\myop{\pi}_{ij}(V_{ij}^n , V_{ji}^n)$\\ Terminals $i$ and $j$ \textbf{do} save the first $\ell$ bits of $S_{ij}$ in $S_{ij}^\prime\leftarrow S_{ij}\vert_{\ell}$ \label{line:cut_first_bits} }  }   }
		
		\BlankLine\BlankLine%
		
		\tcp*[h]{XOR Key Distribution}	\\
		\MyFor%
		{$j\geq 2$}{
			\If(\tcp*[f]{Node (terminal) $j$ has more than one neighbor})
			{$|\Gamma(j)|>1$}{ Terminal $j$
				\textbf{do} find node $j^* \in\Gamma(j)$ s.t. $d(1,j^*)<d(1,i)~\forall i\in\Gamma(j)\setminus\{j^*\}$, \textbf{and}~ \\
				\lForEach{$i\in\Gamma(j)\setminus\{j^*\}$, terminal $j$}{broadcasts $F_{ji}=S^\prime_{j j^*} \oplus S^\prime_{ji}$} \label{line:broadcast}
			}
		}
		
		\BlankLine\BlankLine%

		\tcp*[h]{Local Final Key Calculation}\\
		Terminals $1$ and $2$ set their keys to $K_1=K_2=S^\prime_{12}$.
		
		\MyFor{$j\geq 3$}{
			Terminal $j$
			\textbf{do} find node $j^*\in\Gamma(j)$ s.t. $d(2,j^*)<d(2,i)~\forall i\in\Gamma(j)\setminus\{j^*\}$, \textbf{then}~ \\
			\textbf{do} find $\mathrm{Path}(j\rightarrow2)$, the path from node $j$ to node $2$, \textbf{then}~ \\
			\textbf{do} compute $K_j = S^\prime_{j j^*} \bigoplus\limits_{\substack{i_a,i_b~ \in\mc M \\ \text{~s.t.~} e_{i_a i_b}\in \mathrm{Path}(j\rightarrow2)  }} F_{i_a i_b}$\label{line:final_calculation}
		}

	\end{algorithm}

	\paragraph*{\textbf{The~achievability}}
	We show that the upper bound given in Lemma~\ref{lemma:Upper-bound-Tree} is achievable. 
	More precisely, we  prove that for every $n$, Protocol~\ref{prot:SKA-Tree-PIN} 
	generates an $(\eps_n,\sigma_n)-$SK $K$  %
	with length $\ell$, such that 
	$\lim_{n\to\infty} \eps_n = \lim_{n\to\infty} \sigma_n = 0$,  
	and 
	\begin{align*}
		r_{K}(\vect{\Pi_{TP}}) =  \lim_{n \rightarrow \infty} \frac{\ell}{n} 
		= \min_{\substack{i,j\in\mc M                                        \\ \mathrm{~s.t.~}e_{ij}\in\mc E_{\mc A}}}  I(V_{ij};V_{ji}|Z_{ij}).
	\end{align*}
	
	The protocol works by using the public communication channel in two rounds. First, each pair of connected terminals $i$ and $j$ execute two-party SKA protocols $\pi_{ij}$ (in parallel) to establish  pairwise keys $S'_{ij}$ of length $\ell$, where for each $e_{ij}$ the pairwise key length $\ell \approx n I(V_{ij},V_{ji}|Z_{ij}) - o(n)$ is achievable due to \cite[Theorem 1]{Ahlswede1993}. 
	In the second round, 
	terminals use the public channel to reconcile on one of the pairwise keys, namely $S'_{12}$.  
	In this step, non-leaf nodes (terminals) send enough messages that enables all terminals 
	to calculated $K=S'_{12}$ while keeping the leakage of information to Eve to a minimum amount.

	A complete description of this SKA protocol is given in Protocol~\ref{prot:SKA-Tree-PIN}. 
	In Protocol~\ref{prot:SKA-Tree-PIN}, without loss of generality, we assume that terminal $2$ is the only terminal connected to terminal $1$; i.e., $\Gamma(1)=\{2\}$. In line~\ref{line:final_calculation} of Protocol~\ref{prot:SKA-Tree-PIN}  $\mathrm{Path}(i_1\rightarrow i_f) = (e_{i_1 i_2},e_{i_2 i_3},\cdots, e_{i_{f-1} i_f})$ denotes the path from terminal $i_1$ to $i_f$. Since $G$ is an undirected tree, between each terminal $i\in\mc M$ and $j\in\mc M$ there is always a unique path.  We show in the proof of Lemma \ref{lemma:Ach-Tree} that if pairwise keys are $(\eps,\sigma)-$SKs established by executing two-party SKA protocols $\pi_{ij}$, then the final key of Protocol~\ref{prot:SKA-Tree-PIN} is an $(|\mc E|\eps,2|\mc E|\sigma)-$SK.
	The full proof of achievability is
	in Appendix \ref{sec:app:Ach}.

	\begin{example}\label{sec:ex1}
		
		In the following, we revisit the example of 	Figure \ref{fig:ex1}, and  
		illustrate how protocol~\ref{prot:SKA-Tree-PIN} works. 
		This 
		wiretapped 
		Tree-PIN with $\mc M=\{1,2,3,4\}$ is a simple path from terminal $1$ to terminal $4$. %
		
		Protocol \ref{prot:SKA-Tree-PIN} works as follows. First, each pair of connected terminals establish  pairwise secret keys $S_{ij}$ by employing two-party SKA protocols $\pi_{ij}$. Then, let $\ell$ be the  length of the smallest pairwise key. All parties then keep only the first $\ell$ bits of their pairwise keys. Let $S'_{ij}$ denote the first $\ell$ bits of $S_{ij}$. Note that in this example terminal $2$ has two pairwise keys $\{S'_{12},S'_{23}\}$ and terminal $3$ also has two pairwise keys $\{S'_{23},S'_{34}\}$. In the next phase of the protocol, terminal $2$ broadcasts $F_{23}=S'_{12}\oplus S'_{23}$ and terminal $3$ broadcasts $F_{34}=S'_{23} \oplus S'_{34}$. In the last phase, each terminal $j$ computes the key $K_j$ according to the following
		\begin{align*}
			K_1 & = S'_{12},                            \\
			K_2 & = S'_{12},                            \\
			K_3 & = S'_{23} \oplus F_{23},              \\
			K_4 & = S'_{34} \oplus F_{34}\oplus F_{23}. 
		\end{align*}
		One can easily see that above equations imply that we have $K_1=K_2=K_3=K_4=S'_{12}$.

	\end{example}

	\subsection{Public Communication Cost of Protocol \ref{prot:SKA-Tree-PIN}}

	The Protocol~\ref{prot:SKA-Tree-PIN} 
	is the only known protocol that 
	achieves the WSK capacity of Tree-PIN; however, 
	when $Z$ is known, it can be compared with other protocols that 
	achieve the PK capacity. 
	This protocol 
	is interactive with two rounds of public communication but does not require omniscience.   
	We show that the public communicate cost of Protocol~\ref{prot:SKA-Tree-PIN}, that is
	the asymptotic rate of its public communication, %
	is no larger than other  protocols that require omniscience for achieving the PK capacity.  
	
	\begin{remark}%
		\label{remark:R_SK-vs-Capacity-Tree}
		Let %
		$R_{SK}(X_\mc{M})$ denote the minimum public communication rate
		required for achieving %
		$C_{SK}^{\mc M}(P_{X_\mc{M}})$. That is 
		$R_{SK}(X_\mc{M}) = \min \{ r_{PC}(\Pi) |~ \Pi \text{~achieves~} C_{SK}^{\mc M}(P_{X_\mc{M}}) \}.$  
		It was proved in~\cite{Chan2017} that for PIN model with $V_{ij}=V_{ji}$, %
		we have $R_{SK}(X_\mc{M})= (m-2)C_{SK}^{\mc M}(P_{X_\mc{M}})$.  %
		Similarly, define $R_{WSK}(X_\mc{M}|Z) = \min \{ r_{PC}(\Pi) |~ \Pi \text{~achieves~} C_{WSK}^{\mc M}(P_{X_\mc{M} Z}) \}$.
		We show that for any wiretapped Tree-PIN,  when $V_{ij}\neq V_{ji}$, %
		we have %
		\begin{equation*}
			R_{WSK}(X_\mc{M}|Z)\leq \big(\sum_{i,j} H(V_{ij}|V_{ji})\big) + (m-2)C_{WSK}^{\mc M}(P_{X_\mc{M} Z}) \leq R_{CO}(X_{\mc M}|Z), 
		\end{equation*}
		where $R_{CO}(X_{\mc M}|Z)$ is defined in Theorem~\ref{thm:PK-Cap-CN04}. 
		It is not known whether 
		the left bound is tight. %
		When $Z$ is known, both Protocol \ref{prot:SKA-Tree-PIN} 
		and protocol of \cite{Csiszar2004a} achieve the PK capacity of 
		Tree-PIN.  
		Protocol \ref{prot:SKA-Tree-PIN} 
		does not require achieving omniscience while 
		protocol of \cite{Csiszar2004a} does.
		The above inequality shows that Protocol \ref{prot:SKA-Tree-PIN} %
		uses less public communication than the protocol of \cite{Csiszar2004a} (Also see \cite[Example 7]{Csiszar2004a}).
	\end{remark}
	\begin{IEEEproof}[Proof of Remark~\ref{remark:R_SK-vs-Capacity-Tree}] 
		First we prove the first bound by noting the fact that 
		$R_{WSK}(X_\mc{M}|Z)\leq r_{PC}(\vect{\Pi_{TP}})$ as Protocol \ref{prot:SKA-Tree-PIN} 
		($\vect{\Pi_{TP}}$) achieves the WSK capacity. 
		We now calculate $r_{PC}(\vect{\Pi_{TP}})$. 
		Protocol~\ref{prot:SKA-Tree-PIN} has two rounds of public communication. 
		In the first round terminals agree on their pairwise keys. 
		For each $e_{ij}$ either $V_{ij}-V_{ji}-Z_{ij}$ or $V_{ji}-V_{ij}-Z_{ij}$ holds. 
		With an abuse of notation, assume that $V_{ij}-V_{ji}-Z_{ij}$ for all $e_{ij}$.  
		Then public communication rate of the first round for each $e_{ij}$ is given by $H(V_{ij}|V_{ji})$ \cite{Ahlswede1993}. 
		Since in the first round, pairwise keys are generated in parallel and independently,  
		the total amount of public communication rate of this round is given by
		$\sum_{i,j} H(V_{ij}|V_{ji})$.  
		In the second round, 
		any terminal $j\in\mc M$ finds its {unique\footnote{Exists because of tree structure of the variables.} } neighbour $j^*$ that is closest to the node $1$ and broadcasts $|\Gamma(j)|-1$ encoded messages $\{F_{ji}|~\forall i\in\Gamma(j)\setminus\{j^* \}\}$, where each message has the same length $\ell$ as the final key $K$. Thus, the public communication rate of the protocol \ref{prot:SKA-Tree-PIN} is
		\begin{IEEEeqnarray*}{rl}
			r_{PC}(\vect{\Pi_{TP}}) 
			& = \sum_{i,j} H(V_{ij}|V_{ji}) + \lim_{n\to\infty} \frac{1}{n} \sum_{j=1}^{m} \ell \times \left( |\Gamma(j)|-1  \right) \\
			& = \sum_{i,j} H(V_{ij}|V_{ji}) + \lim_{n\to\infty} \left(\ell/n\right) \times \left(\sum_{j=1}^{m}  |\Gamma(j)| - m\right)\\
			& = \sum_{i,j} H(V_{ij}|V_{ji}) + \lim_{n\to\infty} \left(\ell/n\right) \left( 2|\mc E| -m\right) \\
			& = \sum_{i,j} H(V_{ij}|V_{ji}) + \lim_{n\to\infty} (m-2)\ell/n, 
		\end{IEEEeqnarray*}
		where we used the facts that for a graph $G=(\mc M,\mc E)$, we have $\sum_{j\in\mc M}  |\Gamma(j)| = 2|\mc E| $, and for an undirected tree with $m$ vertexes we have ${|\mc E|=m-1}$. 
		By $\ell\approx n \min_{i,j} I(V_{ij};V_{ji}|Z_{ij}) - o(n)$, and the fact that Protocol~\ref{prot:SKA-Tree-PIN} achieves the WSK capacity of a Tree-PIN, namely $\min_{i,j} I(V_{ij};V_{ji}|Z_{ij})$, proves  the first (left) inequality for any given Tree-PIN $G=(\mc M,\mc E)$. 
		Next, we prove the second (right) inequality by showing 
		that $r_{PC}(\vect{\Pi_{TP}}) \leq R_{CO}(X_{\mc M}|Z)$. 
		\begin{IEEEeqnarray*}{rl}
			r_{PC}(\vect{\Pi_{TP}}) 
			& = \sum_{i,j} H(V_{ij}|V_{ji}) + (m-2) C_{WSK}^{\mc M}(P_{X_\mc{M} Z}) \\
			& \leq  \sum_{i,j} H(V_{ij}|V_{ji}) + (m-1) C_{WSK}^{\mc M}(P_{X_\mc{M} Z}) - C_{WSK}^{\mc M}(P_{X_\mc{M} Z}) \\
			& \leq \sum_{i,j} H(V_{ij}|V_{ji}) + \sum_{i,j} H(V_{ij}|Z_{ij}) - H(V_{ij}|V_{ji})  - C_{WSK}^{\mc M}(P_{X_\mc{M} Z}) \\
			& = \sum_{i,j} H(V_{ij}|Z_{ij}) - C_{WSK}^{\mc M}(P_{X_\mc{M} Z}) \\
			& \leq H(X_{\mc M}|Z) - C_{WSK}^{\mc M}(P_{X_\mc{M} Z}) \\
			& = R_{CO}(X_{\mc M}|Z)
		\end{IEEEeqnarray*}
		where the last equality is due to \eqref{eq:Rco-Tree-app}. 	
	\end{IEEEproof}

	\section{Finite-length Bounds for Wiretapped Tree-PIN}\label{sec:FLA-notsym}
	
	{\color{black}
		Finite-length analysis of information theoretic tasks such as SKA is important in practice, as in real-life deployment of SKA protocols the number of samples, $n$, accessible  to each terminal is finite. In this case, better estimations and bounds on the maximum achievable key length (i.e., $S_{\eps,\sigma}(X_{\mc A}^n|Z^n)$)
		are desired (see Definition~\ref{def:SK-length}). 
		In this section, we give a finite-length upper bound, and  three finite-length lower bounds for the maximum achievable key length in a wiretapped Tree-PIN.

		\subsection{The Finite-length Upper Bound}

		\begin{theorem}\label{thm:SOA-upper}
			For any given wiretapped Tree-PIN ${(X_{\mc M},Z)}$, described by  $P_{ZX_{\mc M}}$, and for every $n\in\mathds{N}$, every $\eps,\sigma>0$, with $\eps+\sigma<1$, and any subset $\mc A\subseteq\mc M$, we have that $S_{\eps,\sigma}(X_{\mc A}^n|Z^n)$ is upper bounded by
			\begin{align}\label{eq:UP-1}
				\min_{\substack{i,j\in\mc M \\ \mathrm{~s.t.~}e_{ij}\in\mc E_{\mc A}}} \left\{ nR_{ij}- \sqrt{n \Delta_{ij} }Q^{-1}({\eps+\sigma}) \right\} + \frac{3}{2}\log n + \mc O(1),
			\end{align}
			where $R_{ij} = I(V_{ij};V_{ji}|Z_{ij})$.
		\end{theorem}

		For the proof of Theorem \ref{thm:SOA-upper} we use the Hypothesis testing upper bound of Tyagi and Watanabe \cite{Tyagi2015} which is a general single-shot bound for any wiretapped multiterminal source model. Hayashi et al.\ used the upper bound of \cite{Tyagi2015} to prove a finite-length upper bound for the case of two-party SKA. To our knowledge, Theorem~\ref{thm:SOA-upper} is the first multiterminal finite-length upper bound based on the Hypothesis testing upper bound.

		To prove Theorem \ref{thm:SOA-upper}, we first recall the notion hypothesis testing and a couple of lemmas. %

		The binary hypothesis testing problem is defined as follows. For a random variable $X$, there are two possible distributions $P_X$ and $Q_X$. Using a test algorithm $\mymtrx{T}$ we shall decide between $P_X$ or $Q_X$. Let the null hypothesis be $H_0=P_X$. If we reject the null hypothesis $P_X$ when the actual distribution is $P_X$ then type I error is occurred, and if we  accept the null hypothesis when the actual distribution is $Q_X$ then type II error is occurred. Let $\beta_{\eta}(P_X,Q_X)$ denote the infimum of  type II error probability given that type I error probability is less than $\eta$. That is, 
		\begin{equation*}
			\beta_{\eta}(P_X,Q_X) = \inf\limits_{\mymtrx{T}: E_1(\mymtrx{T})\leq \eta} E_2(\mymtrx{T}),
		\end{equation*}
		where $E_1(\mymtrx{T}) = \sum_{x\in\mc X} P_X(x)\pr{\mathrm{Rej~} H_0|x},$ and $E_2(\mymtrx{T}) = \sum_{x\in\mc X} Q_X(x)\pr{\mathrm{Acc~} H_0|x},$ are respectively the type I and type II errors of a given hypothesis testing algorithm $\mymtrx{T}$.

		\begin{lemma}[Hypothesis testing upper bound \cite{Tyagi2015}]\label{thm:Tyagi-HT-upper}
			Given an arbitrary multiterminal source model $(X_{\mc M}, Z)$, and any given partition $\mc P=\{\mc P_1,\ldots, \mc P_l\}$ of $\mc M$, for every $\eps,\sigma>0$, with $\eps+\sigma<1$, and every $0<\eta<1-\eps-\sigma$, we have
			\begin{equation*}
				S_{\eps,\sigma}(X_{\mc M}|Z) \leq  \frac{1}{|\mc P|-1} \left[-\log \beta_{\eps+\sigma+\eta}\left(P_{X_{\mc M}Z} , Q_{X_{\mc M}Z}^{\mc P} \right) + |\mc P|\log \frac{1}{\eta}  \right],
			\end{equation*}
			where $Q_{X_{\mc M}Z}^{\mc P}$ is any probability distribution for which $Q_{X_{\mc M}|Z}^{\mc P} = \Pi_{j=1}^{l} Q_{X_{\mc P_{j}}|Z}$ holds. 
		\end{lemma}

		\begin{lemma}[Also see Lemma 4.1.2 of \cite{Han2003}]\label{thm:Han}
			Consider  a hypothesis testing problem where $P_X$ and $Q_X$ are respectively the null and alternative hypotheses. For any $\lambda>0$, we have
			\begin{equation*}
				-\log\beta_{\eps}(P_X,Q_X) \leq \lambda - \log \left( P_{X}\left(\left\{x:~\log\frac{P_X(x)}{Q_X(x)} \leq \lambda\right\} \right) - \eps  \right).
			\end{equation*}
		\end{lemma}
		
		\begin{IEEEproof}
			Let $$\mc C=\left\{x:~ \log\frac{P_X(x)}{Q_X(x)} \geq \lambda \right\}.$$
			Suppose that the hypothesis testing algorithm $\mymtrx{T}$ is such that  accepts the null hypothesis $P_X$ if the observed value $x$ belongs to $\mc C$. Also, let $\eps$ denote the type I error of test $\mymtrx{T}$. That is,
			\begin{equation*}
				\eps=E_1(\mymtrx{T})=P_X\left(\left\{x:~ \log\frac{P_X(x)}{Q_X(x)} < \lambda \right\} \right) = \sum_{x\in\mc X} P_X(x)\mathds{1}\left(x\notin \mc C\right).
			\end{equation*}
			Due to the Neyman-Pearson lemma, $\mymtrx{T}$ gives the least type II error of all tests with type I error of at most $\eps$. To simplify the proof, let $$\mc S = \left\{x:~ \log\frac{P_X(x)}{Q_X(x)} \leq \lambda \right\}.$$ Using the Neyman-Pearson lemma we have,
			\begin{align*}
				P_X\left(\left\{x:~ \log\frac{P_X(x)}{Q_X(x)} \leq \lambda \right\}\right) & = \sum_{x\in\mc X} P_X(x)\mathds{1}\left(x\in \mc S \cap \mc C^c \right) + \sum_{x\in\mc X} P_X(x)\mathds{1}\left(x\in \mc S \cap \mc C \right)                   \\
				& \stackrel{(a)}{\leq} \sum_{x\in\mc X} P_X(x)\mathds{1}\left(x\notin\mc C \right) + \sum_{x\in\mc X} 2^\lambda Q_X(x)\mathds{1}\left(x\in \mc S \cap \mc C \right) \\
				& \leq \sum_{x\in\mc X} P_X(x)\mathds{1}\left(x\notin\mc C \right) + \sum_{x\in\mc X} 2^\lambda Q_X(x)\mathds{1}\left(x\in \mc C \right)                            \\
				& \stackrel{(b)}{=} \eps + 2^\lambda \beta_\eps(P_X,Q_X),                                                                                                           
			\end{align*}
			where in (a) we use that  $P_X(x)\leq 2^\lambda Q_X(x)~\forall x\in \mc S$, and in (b) we use Neyman-Pearson lemma. The proof is complete by taking logarithm from both sides of the inequality.
		\end{IEEEproof}

		\begin{theorem}[Berry-Esseen, see \cite{Feller1982} Theorem 1, Chapter~XVI, Section 5] \label{thm:Berry-Esseen}
			
			Let $W^n$ be an $n-$IID variable, and $-\infty<\alpha<\infty$, then
			\begin{equation*}
				\left\vert \pr{\sum_{j=1}^{n} W_j \leq n\mu - \alpha\sqrt{\Delta n}} - Q(\alpha) \right\vert \leq \frac{3\rho}{\Delta^{3/2} \sqrt{n}} ,
			\end{equation*}
			where $\mu=\ev{W}, \Delta=\myvar{W}, \rho=\ev{|W-\mu|^3}$, and $Q(\cdot)$ is the tail probability of the standard Gaussian distribution given by
			\begin{equation*}
				Q(\alpha) = \frac{1}{\sqrt{2\pi}} \int_{\alpha}^{\infty} \exp{\big( -\frac{t^2}{2} \big)} \mathrm{d}t .
			\end{equation*}
			
		\end{theorem}
		
		We now prove the upper bound of Theorem~\ref{thm:SOA-upper}.
		
		\begin{IEEEproof}
			Denote the set of all terminals in $G_{\mc A}$ by $\mc B = \mc M_{\mc A} \subseteq\mc M$. For SKA in the Tree-PIN $G_{\mc A}$, lemma~\ref{thm:Tyagi-HT-upper} implies that for an arbitrary partition $\mc P$ of $\mc B$, we have
			\begin{equation}\label{eq:ht-bound-B}
				S_{\eps,\sigma}(X^n_{\mc A}|Z) = S_{\eps,\sigma}(X^n_{\mc B}|Z) \leq  \frac{1}{|\mc P|-1} \left[-\log \beta_{\eps+\sigma+\eta}\left(P_{X^n_{\mc B}Z^n} , Q_{X^n_{\mc B}Z^n}^{\mc P} \right) + |\mc P|\log \frac{1}{\eta}  \right].
			\end{equation}
			Fix an edge $e_{i'j'}\in\mc E_{\mc A}$ of $G_{\mc A}$ that connects nodes (terminals) $i'$ and $j'$. Cutting this edge induces a partition $\mc P_{i'j'}=\{\mc P_1, \mc P_2\}$, such that $i'\in\mc P_1$ and $j'\in\mc P_2$. By applying \eqref{eq:ht-bound-B} and lemma \ref{thm:Han}, with $\mc P=\mc P_{i'j'}$, $P_{X^n_{\mc B}Z^n} = \prod_{e_{ij}} P_{V_{ij}^n V_{ji}^n Z_{ij}^n}$, $Q^{\mc P}_{X^n_{\mc B}Z^n} = P_{V_{i'j'}^n|Z_{i'j'}^n} P_{V_{j'i'}^n Z_{i'j'}^n} \prod_{e_{ij}\neq e_{i'j'}} P_{V_{ij}^n V_{ji}^n Z_{ij}^n}$, and $\eta=\frac{1}{\sqrt{n}}$, we get
			\begin{equation}\label{eq:upper-w-lam}
				S_{\eps,\sigma}(X^n_{\mc A}|Z) \leq \lambda - \log \left(\pr{\log\frac{P_{X^n_{\mc B}Z^n}}{Q^{\mc P_{i'j'}}_{X^n_{\mc B}Z^n}} \leq \lambda} - \eps -\sigma - \frac{1}{\sqrt{n}}  \right) + \log n .
			\end{equation}
			Let $$\theta_n=\frac{2}{\sqrt{n}} + \frac{3\rho_{i'j'}}{\Delta_{i'j'}^{3/2} \sqrt{n}},$$ where
			\begin{equation*}
				\Delta_{ij}=\myvar{\log \frac{P_{V_{ij}V_{ji}|Z}(V_{ij},V_{ji}|Z) }{P_{V_{ij}|Z}(V_{ij}|Z) P_{V_{ji}|Z}(V_{ji}|Z) }},
			\end{equation*}
			and
			\begin{equation*}
				\rho_{ij}=\ev{\left\vert {\log \frac{P_{V_{ij}V_{ji}|Z}(V_{ij},V_{ji}|Z) }{P_{V_{ij}|Z}(V_{ij}|Z) P_{V_{ji}|Z}(V_{ji}|Z) }} - I(V_{i'j'};V_{j'i'}|Z)  \right\vert^3}.
			\end{equation*}
			By choosing
			\begin{equation*}
				\lambda = nI(V_{i'j'};V_{j'i'}|Z)- \sqrt{n \Delta_{i'j'} }Q^{-1}(\eps+\sigma+\theta_n),
			\end{equation*}
			and by the Berry-Esseen theorem we get
			\begin{align*}
				\pr{\log\frac{P_{X^n_{\mc B}Z^n}}{Q^{\mc P_{i'j'}}_{X^n_{\mc B}Z^n}} \leq \lambda} &= \pr{\log\frac{P_{V_{i'j'}^n V_{j'i'}^n Z_{i'j'}^n}}{P_{V_{i'j'}^n|Z_{i'j'}^n} P_{V_{j'i'}^n Z_{i'j'}^n} } \leq \lambda}\\
				& = \pr{\log\frac{P_{V_{i'j'}^n V_{j'i'}^n | Z^n}}{P_{V_{i'j'}^n|Z^n} P_{V_{j'i'}^n| Z^n} }\leq \lambda} \\ &\geq \eps+\sigma+\frac{2}{\sqrt{n}}.
			\end{align*}
			Note that $\ev{\log \frac{P_{V_{ij}V_{ji}|Z}(V_{ij},V_{ji}|Z) }{P_{V_{ij}|Z}(V_{ij}|Z) P_{V_{ji}|Z}(V_{ji}|Z) }} =  I(V_{ij};V_{ji}|Z) $. Applying the above inequality in \eqref{eq:upper-w-lam} gives
			\begin{align*}%
				S_{\eps,\sigma}(X^n_{\mc A}|Z) & \leq nI(V_{i'j'};V_{j'i'}|Z)- \sqrt{n \Delta_{i'j'} }Q^{-1}(\eps+\sigma+\theta_n) - \log \left( \frac{1}{\sqrt{n}} \right) + \log n . 
			\end{align*}
			By using Taylor approximation of $Q(\cdot)$ to remove $\theta_n$ we get
			\begin{align*}
				S_{\eps,\sigma}(X^n_{\mc A}|Z) & \leq nI(V_{i'j'};V_{j'i'}|Z)- \sqrt{n \Delta_{i'j'} }Q^{-1}(\eps+\sigma)  + \frac{3}{2}\log n + \mc{O}(1), 
			\end{align*}
			that holds for any edge $e_{i'j'}$ of $G_{\mc A}$. The proof is complete by minimizing over all  $e_{i'j'}$'s.
		\end{IEEEproof}

		\subsection{Finite-length Lower Bounds}

		The achievability (lower) bounds are based on %
		variations of the SKA protocol that achieves the WSK capacity of wiretapped Tree-PIN given in Theorem~\ref{thm:Tree-PIN}.
		This protocol has two main steps. In the first step, each pair of connected terminals $i$ and $j$ (i.e., $e_{ij}\in\mc E$) preform a two-party SKA protocol to obtain a pairwise secret key. For this task, terminals can use, for example, the two-party SKA protocols of \cite{Sharifian2020} or \cite{Poostindouz2021} -- see also \cite{Ahlswede1993,Maurer1993,Hayashi2016}. %
		In the second step, terminals use their pairwise keys and public communication to agree on the final shared secret key. See the details of this SKA protocol in Appendix~\ref{sec:app:Ach}. %

		For the case of two-party SKA, Hayashi et al.\ \cite{Hayashi2016} proved that for a given source model $(V_{ij}, V_{ji}, Z_{ij})$, if $V_{ij}- V_{ji}- Z_{ij}$, and for every $n\in\mathds{N}$ and $\eps,\sigma>0$, with $\eps+\sigma<1$,  we have
		\begin{align*}
			S_{\eps,\sigma}(V_{ij}^n, V_{ji}^n|Z_{ij}^n) =  nR_{ij}- \sqrt{n \Delta_{ij} }Q^{-1}(\eps+\sigma) 
			\pm \mc{O}(\log n),                                                                               
		\end{align*} where $S_{\eps,\sigma}(\cdot)$ denotes the maximum achievable key length,
		\begin{equation*}
			\Delta_{ij}=\myvar{\log \frac{P_{V_{ij}V_{ji}|Z_{ij}}(V_{ij},V_{ji}|Z_{ij}) }{P_{V_{ij}|Z_{ij}}(V_{ij}|Z_{ij}) P_{V_{ji}|Z_{ij}}(V_{ji}|Z_{ij}) }},
		\end{equation*}
		$R_{ij} = I(V_{ij};V_{ji}|Z_{ij})$ is the two-party WSK capacity of $(V_{ij},V_{ji},Z_{ij})$, and $Q(\cdot)$ is the tail probability of the standard Gaussian distribution. This second-order approximation of the key length is achievable by the interactive protocol of \cite{Hayashi2016}.
		Sharifian et al.\ \cite{Sharifian2020}  gave also two finite-length approximations corresponding to a one-way two-party SKA protocol. 
		One-way SKA protocols are more efficient in terms of the public communication than the interactive  construction of  \cite{Hayashi2016}, while in finite-length regime, the SKA protocol of \cite{Hayashi2016} is closer to the two-party capacity ($R_{ij}$) than the SKA protocol of \cite{Sharifian2020}. However, by a numerical example in Section~\ref{sec:comp_num} we illustrate that the lower bound that is based on \cite{Sharifian2020} can be very close to the lower bound which is based on \cite{Hayashi2016}.

		By using the SKA protocols of \cite{Hayashi2016} and \cite{Sharifian2020} in the first step of our SKA protocol for obtaining pairwise keys, we prove the following lower bounds for  wiretapped Tree-PIN.

		\begin{proposition}[Lower bounds]\label{thm:SOA-lower}
			For any given wiretapped Tree-PIN, described by  $P_{Z X_{\mc M}}$, and for every $n\in\mathds{N}$, every $\eps,\sigma>0$, with $\eps+\sigma<1$, and any subset $\mc A\subseteq\mc M$, we have
			\begin{align}\label{eq:LB-1}
				S_{\eps,\sigma}(X_{\mc A}^n|Z^n) & \geq 
				F_1(X^n_{\mc A}|Z^n) -\frac{11}{2}\log n +  \mc O(1) \\
				\label{eq:LB-2}
				S_{\eps,\sigma}(X_{\mc A}^n|Z^n) & \geq 
				F_2(X^n_{\mc A}|Z^n) -\log n +  \mc O(1) \\ %
				\label{eq:LB-3}
				S_{\eps,\sigma}(X_{\mc A}^n|Z^n) & \geq 
				F_3(X^n_{\mc A}|Z^n) -\log n +  \mc O(1)
			\end{align}
			where
			\begin{align*}
				F_1(X^n_{\mc A}|Z^n) & = \min_{\substack{i,j\in\mc M \\ \mathrm{~s.t.~}e_{ij}\in\mc E_{\mc A}}} \left\{ nR_{ij}- \sqrt{n \Delta_{ij} }Q^{-1}(\frac{2\eps + \sigma}{2|\mc E_{\mc A}|}) \right\}, \\
				F_2(X^n_{\mc A}|Z^n) & = \min_{\substack{i,j\in\mc M \\ \mathrm{~s.t.~}e_{ij}\in\mc E_{\mc A}}} \left\{ nR_{ij}-  Q^{-1}(\frac{\eps}{|\mc E_{\mc A}|}) \sqrt{n\Delta'_{ij}}  
				-Q^{-1}(\frac{\sigma}{2|\mc E_{\mc A}|}) \sqrt{n\Delta''_{ij}}  \right\},\\
				F_3(X^n_{\mc A}|Z^n) & = \min_{\substack{i,j\in\mc M \\ \mathrm{~s.t.~}e_{ij}\in\mc E_{\mc A}}} \left\{ nR_{ij} \right\} 
				-\sqrt{2n}\log({|\mc X|+3})(\sqrt{\log \frac{|\mc E_{\mc A}|}{\eps}} + \sqrt{\log \frac{2|\mc E_{\mc A}|}{\sigma}}),
			\end{align*}
			with $R_{ij} = I(V_{ij};V_{ji}|Z_{ij})$, $\Delta'_{ij} = \myvar{-\log P_{V_{ij}|V_{ji}}}$, $ \Delta''_{ij} = \myvar{-\log P_{V_{ij}|Z_{ji}}},$ and $|\mc E_{\mc A}|$ is the number of edges in the sub-tree $G_{\mc A}$. 
			
		\end{proposition}

		For Theorem \ref{thm:SOA-lower} we note that according to the proof of Theorem~\ref{thm:Tree-PIN}, obtaining pairwise $(\eps,\sigma)-$SKs leads to a finial $(|\mc E_{\mc A}|\eps,2|\mc E_{\mc A}|\sigma)-$SK.
		Thus, for all of the above achievability (lower) bounds,  parties first establish pairwise $(\frac{\eps}{|\mc E_{\mc A}|}, \frac{\sigma}{2|\mc E_{\mc A}|})$ secret keys, and then use %
		Protocol \ref{prot:SKA-Tree-PIN} to agree on the final key.
		None of the bounds  require %
		omniscience. %
		Lower bound of \eqref{eq:LB-1} is based on Protocol \ref{prot:SKA-Tree-PIN} which uses the two-party protocol of 
		\cite{Hayashi2016} for generating pairwise keys, and lower bounds in \eqref{eq:LB-2} and \eqref{eq:LB-3} are 
		based on Protocol \ref{prot:SKA-Tree-PIN} when the one-way two-party protocol of \cite{Sharifian2020} is used for pairwise key generation. 
		Lower bounds in \eqref{eq:LB-1} and \eqref{eq:LB-2} assume that samples are IID and lower bound of \eqref{eq:LB-3} only assumes that samples are independent (and not necessarily IID.)   
		The full proof of Theorem \ref{thm:SOA-lower} is given in Appendix \ref{sec:app:FL-lower}. 

		Note that the second-order terms (in $\mc O(\sqrt{n})$) of the upper and lower bounds do not match. %
		Finding tighter bounds with matching second-order terms %
		is  an interesting open problem.

		\begin{example} \label{sec:comp_num}
			
			\begin{figure}[t]
				\centering
				\includegraphics[width=0.6\linewidth]{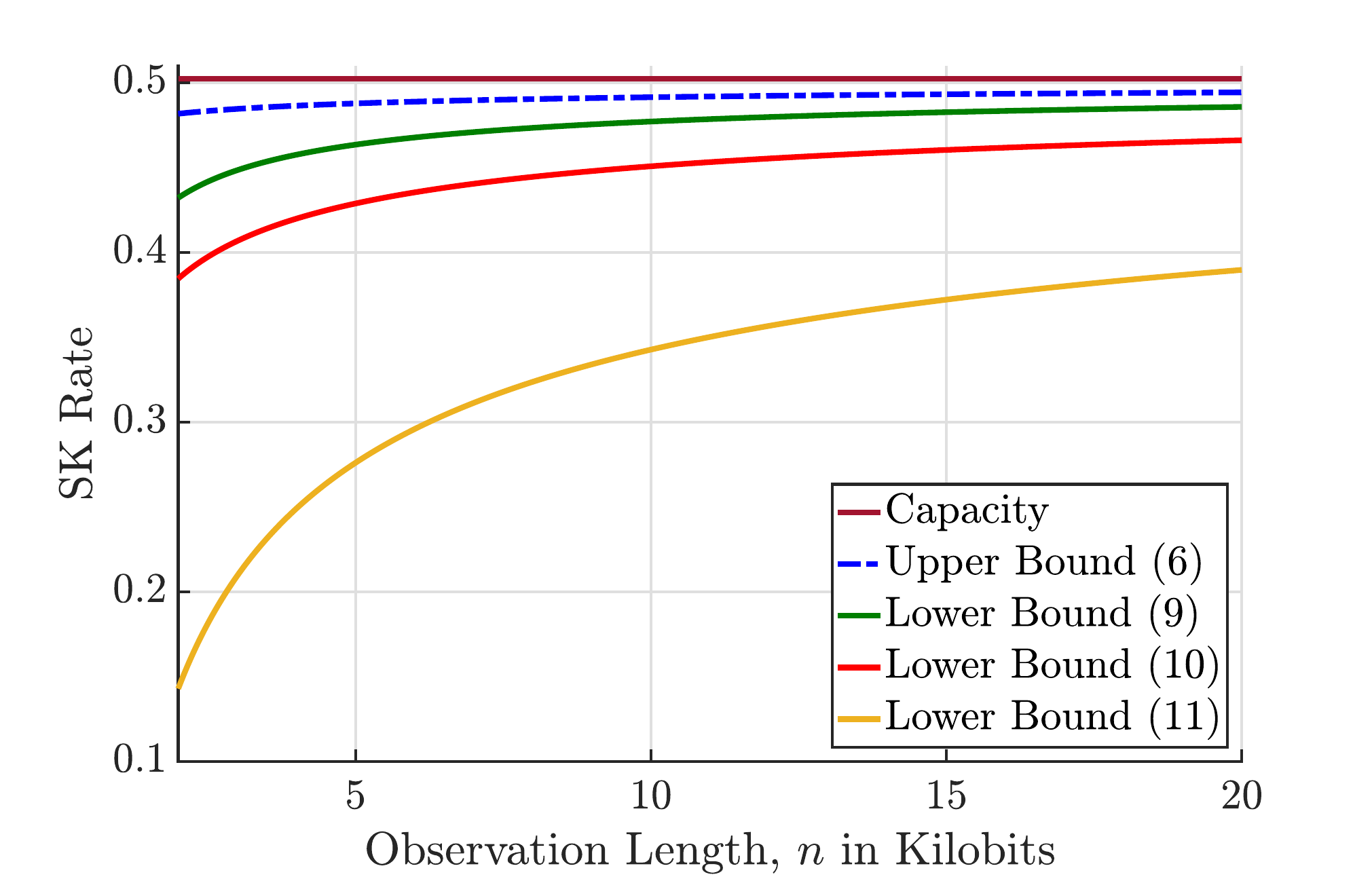}
				\caption[Comparing the proven finite-length bounds for the Tree-PIN source model.]{Comparing finite-length bounds of %
					the example in Section \ref{sec:comp_num}. Here, $m=3$,  $\eps=\sigma=0.05$, and the WSK capacity is $0.502$. Lower bound of \eqref{eq:LB-1} is the tightest lower bound and is by Protocol \ref{prot:SKA-Tree-PIN} if the two-party interactive SKA of \cite{Hayashi2016} is used for pairwise key generation. Lower bounds in \eqref{eq:LB-2} and \eqref{eq:LB-3} are based on Protocol \ref{prot:SKA-Tree-PIN} if the two-party one-way SKA protocol of \cite{Sharifian2020} is used for pairwise key generation. 
				}\label{fig:comp_treepin}
			\end{figure}

			The following %
			numerical example  %
			compares the finite-length bounds given in \eqref{eq:UP-1} and \eqref{eq:LB-1}-\eqref{eq:LB-3}. Consider a source model with $m=3$ terminals, $\mc M=\{1,2,3\}$, and $\mc A=\mc M$. Let $X_1 = (V_{12}, V_{13})$ such that $V_{1j}$'s are binary uniform variables. Also for $p,q\in (0,1)$ and for $j\in\{2,3\}$, let $X_j = V_{j1} = \mathrm{BSC}_p(V_{1j})$ and $Z_{1j} = \mathrm{BSC}_q(V_{j1})$. Here, $\mathrm{BSC}_p(\cdot)$ denotes a binary symmetric channel with crossover probability of $p$. For this example, the WSK capacity is  $C_{WSK} = h_2(p*q) - h_2(p)$, where $p*q = p(1-q) + (1-p)q$, and $h_2$ is the binary entropy given by $h_2(p) = -p\log p - (1-p)\log(1-p)$. Consider, $p=0.0093,$ $q=0.13$, and $\eps=\sigma=0.05$. Then, $C_{WSK}= 0.502$, and the finite-length approximations of \eqref{eq:UP-1} and \eqref{eq:LB-1}-\eqref{eq:LB-3} calculated for this example are depicted in Figure~\ref{fig:comp_treepin} for $n\in[2000,20000]$. The bounds
			are converted to rate (both sides are divided by $n$) to show the gap to the WSK capacity. Note that \eqref{eq:LB-1} is the tightest lower bound.
			Though, we also observe that  \eqref{eq:LB-2} is very close to \eqref{eq:LB-1}.

		\end{example}

		\subsection{A Lower Bound for a Special Case}

		In this section, we consider the wiretapped  Tree-PIN  with ${V_{ij} = V_{ji}}$ that is studied in  \cite{Poostindouz2019}.
		For this case, it was proved  that the WSK capacity is $C_{WSK}^{\mc A}(P_{X_\mc{M} Z}) = \min_{\substack{i,j\in\mc M \\ \mathrm{~s.t.~}e_{ij}\in\mc E_{\mc A}}} H(V_{ij}|Z_{ij})$ \cite{Poostindouz2019}.
		We use the lower bound in \cite[Theorem 1]{Hayashi2019}, and give the following finite-length lower bound for $S_{\eps,\sigma}(X^n_{\mc A}|Z^n)$.

		\begin{proposition}\label{thm:semi-lower}
			For wiretapped Tree-PIN ${(X_{\mc M},Z)}$ described by $P_{ZX_{\mc M}}$, with $V_{ij}= V_{ji}$ and for every $n\in\mathds{N}$, every $\eps,\sigma>0$, with $\eps+\sigma<1$, and any subset $\mc A\subseteq\mc M$, we have
			\begin{align}\label{eq:LB-4}
				S_{\eps,\sigma}(X_{\mc A}^n|Z^n) \geq  F_4(X^n_{\mc A}|Z^n) - \frac{1}{2}\log n + \mc O(1), 
			\end{align}
			where
			$$ F_4(X^n_{\mc A}|Z^n) = \min_{\substack{i,j\in\mc M \\ \mathrm{~s.t.~}e_{ij}\in\mc E_{\mc A}}} \left\{ nR_{ij} - \sqrt{n \Delta''_{ij}} Q^{-1}(\frac{\sigma}{2|\mc E_{\mc A}|}) \right\}, $$
			$R_{ij}= H(V_{ij}|Z_{ij})$, $ \Delta''_{ij} = \myvar{-\log P_{V_{ij}|Z_{ji}}},$ and $|\mc E_{\mc A}|$ is the number of edges of $G_{\mc A}$.
		\end{proposition}
		
		Note that \eqref{eq:LB-4} does not depend on $\eps$ as the reconciliation phase is not required for obtaining pairwise keys, and for $\eps=0$, the lower bounds in \eqref{eq:LB-1}  and \eqref{eq:LB-4} are equal up to their second-order term.
		The proof is in Appendix~\ref{sec:app:FL-lower}. %
		
	}%

	{\color{black}
		\section{Extended Models}\label{sec:disc}
		
		In this section, we  extend our capacity result of wiretapped Tree-PIN.  
		While doing so, we compare our results with some important related previous works. 
		We give an upper and a lower bound for the WSK capacity of wiretapped PIN, which is a generalization of the bounds given in \cite{Nitinawarat2010c} for (non-wiretapped) PIN. 
		More importantly, these bounds lead to capacity results for the case wiretapped PIN when $\mc A=\mc M$ or $|\mc A|=2$. 
		We then, review the notion of \emph{wiretapped Markov Trees} which was introduced in \cite{Csiszar2004a}. 
		The WSK capacity of wiretapped Markov Trees is an open problem. 
		We show that a wiretapped PIN is a wiretapped Markov Tree but the converse is not true. 
		Thus, Theorem \ref{thm:Tree-PIN} resolves the capacity problem for a large class of wiretapped Markov Trees -- i.e., wiretapped PIN. 
		Moreover, we show that Theorem \ref{thm:Tree-PIN} can be extended furthermore and gives WSK capacity  for an even larger class of wiretapped Markov Trees. 
		Finally, we consider the case when in a wiretapped PIN there is a non-cooperative compromised terminal. 
		For this case we show that WSK capacity is equal to the PK capacity of the same wiretapped PIN in which  the compromised terminal is cooperative. 
		In fact, this result generalizes Proposition 4.1 of \cite{Nitinawarat2010c}. 
		
	}
	
	\subsection{WSK Capacity of Wiretapped PIN}\label{sec:PIN}

	For the case of  wiretapped PIN (as defined in \ref{def:Tree-PIN}), we give a lower bound and an upper on the WSK capacity. These bounds are tight for the special cases of $\mc A=\mc M$ and $|\mc A|=2$.
	Finding the WSK capacity of a wiretapped PIN as defined in Definition~\ref{def:Tree-PIN} for any given $\mc A$ remains an open problem.

	\begin{proposition}\label{thm:PK-upbound-steiner}
		For any given wiretapped PIN ${(X_{\mc M},Z)}$, described by $G=(\mc M,\mc E)$ and $P_{ZX_{\mc M}}$, and for any $\mc A\subseteq \mc M$, let $R_{ij}=I(V_{ij};V_{ji}|Z_{ij})$, then we have
		\begin{align*}
			C_{WSK}^{\mc A}(P_{X_\mc{M} Z})\leq \min_{\mc P} \left(\frac{1}{|\mc P|-1} \right)\left[ \sum_{\substack{i<j ~\text{s.t.} \\ (i,j) \text{ crosses } \mc P }}  R_{ij} \right],
		\end{align*}
		where  the minimization is over all partitions of $\mc M$ such that for every part of the partition there exists a node in that part that is also in $\mc A$. %
		In a partition $\mc P$ a pair of nodes $(i,j)$ crosses $\mc P$, if $i$ and $j$ are in different parts of $\mc P$.
	\end{proposition}

	\begin{IEEEproof}
		The proof goes along the same lines as the proof in \cite[Example 4]{Csiszar2004a}.
		According to Lemma~\ref{thm:PKisUpp} we know
		$C_{WSK}^{\mc A}(P_{X_\mc{M} Z})\leq C_{PK}^{\mc A|\{m+1\}}(P_{X_\mc{M} Z})$,  and for any $\mc B\subset \mc M$ we have
		\begin{align}\label{eq:Ineq-rate-assignment-app2}
			\sum\limits_{j\in\mc B} R_j & \geq \sum\limits_{\substack{i<j \\ \text{~s.t.~} e_{ij}\in\mc{E}_{\mc{B}}  }} H(V_{ij},V_{ji}|Z_{ij}) + \sum\limits_{\substack{i<j \\ \text{~s.t.~} i\in{\mc{B}} , j\notin  {\mc{B}} }} H(V_{ij}|V_{ji},Z_{ij}).
		\end{align}
		Consider a partition  $\mc P =\{B_1,\ldots, B_{|\mc P|} \}$ of $\mc M$. Then, corresponding to each part of $\mc P$ we have
		\begin{align*}
			\sum\limits_{j\in\mc B_k^c} R_j & \geq H(X_{\mc M}|Z) -\sum\limits_{\substack{i<j \\ \text{~s.t.~} e_{ij}\notin\mc{E}_{\mc{B}_k}  }} H(V_{ij},V_{ji}|Z_{ij}) + \sum\limits_{\substack{i<j\\ \text{~s.t.~} (i,j) \text{~crosses~} \{\mc{B}_k, \mc{B}_k^c \}}} H(V_{ij}|V_{ji},Z_{ij}).
		\end{align*}
		By adding all $|\mc P|$ inequalities, and remembering the fact that $H(V_{ij},V_{ji}|Z_{ij}) = H(V_{ij}|V_{ji},Z_{ij}) + H(V_{ji}|V_{ij},Z_{ij}) + I(V_{ij};V_{ji}|Z_{ij})$,  we get
		\begin{align*}
			(|\mc P| -1)\sum\limits_{j\in\mc M} R_j & \geq |\mc P| H(X_{\mc M}|Z) -\sum_{k=1}^{|\mc P|}  \sum\limits_{\substack{i<j \\ \text{~s.t.~} e_{ij}\notin\mc{E}_{\mc{B}_k}  }} H(V_{ij},V_{ji}|Z_{ij})  + \sum\limits_{\substack{i,j\\ \text{~s.t.~} (i,j) \text{~crosses~} \mc P}} H(V_{ij}|V_{ji},Z_{ij}).
			\\
			& = (|\mc P| -1) H(X_{\mc M}|Z) - \sum_{\substack{i<j ~\text{s.t.}              \\ (i,j) \text{ crosses } \mc P }}  I(V_{ij};V_{ji}|Z_{ij}),
		\end{align*}
		which implies,
		\begin{align*}
			R_{CO}(X_{\mc A}|Z)\geq H(X_{\mc M}|Z) - \frac{1}{|\mc P|-1} \sum_{\substack{i<j ~\text{s.t.} \\ (i,j) \text{ crosses } \mc P }}  I(V_{ij};V_{ji}|Z_{ij}),
		\end{align*} and thus due to Theorem~\ref{thm:PK-Cap-CN04}
		\begin{IEEEeqnarray}{c}
			C_{PK}^{\mc A|\{m+1\}}(P_{X_\mc{M} Z}) \leq \frac{1}{|\mc P|-1} \sum_{\substack{i<j ~\text{s.t.} \\ (i,j) \text{ crosses } \mc P }}  I(V_{ij};V_{ji}|Z_{ij}).  \IEEEnonumber
		\end{IEEEeqnarray} %
		Which is also an upper on the WSK capacity $C_{WSK}^{\mc A}(P_{X_\mc{M} Z})$. \hfill\IEEEQEDhere
	\end{IEEEproof}

	We show that the Steiner tree packing methods of \cite{Nitinawarat2010c} for key agreement, leads to the following lower bound on the WSK capacity of PIN.
	A Steiner tree of $G$ for terminals of $\mc A$ is a subtree of $G$ that spans (connects) all terminals in $\mc A$. A family of edge-disjoint Steiner trees is called a Steiner tree packing~\cite{Diestel2017}. We show that for each family with $\ell$ Steiner trees, a secret key of length $\ell$ can be generated. Let $\mu(G,\mc A)$ denote the maximum cardinality of such family. Therefore, for a general wiretapped PIN %
	we have the following.

	\begin{proposition}\label{thm:Steiner}
		The WSK capacity
		of a wiretapped PIN ${(X_{\mc M},Z)}$ defined by ${G=(\mc M,\mc E)}$ and $P_{Z X_{\mc M}}$
		for any $\mc A\subseteq \mc M$ is lower-bounded by %
		\begin{equation*}
			C_{WSK}^{\mc A}(P_{X_\mc{M} Z})\geq	\sup_{n\in\mc N} \frac{1}{n} \mu(G^n,\mc A),
		\end{equation*}
		where $\mc N$ is the %
		set of $n$'s such that $nI(V_{ij};V_{ji}|Z_{ij})$ for any $(i,j)$ is integer-valued and for each $n$, we define a multigraph $G^{n}=(\mc M,\mc E^n)$ such that for any $e_{ij}\in\mc E$ of $G$ there exists $nI(V_{ij};V_{ji}|Z_{ij})$ edges between nodes $i$ and $j$ in $\mc E^n$.
	\end{proposition}
	\begin{IEEEproof}
		For a given $n\in\mc N$, each pair of connected nodes $(i,j)$ establish  a pairwise key $S_{ij}$ of length approximately equal to $nI(V_{ij};V_{ji}|Z_{ij})$. There exists a Steiner tree packing with cardinality $\mu(G^n,\mc A)$; thus, for any Steiner tree of this Steiner packing, the terminals in $\mc A$ can establish one bit of shared secret key due to Theorem~\ref{thm:Tree-PIN}. Thus, the asymptotic SK rate is  $\sup_{n\in\mc N} \frac{1}{n} \mu(G^n,\mc A)$. 
		Let pairwise keys $S_{ij}$ be all $(\eps_n,\sigma_n)$-SK's such that $\eps_n , \sigma_n \in \mc O(2^{-n})$. 
		We prove that the final key $K$ is an $\eps'_n,\sigma'_n$-SK such that 
		$\lim_{n\to\infty} \eps'_n = \lim_{n\to\infty} \sigma'_n = 0$. 
		The reliability of the final key follows similar to the proof of Theorem~\ref{thm:Tree-PIN}, 
		and $\eps'_n = |\mc E| \eps_n$.
		The security of the final key is as follows.  
		By Corollary~\ref{thm:SD_cor_1} each bit of pairwise keys is also $\sigma_n$ secure. 
		By Lemma~\ref{lemma:Ach-Tree} each bit of the final key is $2(m-1)\sigma_n$ secure, and 
		by Corollary~\ref{thm:SD_cor_3} the final key is $\sigma'_n = 2 (m-1)(\log |\mc K|)\sigma_n$ 
		secure\footnote{We note that one can use our techniques 
			presented in the security part of the proof of Lemma~\ref{lemma:Ach-Tree} 
			to show a tighter secrecy bound,  that is $\sigma'_n = 3|\mc E|\sigma_n$,  
			without requiring $\sigma_n$ to decay exponentially in $n$. 
			However, the presented 
			proof here is more straightforward and suffices for the capacity results in Corollary~\ref{thm:tightness-AeqM}.  
		}. 
		Since we chose $\eps_n , \sigma_n \in \mc O(2^{-n})$, we have $\lim_{n\to\infty} \eps'_n = \lim_{n\to\infty} \sigma'_n = 0$. 
	\end{IEEEproof}

	{\color{black}
		\begin{corollary}%
			\label{thm:tightness-AeqM}
			For the special case of $\mc A=\mc M$ or $|\mc A|=2$, the problem of calculating $\mu(G^n,\mc A)$ is
			efficiently solvable~\cite{Diestel2017}; rendering the above lower bound of Preposition \ref{thm:Steiner} achieving the upper bound of Preposition \ref{thm:PK-upbound-steiner} if $\mc A=\mc M$ or $|\mc A|=2$.
		\end{corollary}

		\begin{IEEEproof}
			It has been proven \cite[See Menger's theorem in Section 3.3]{Diestel2017} that When $|\mc A|=2$ then the problem of maximal Steiner Tree Packing in multigraph $G^{n}=(\mc M, \mc E^n)$ will reduce to the problem of finding maximum number of edge-disjoint paths connecting the two terminals in $\mc A$. Thus, for any multigraph $G^{n}=(\mc M, \mc E^n)$ and any arbitrary subset $\mc A\subseteq \mc M$ with $|\mc A|=2$ we have
			\begin{align*}
				\mu(G^{n},\mc A) = \min_{\substack{\mc B\subsetneq \mc M \\ \text{s.t. } \mc A\nsubseteq \mc B }} \left\vert \left\{e_{ij}\in \mc E^n | (i,j) \text{ crosses } \mc P =\{\mc B,\mc B^c\}  \right\} \right \vert .
			\end{align*}
			Therefore, we will have the following lower bound.
			\begin{align*}
				C_{WSK}^{\mc A}(P_{X_\mc{M} Z}) & \stackrel{(a)}{\geq} \sup_{n\in\mc N} \frac{1}{n} \mu(G^n,\mc A) \\
				& \stackrel{(b)}{=} \min_{\substack{\mc B\subsetneq \mc M          \\ \text{s.t. } \mc A\nsubseteq \mc B }}  \left[ \sum_{\substack{i<j ~\text{s.t.} \\ (i,j) \text{ crosses } \mc P=\{\mc B,\mc B^c\}   }}  I(V_{ij};V_{ji}|Z_{ij}) \right] \\
				& \stackrel{(c)}{=} C_{PK}^{\mc A|\{m+1\}}(P_{X_\mc{M} Z}),        
			\end{align*} where (a) is due to Corollary~\ref{thm:Steiner}, (b) is due to Menger's Theorem, and (c) is due to Lemma~\ref{thm:PK-upbound-steiner}. This proves the tightness of the bound in Corollary~\ref{thm:Steiner} for $|\mc A|=2$.
			
			For the special case of $\mc A=\mc M$, in the problem of maximal Steiner Tree Packing in multigraph $G^{n}=(\mc M, \mc E^n)$ the exact value of $\mu(G^n,\mc M)$ is known due to the Tutte/Nash-Williams Theorem~\cite[Section 3.5]{Diestel2017}, which is
			\begin{align*}
				\mu(G^n,\mc M) = \min_{\mc P} \floor{\frac{  \left\vert \left\{e_{ij}\in \mc E^n | (i,j) \text{ crosses } \mc P \right\} \right \vert    }{|\mc P|-1} }. 
			\end{align*}
			Therefore, we have %
			\begin{align*}
				C_{WSK}^{\mc M}(P_{X_\mc{M} Z}) & \stackrel{(a)}{\geq} \sup_{n\in\mc N} \frac{1}{n} \mu(G^n,\mc M)                                        \\
				& \stackrel{(b)}{=} \min_{\mc P} \left(\frac{1}{|\mc P|-1} \right)\left[ \sum_{\substack{i<j ~\text{s.t.} \\ (i,j) \text{ crosses } \mc P }}  I(V_{ij};V_{ji}|Z_{ij}) \right],\\
				& \stackrel{(c)}{=} C_{PK}^{\mc M|\{m+1\}}(P_{X_\mc{M} Z}),                                               
			\end{align*}  where  (a) is due to Corollary~\ref{thm:Steiner}, (b) is due to Tutte/Nash-Williams Theorem, and (c) is due to Lemma~\ref{thm:PK-upbound-steiner}.  This proves the tightness of the bound in Corollary~\ref{thm:Steiner} for  $\mc A=\mc M$.
		\end{IEEEproof}

		\begin{figure}[t]
			\centering
			\begin{adjustbox}{width = 0.4\linewidth}
				\begin{tikzpicture}[font=\bfseries\Huge]
					\tikzstyle{place}=[circle,draw=black!90,fill=blue!10,very thick,inner sep=3pt, minimum size=42pt,line width=2pt,font=\bfseries\Huge]
					\tikzstyle{eve}=[circle,draw=black!90,fill=red!10,very thick,inner sep=3pt, minimum size=52pt,line width=2pt]
					
					\node (n1) at ( 10,10) [place] {1};
					\node (n2) at ( 0,10) [place] {2};
					\node (n3) at ( 0,0) [place] {3};
					\node (n4) at ( 10,0) [place] {4};

					\node (eve) at (5,5) [eve] {Eve};

					\draw[red!60!orange,line width=7pt,round cap-round cap,shorten <=1mm,shorten >=1mm] (n1.west) node [above left] {$V_{12}$} -- (n2.east) node [above right] {$V_{21}$};
					\draw[snake=snake, line before snake=10mm, line after snake=10mm,segment aspect=5, segment amplitude=5pt,->,red!60!orange,line width=3pt,round cap-round cap,shorten <=2mm,shorten >=1mm] (n2.east) -- (eve) node [pos=0.8, above right] {$Z_{12}$};

					\draw[blue!70!green,line width=7pt,round cap-round cap,shorten <=1mm,shorten >=1mm] (n2.south) node [below left] {$V_{23}$} -- (n3.north) node [ above left] {$V_{32}$} ;
					\draw[snake=snake, line before snake=10mm, line after snake=10mm,segment aspect=5, segment amplitude=5pt,->,blue!70!green,line width=3pt,round cap-round cap,shorten <=2mm,shorten >=1mm] (n2.south) -- (eve) node [pos=0.8, below left] {$Z_{23}$};

					\draw[black!50!red,line width=7pt,round cap-round cap,shorten <=1mm,shorten >=1mm] (n4.west) node [below left] {$V_{43}$} -- (n3.east) node [below right] {$V_{34}$} ;
					\draw[snake=snake, line before snake=10mm, line after snake=10mm,segment aspect=5, segment amplitude=5pt,->,black!50!red,line width=3pt,round cap-round cap,shorten <=2mm,shorten >=1mm] (n4.west) -- (eve) node [pos=0.8, below left] {$Z_{34}$};

					\draw[black!70!green,line width=7pt,round cap-round cap,shorten <=1mm,shorten >=1mm] (n4.north) node [above right] {$V_{41}$} -- (n1.south) node [ below right] {$V_{14}$} ;
					\draw[snake=snake, line before snake=10mm, line after snake=10mm,segment aspect=5, segment amplitude=5pt,->,black!70!green,line width=3pt,round cap-round cap,shorten <=2mm,shorten >=1mm] (n1.south) -- (eve) node [pos=0.8, below  right] {$Z_{14}$};

				\end{tikzpicture}
			\end{adjustbox}
			
			\caption[The wiretapped PIN of Example~\ref{sec:example-STP}.]{The wiretapped PIN of Example~\ref{sec:example-STP}. Here we have 4 terminals, $\mc M=\mc A=\{1,2,3,4\}$ and the connectivity graph is given by $G=(\mc M,\mc E)$, where $\mc E=\{e_{12}, e_{23}, e_{34}, e_{41}\}$. Terminals variables are $X_1=(V_{12},V_{14})$, $X_2=(V_{21},V_{23})$, $X_3 = (V_{32},V_{34})$, and $X_4=(V_{41},V_{43})$. Eve's side information is $Z=(Z_{12},Z_{23}, Z_{34}, Z_{41})$, where the  following Markov relations hold: $V_{12} - V_{21} - Z_{12}$, $V_{32} - V_{23} - Z_{23}$, $V_{34} - V_{43} - Z_{34}$, and $V_{41} - V_{14} - Z_{14}$.}
			\label{fig:sq}
		\end{figure}

		\begin{example}\label{sec:example-STP}

			To illustrate the result of Corollary~\ref{thm:tightness-AeqM}, we give the following simple example. Let $m=4$, and $\mc A = \mc M = \{1,2,3,4\}$ and assume that $G=(\mc M,\mc E)$ is a square as depicted in Figure~\ref{fig:sq}. We also assume that for any $e_{ij}\in\mc E$, $V_{ij}-V_{ji}-Z_{ij}$, such that $R_{ij} = I(V_{ij};V_{ji}|Z_{ij})=1/2$. According to Corollary~\ref{thm:tightness-AeqM} and Preposition~\ref{thm:PK-upbound-steiner},
			for this example we have
			\begin{equation*}
				C_{WSK}^{\mc M}(P_{X_\mc{M} Z}) = \frac{1}{3} \sum_{(i,j) \text{~crosses~} \mc P} R_{ij} = \frac{2}{3},
			\end{equation*}
			where the minimizing partition is $\mc P = \{\{1\},\{2\},\{3\},\{4\} \}$. To see how the Steiner tree packing method attains this WSK capacity, we first note that if according to each edge $e_{ij}\in\mc E$, terminals have obtained pairwise SKs of length $3$ bits, then a group secret key of length $4$ bits can be generated. The reason is that, $G^3=(\mc M,\mc E^3)$ of the square can be decomposed by $4$ edge-disjoint trees, and corresponding to each tree one bit of group SK can be generated. This Steiner tree packing is demonstrated in Figure~\ref{fig:3sq-packing}.
			
			\begin{figure}[b]
				\centering
				\includegraphics[width=0.95\linewidth]{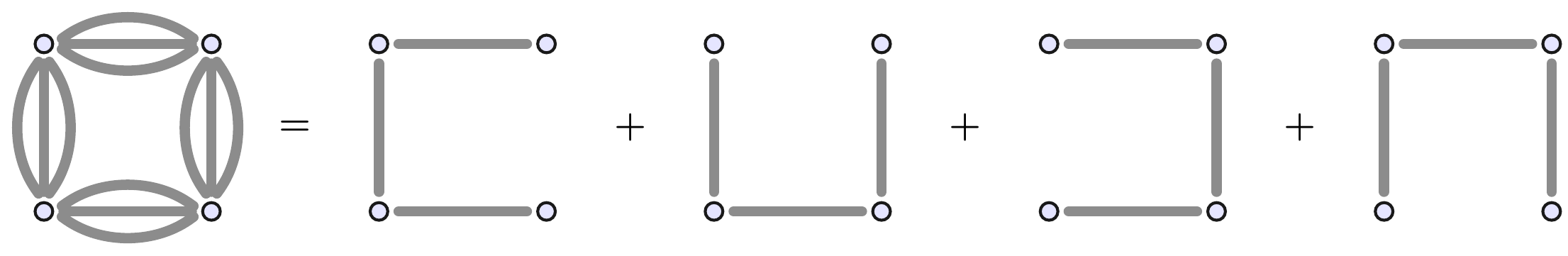}
				\caption{Steiner packing of $G^3$ into $4$ edge-disjoint trees.}
				\label{fig:3sq-packing}
			\end{figure}
			
			Recall that for any large enough $n$,  each pair of connected terminals can obtain pairwise keys of length $\ell_{ij}(n) \approx n\times R_{ij} = n/2$ for all $e_{ij}\in\mc E$. Thus, for any $n$ we find $a$ and $b$ such that $\ell_{ij}(n) = 3\times b + a$, and thus the final group key will have length of $\ell(n)=4\times b + a$. As $n\to\infty$, we will have $\ell(n)/n \to 2/3,$ that is the WSK capacity given by Preposition~\ref{thm:PK-upbound-steiner}. 

		\end{example}

		\subsection[A Generalization of Wiretapped Tree-PIN]{Comparison with Wiretapped Markov Trees and Generalizing Wiretapped Tree-PIN}\label{sec:Markov_tree}

		As examples of the general source model, \citeauthor{Csiszar2004a} introduced the notion of Markov chain on a tree and its wiretapped analogue. 
		We first define the notion of Markov chain on a tree (or Markov Tree in short) as defined in \cite{Csiszar2004a}. 

		\begin{definition}[Markov Tree]\label{def:Markov_Tree}
			Let $\mc M=[m]$ be a set of $m$ terminals, and let $G=(\mc M, \mc E)$  be an undirected tree. Note that for any $e_{ij}\in\mc E$ we can partition $\mc M$ into two sets $\mc B_i$ and $\mc B_j$ such that $\mc M=\mc B_i \cup \mc B_j$, $i\in\mc B_i$, and $j\in\mc B_j$. A source model $P_{X_{\mc M}}$ forms a Markov chain on $G$ if for any $e_{ij}\in\mc E$ we have $\pr{X_i|X_{\mc B_j}} = \pr{X_i|X_j}.$ 
			A special case of such source models is the case when we have $X_1-X_2-X_3-\cdots - X_m$. 
		\end{definition}

		For any Markov Tree described by $P_{X_{\mc M}}$, it is proved that 
		\begin{equation}\label{eq:SK_Cap_Markov_Tree}
			C_{SK}^{\mc A}(P_{X_{\mc M}}) = \min_{\substack{i,j\in\mc M \\ \mathrm{~s.t.~}e_{ij}\in\mc E_{\mc A}}} I(X_i;X_j),
		\end{equation}
		where $G_{\mc A} = (\mc M_{\mc A}, \mc E_{\mc A})$ is the smallest subtree connecting all nodes of $\mc A$. See Example 7, Equation (36) of \cite{Csiszar2004a}. The same equation also holds for any given non-wiretapped Tree-PIN -- that is implied by Theorem~\ref{thm:Tree-PIN} when $Z=\text{constant}$. In fact we observe that any Tree-PIN is also a Markov Tree but the converse is not true.

		Next, we define the notion of wiretapped Markov chain on a Tree (or wiretapped Markov Tree for short), which was defined first in \cite{Csiszar2004a}. 
		\begin{definition}[Wiretapped Markov Tree]\label{def:Wirtp_Markov_Tree}
			Consider a model $P_{ZX_{\mc M}}$ where $\mc M=[m]$ is the set of $m$ terminals and $Z$ is Eve's side information. If $Z$ is of the form $Z=\{Z_1,Z_2,\ldots,Z_m\}$ then we can define an auxiliary model as follows. Let %
			$\mc M'=\{m+1,\ldots,2m\}$ be the set of $m$ dummy terminals. Let terminals in $\mc M$ have access to RVs $X_j$ for all $j\in \mc M$, and let dummy terminals in $\mc M'$ have access to RVs $Z_{j-m}$ for all $j\in \mc M'$. Thus the probability distribution of the auxiliary model defined over $\overline{\mc M} = \mc M \cup \mc M'= \{1,2,\ldots,2m\}$, is $P_{X_{\mc M} Z_{\mc M}} = P_{X_{\mc M} Z}$. 
			Any wiretapped SKA model with distribution $P_{X_{\mc M}Z}$ is called a wiretapped Markov chain on a Tree if, Eve's side information $Z$ is of the form $Z=\{Z_1,Z_2,\ldots,Z_m\}$ such that $\pr{Z_\mc M|X_\mc M} = \Pi_{j\in\mc M} \pr{Z_j|X_j}$, and if its corresponding auxiliary model defined over $\overline{\mc M} = \{1,2,\ldots,2m\}$ forms a Markov chain on a tree (according to definition \ref{def:Markov_Tree}).  See an example of such model in the figure \ref{fig:Wiretapped_Markov_tree} below. 
		\end{definition}

		\begin{figure}[t]
			\centering
			\begin{adjustbox}{ width =0.6\linewidth}
				
				\begin{tikzpicture}[font=\bfseries\Huge]
					\tikzstyle{place}=[circle,draw=black!90,fill=blue!10,very thick,inner sep=3pt, minimum size=42pt,line width=2pt,font=\bfseries\Huge]
					\tikzstyle{eve}=[circle,draw=black!90,fill=red!10,very thick,inner sep=3pt, minimum size=42pt,line width=2pt]

					\node (n1) at ( 0,0) [place] {1};
					\node (n2) at ( 5,0)   [place] {2};
					\node (n3) at ( 5,-5)  [place] {3};
					
					\node (e1) at ( -8,0) [eve] {4};
					\node (e2) at ( 13,0)   [eve] {5};
					\node (e3) at ( 13,-5)  [eve] {6};

					\node [above] at (n1.north) {$X_1$};
					\node [above] at (n2.north) {$X_2$};
					\node [below left] at (n3.west) {$X_3$};
					
					\node [above] at (e1.north) {$Z_1$};
					\node [above] at (e2.north) {$Z_2$};
					\node [above] at (e3.north) {$Z_3$};

					\draw[draw=gray!90!,line width=7pt,round cap-round cap,shorten <=1mm,shorten >=1mm] (n1) -- (n2);
					\draw[draw=gray!90!,line width=7pt,round cap-round cap,shorten <=1mm,shorten >=1mm] (n2) -- (n3);

					\draw[snake=snake, line before snake=10mm, line after snake=10mm,segment aspect=5, segment amplitude=5pt,draw=gray!90!,line width=3pt,round cap-round cap,shorten <=1mm,shorten >=1mm] (n1) -- (e1);
					
					\draw[snake=snake, line before snake=10mm, line after snake=10mm,segment aspect=5, segment amplitude=5pt,draw=gray!90!,line width=3pt,round cap-round cap,shorten <=1mm,shorten >=1mm] (n2) -- (e2);
					
					\draw[snake=snake, line before snake=10mm, line after snake=10mm,segment aspect=5, segment amplitude=5pt,draw=gray!90!,line width=3pt,round cap-round cap,shorten <=1mm,shorten >=1mm] (n3) -- (e3);

				\end{tikzpicture}
			\end{adjustbox}
			\caption[A simple wiretapped Markov Chain on a Tree with three terminals.]{A simple wiretapped Markov Chain on a Tree with three terminals. Here, the terminals variables $X_1,X_2, X_3$ and Eve's side information components $Z_1, Z_2, Z_3$ form a Markov Tree. The WSK capacity of this model is still unknown.}
			\label{fig:Wiretapped_Markov_tree}
		\end{figure}

		\begin{figure}[h!]
			\centering
			\begin{adjustbox}{ width =0.6\textwidth}
				
				\begin{tikzpicture}[font=\bfseries\Huge]
					\tikzstyle{place}=[circle,draw=black!90,fill=blue!10,very thick,inner sep=3pt, minimum size=42pt,line width=2pt,font=\bfseries\Huge]
					\tikzstyle{eve}=[circle,draw=black!90,fill=red!10,very thick,inner sep=3pt, minimum size=42pt,line width=2pt]

					\node (n1) at ( 0,0) [place] {1};
					\node (n2) at ( 5,0)   [place] {2};

					\node (e1) at ( -8,0) [eve] {3};
					\node (e2) at ( 13,0)   [eve] {4};

					\node [above] at (n1.north) {$X_1$};
					\node [above] at (n2.north) {$X_2$};

					\node [above] at (e1.north) {$Z_1$};
					\node [above] at (e2.north) {$Z_2$};

					\draw[draw=gray!90!,line width=7pt,round cap-round cap,shorten <=1mm,shorten >=1mm] (n1) -- (n2);

					\draw[snake=snake, line before snake=10mm, line after snake=10mm,segment aspect=5, segment amplitude=5pt,draw=gray!90!,line width=3pt,round cap-round cap,shorten <=1mm,shorten >=1mm] (n1) -- (e1);
					
					\draw[snake=snake, line before snake=10mm, line after snake=10mm,segment aspect=5, segment amplitude=5pt,draw=gray!90!,line width=3pt,round cap-round cap,shorten <=1mm,shorten >=1mm] (n2) -- (e2);

				\end{tikzpicture}
			\end{adjustbox}
			\caption[A simple wiretapped Markov Chain on a Tree with two terminals.]{A simple wiretapped Markov Chain on a Tree with two terminals. Here, the terminals variables and Eve's side information satisfy the Markov relation of $Z_1 - X_1 - X_2 - Z_2$. The WSK capacity of this model is still unknown.}
			\label{fig:Wiretapped_Markov_tree_m_is_2}
		\end{figure}

		Unfortunately, for the wiretapped Markov Tree model defined in definition \ref{def:Wirtp_Markov_Tree},
		where all terminals are wiretapped, the WSK capacity is not known\footnote{In \cite{Csiszar2004a} the authors mistakenly claim to prove the WSK capacity of all wiretapped Markov Trees. See the remark after Theorem 5.1 in \cite{Csiszar2008}.}, even for the special case when $m=2$ (see Figure \ref{fig:Wiretapped_Markov_tree_m_is_2} below). 
		The WSK capacity is proved \cite{Csiszar2008} for wiretapped Markov Trees where only one terminal (say terminal 1) is wiretapped, that is $Z=Z_1$ and $Z_j=\text{constant}~\forall j\neq 1$. 
		We observe that every wiretapped Tree-PIN is a wiretapped Markov Tree  but the converse is not true. 
		Even though the WSK capacity is not known  for all wiretapped Markov Trees, Theorem \ref{thm:Tree-PIN}, proves the WSK capacity for a large subset of wiretapped Markov Trees. For the special case of $m=2$ our wiretapped Tree-PIN model and our main result reduces to the well-known case of $X_1 - X_2 - Z$ \cite{Ahlswede1993,Maurer1993}, where $X_1 = V_{12}$, $X_2=V_{21}$, and $Z=Z_{12}$.

		We can extend our model of wiretapped Tree-PIN and obtain a generalized version of Theorem \ref{thm:Tree-PIN}.  
		In this case for each pair of connected terminals $i$ and $j$ we assume two sets of correlated variables $(V_{ij}^a,V_{ji}^a,Z_{ij}^a)$ and $(V_{ij}^b,V_{ji}^b,Z_{ji}^b)$. 
		
		\begin{definition}[General Wiretapped Tree-PIN]\label{def:Tree-PIN_gen}
			A set of $m$ terminals form a ``General Wiretapped Tree-PIN'' if there exists a tree $G=(\mc M,\mc E)$ with $\mc M=[m]$ such that the RV of any terminal $j\in \mc M$ can be represented by $X_j=(V_{ji}^\theta|~i \in\Gamma(j), \theta\in\{a,b\})$, %
			where Eve's side information is of the form $Z=(Z_{ij}^\theta, i\in\mc M, j\in\mc M, \theta\in\{a,b\})$ and
			all pairs of RVs in $\{(V_{ij}^\theta,V_{ji}^\theta,Z_{ij}^\theta) |~\theta\in\{a,b\}, i<j \text{~and~} e_{ij}\in\mc E \}$ are mutually independent, such that $V_{ij}^\theta- V_{ji}^\theta - Z_{ij}^\theta$ for all $i,j\in\mc M$ and any $\theta\in\{a,b\}$. 
		\end{definition}

		Note that any general wiretapped Tree-PIN is a wiretapped Markov Tree, but the converse is not true.

		\begin{example}\label{exm:two-way}

			For the two-party SKA, the general wiretapped Tree-PIN model of definition \ref{def:Tree-PIN_gen} reduces to a case where both terminals are wiretapped. See figure below.

			\begin{figure}[t]
				\centering
				\begin{adjustbox}{ width =0.6\textwidth}
					
					\begin{tikzpicture}[font=\bfseries\Huge]
						\tikzstyle{place}=[circle,draw=black!90,fill=blue!10,very thick,inner sep=3pt, minimum size=42pt,line width=2pt,font=\bfseries\Huge]
						\tikzstyle{eve}=[circle,draw=black!90,fill=red!10,very thick,inner sep=3pt, minimum size=42pt,line width=2pt]

						\node (n1) at ( 0,0) [place] {1};
						\node (n2) at ( 5,0)   [place] {2};
						
						\node (e1) at ( -8,0) [eve] {E};
						\node (e2) at ( 13,0)   [eve] {E};

						\node [above] at (n1.north) {$V_{12}^a$};
						\node [above] at (n2.north) {$V_{21}^a$};
						
						\node [above] at (e2.north) {$Z_{12}^a$};

						\node [below] at (n1.south) {$V_{12}^b$};
						\node [below] at (n2.south) {$V_{21}^b$};
						
						\node [below] at (e1.south) {$Z_{12}^b$};

						\draw[draw=blue!40!red,line width=7pt,round cap-round cap,shorten <=1mm,shorten >=1mm] (n1.north east) -- (n2.north west);
						
						\draw[draw=yellow!40!red,line width=7pt,round cap-round cap,shorten <=1mm,shorten >=1mm] (n1.south east) -- (n2.south west);

						\draw[snake=snake, line before snake=10mm, line after snake=10mm,segment aspect=5, segment amplitude=5pt,draw=yellow!40!red,line width=3pt,round cap-round cap,shorten <=1mm,shorten >=1mm] (n1.south west) -- (e1.south east);
						
						\draw[snake=snake, line before snake=10mm, line after snake=10mm,segment aspect=5, segment amplitude=5pt,draw=blue!40!red,line width=3pt,round cap-round cap,shorten <=1mm,shorten >=1mm] (n2.north east) -- (e2.north west);

					\end{tikzpicture}
				\end{adjustbox}
				\caption[A general wiretapped Tree-PIN  with two terminals.]{A general wiretapped Tree-PIN  with two terminals. 
					Here both terminals labeled ``E'' represent the adversary Eve. 
					Since the Markov relations $V_{12}^a- V_{21}^a - Z_{12}^a$ and $V_{21}^b- V_{12}^b - Z_{12}^b$ hold, we 
					have $Z_{12}^a - V_1 - V_2 - Z_{12}^b$ which resembles the Markov relation in the Markov Tree example of Figure~\ref{fig:Wiretapped_Markov_tree_m_is_2}. 
				}
				\label{fig:gen_tree_pin_m_is_2}
			\end{figure}

			For this case we prove that 
			\begin{equation}\label{eq:Gen2P_WSK}
				C_{WSK}(P_{X_1,X_2,Z}) = I(V_{12}^a;V_{21}^a|Z_{12}^a) + I(V_{12}^b;V_{21}^b|Z_{12}^b).
			\end{equation}

			\begin{IEEEproof}[Proof of Equation \ref{eq:Gen2P_WSK}] 
				The achievablity follows directly from Lemma \ref{lemma:Ach-Tree} applied two times, once for $\theta=a$   and once for $\theta = b$. The converse follows from lemma \ref{thm:PKisUpp} and Theorem \ref{thm:PK-Cap-CN04}. That is
				\begin{IEEEeqnarray*}{rCl+x*}
					C_{WSK}(P_{X_1,X_2,Z}) &\leq& C_{PK}(P_{X_1,X_2,Z}) \\
					&=& H(V_{12}^a,V_{21}^a|Z_{12}^a) + H(V_{12}^b,V_{21}^b|Z_{12}^b)  - R_{CO}(X_1,X_2|Z)\\
					&=& H(V_{12}^a,V_{21}^a|Z_{12}^a) - H(V_{12}^a|V_{21}^a,Z_{12}^a)  - H(V_{21}^a|V_{12}^a,Z_{12}^a) \\
					&&\quad +H(V_{12}^b,V_{21}^b|Z_{12}^b)  - H(V_{12}^b|V_{21}^b,Z_{12}^b)  - H(V_{21}^b|V_{12}^b,Z_{12}^b)  \\
					&=&  I(V_{12}^a;V_{21}^a|Z_{12}^a) + I(V_{12}^b;V_{21}^b|Z_{12}^b).  & \hfill\IEEEQEDhere
				\end{IEEEeqnarray*}
			\end{IEEEproof}
		\end{example}
		
		Note that the two-party SKA model of Figure \ref{fig:Wiretapped_Markov_tree_m_is_2} is more general than the model in Figure \ref{fig:gen_tree_pin_m_is_2}. The WSK capacity of the model of Figure \ref{fig:Wiretapped_Markov_tree_m_is_2} is still unresolved, while for the case of general wiretapped Tree-PIN models, including the model of Figure \ref{fig:Wiretapped_Markov_tree_m_is_2} can be proved. Moreover, it is easy to see that the following holds -- the proof follows the same argument of the proof of Theorem \ref{thm:Tree-PIN} with considering the independence of (a) and (b) variables.
		
		\begin{proposition}[WSK capacity of general wiretapped Tree-PIN]\label{thm:gen_tree_pin}
			The WSK capacity of a given general wiretapped Tree-PIN ${(X_{\mc M},Z)}$, defined as in Definition~\ref{def:Tree-PIN_gen}, for
			any subset $\mc A\subseteq\mc M$ is 
			\begin{align}\label{eq:WSK_Cap_Tree-PIN_gen}
				C_{WSK}^{\mc A}(P_{X_\mc{M} Z}) = \min_{\substack{i,j\in\mc M \\ \mathrm{~s.t.~}e_{ij}\in\mc E_{\mc A}}} I(V_{ij}^a;V_{ji}^a|Z_{ij}^a) + I(V_{ij}^b;V_{ji}^b|Z_{ij}^b),
			\end{align} where $G_{\mc A} = (\mc M_{\mc A}, \mc E_{\mc A})$ is the smallest subtree connecting all nodes of $\mc A$. 
		\end{proposition}
		
		Note that Preposition \ref{thm:gen_tree_pin} generalizes Theorem \ref{thm:Tree-PIN}
		as it implies the case of Theorem \ref{thm:Tree-PIN} when $\theta\in\{a\}$.

		\subsection{The Case of a Non-cooperative Compromised Terminal}\label{sec:non-coop}
		
		Consider a wiretapped  PIN defined by $G=(\mc M, \mc E)$. 
		Recall that terminals RVs are defined by $X_j = (V_{ji}|~ i\in \Gamma(j))$. 
		Furtherer assume that one terminal (denoted by $\mc D= \{d\}$) is compromised and is not cooperating 
		with the SKA. 
		Thus, Eve's side information is
		given by $Z=(Z_{jk}|~ e_{jk}\in\mc E) \text{~and~} X_{\mc D}$. 
		The following theorem gives the secrecy capacity of this model, which we denote by $C_W(G)$ for simplicity.  
		
		\begin{proposition}
			For a given wiretapped  PIN defined by $G=(\mc M, \mc E)$ with a non-cooperative 
			compromised terminal denoted by $\mc D = \{d\}$, define the following associated model. 
			Let $\tilde G = (\mc{\tilde{M}}, \mc{\tilde{E}})$, where $\mc{\tilde{M}} = \mc M\setminus \mc D$ and 
			$\mc{\tilde{E}} = \mc E \setminus \{ e_{dj}|~j\in\Gamma(d) \}$. 
			Eve's side information of the associated model is also defined by 
			$\tilde Z=(Z_{jk}|~ e_{jk}\in\mc{\tilde{E}})$. Then $C_W(G) = C_{W}(\tilde G)$ where $C_{W}(\tilde G)$ is the 
			WSK capacity of the associated wiretapped PIN model. 
		\end{proposition}
		
		\begin{IEEEproof}
			The proof follows along the same line as for the proof of Proposition 4.1 of \cite{Nitinawarat2010c}. 
			We show that
			\begin{equation*}
				C_W(\tilde G) \stackrel{\text{(a)}}{\leq} 
				C_W(G) \stackrel{\text{(b)}}{\leq} 
				C_P(G) \stackrel{\text{(c)}}{\leq} 
				C_W(\tilde G) ,
			\end{equation*} 
			where $C_P(G)$ denoted the secrecy capacity of model $G$ when compromised terminal is cooperative. 
			To prove (a) we argue that 
			a secrete key for model $\tilde{G}$ also constitutes a valid secret key for model $G$. 
			Let $Z=(\tilde Z , Z_d , X_d)$ where $Z_d = (Z_{dj}|~j\in\Gamma(d))$. 
			Let $K$ be secrete key established for model $\tilde{G}$ by public communication $\vect F$
			By the independence of $(Z_d , X_d)$ from $(K,\vect F, \tilde{Z})$ and due to corollary \ref{thm:SD_cor_2}, we have
			\begin{equation*}
				\sd((K,\vect F, Z) , (U, \vect F, Z)) = \sd((K,\vect F, \tilde Z ), (U, \vect F, \tilde Z)),
			\end{equation*}
			which completes the proof of (a). 
			Relation (b) is due to Lemma~\ref{thm:PKisUpp} and 
			to prove (c) we show that a secret key based on the protocol
			that achieves $C_P(G)$ can be used to generate key for model $\tilde{G}$. 
			In model $\tilde{G}$ one terminal, e.g., terminal 1, can use local randomization and 
			simulate 
			$X_d^n$ (since the source distribution is assumed to be known) and reveal it via public communication. 
			Then all terminals can independently simulate their correlated RVs with respect to the compromised terminal $d$. 
			Therefore, a model is simulated (or emulated) by terminals 
			such that terminal $d$ is compromised and its RV is revealed. 
			Thus, the protocol that achieves $C_P(G)$  can be executed for SKA. 
			Hence,  $C_P(G)$  constitutes a lower bound for $C_W(\tilde G)$.
		\end{IEEEproof}

		The above results can be regarded as a generalization for %
		Proposition 4.1 of \cite{Nitinawarat2010c} in which $(Z_{jk}|~ e_{jk}\in\mc E) = \text{constant}$.

	}%

	\section{Need for Interaction in Source Model SKA}\label{sec:interaction}

	Let $N_{PC}$ denote the number of public communication rounds of an SKA protocol. %
	For noninteractive SKA protocols we have $N_{PC}=1$, and for interactive ones %
	$N_{PC}>1$. 
	For two-party SKA in source model, considering the key capacity achieving protocols that use at least use one public message, 
	the following three types of interactions 
	have been studied \cite{Ahlswede1993,Gohari2010,Hayashi2016}.  
	(We note that, as shown in  \cite{Ahlswede2006,Gacs1973}, for  two-party non-wiretapped source model, achieving the maximum rate of common randomness extraction %
	requires public communication, and  %
	two-party SK capacity $C_{SK} = I(X_1 ; X_2)$  in general is not achievable  without using at least a single public message.) 
	
	First, is ``\emph{one-way}'' in which only one party (terminal 1, or Alice) sends a public message to the other party (terminal 2, or Bob). 
	Second, is  when each party sends a single public message that is independent of other parties' message. 
	Both these %
	are noninteractive. %
	The third type is  ``\emph{interactive}'' SKA  where
	$N_{PC}>1$ and  
	in each round,  each terminal (party) sends a single message  that is a function of the terminal's private samples  
	and previous public messages, and is independent of the other message in the same round.   
	The next round begins when all sent public messages are received by all terminals. 
	See Figure~\ref{fig:int_modes}.  
	The general key capacity of an adversarial  model SK, PK, or WSK  upper bounds  the noninteractive key capacity of the model, and in general we have 
	$C_{XK}^{\rightarrow} \leq C_{XK}^{NI} \leq C_{XK}$, where 
	$C_{XK}^{\rightarrow}$ 
	denotes the one-way key capacity, 
	$C_{XK}^{NI}$ 
	denotes noninteractive key capacity,  
	$C_{XK}$ denotes the key capacity when interaction is allowed,  
	and $XK\in \{SK,PK,WSK\}$. 
	In the following, we review %
	previous results obtained regarding the required %
	interaction to achieve the key capacity.

	\begin{figure}[t]
		\centering
		\begin{adjustbox}{width =0.99\textwidth}
			
			\begin{tikzpicture}[font=\bfseries\Huge]
				\tikzstyle{place}=[circle,draw=black!90,fill=blue!10,very thick,inner sep=3pt, minimum size=42pt,line width=2pt,font=\bfseries\Huge]
				\tikzstyle{eve}=[circle,draw=black!90,fill=red!10,very thick,inner sep=3pt, minimum size=42pt,line width=2pt]

				\node (n1) at ( 0,0) [place] {1};
				\node (n2) at ( 7,0) [place] {2};
				\node[above=of n1] (alice) {Alice};
				\node[above=of n2] (bob) {Bob};
				\node[anchor=base] (title) at (3,4) {One-way SKA};
				\draw[->,black,thick] (0,-1) -- (0,-10);
				\draw[->,black,thick] (7,-1) -- (7,-10);

				\draw[->,color=green!50!black,line width=2pt] (0,-1) -- (7,-2);
				\draw[snake=brace] (7.1,-1) -- (7.1,-2) node[right,midway,font=\large] {$t=1$};

				\def\mywall{(0.01,-2.2) rectangle (6.99,-2.6)}; 
				\fill[color=brown!60!red] \mywall;
				\pattern[pattern color=black,pattern=none] \mywall;
				\node[font=\large] (end) at (3.5,-3) {End of Public Communication};
				
				\node[ ] (npc) at (3.5,-11) {$N_{PC}=1$};
				\node[ ] (fve) at (3.5,-12.5) {$\vect{F}=F_1$};

				\begin{scope}[xshift=12cm]
					\node (n1) at ( 0,0) [place] {1};
					\node (n2) at ( 7,0) [place] {2};
					\node[above=of n1] (alice) {Alice};
					\node[above=of n2] (bob) {Bob};
					\node[anchor=base] (title) at (3.5,4) {Noninteractive SKA};
					\draw[->,black,thick] (0,-1) -- (0,-10);
					\draw[->,black,thick] (7,-1) -- (7,-10);

					\draw[->,color=green!50!black,line width=2pt] (0,-1) -- (7,-2);
					\draw[->,color=green!50!black,line width=2pt] (7,-1) -- (0,-2);
					\draw[snake=brace] (7.1,-1) -- (7.1,-2) node[right,midway,font=\large] {$t=1$};

					\def\mywall{(0.01,-2.2) rectangle (6.99,-2.6)}; 
					\fill[color=brown!60!red] \mywall;
					\pattern[pattern color=black,pattern=none] \mywall;
					\node[font=\large] (end) at (3.5,-3) {End of Public Communication};
					
					\node[ ] (npc) at (3.5,-11) {$N_{PC}=1$};
					\node[ ] (fve) at (3.5,-12.5) {$\vect{F}=(F_{11},F_{12})$}; 
				\end{scope}
				
				\begin{scope}[xshift=24cm]
					\node (n1) at ( 0,0) [place] {1};
					\node (n2) at ( 7,0) [place] {2};
					\node[above=of n1] (alice) {Alice};
					\node[above=of n2] (bob) {Bob};
					\node[anchor=base] (title) at (3.5,4) {Interactive  SKA};
					\draw[->,black,thick] (0,-1) -- (0,-10);
					\draw[->,black,thick] (7,-1) -- (7,-10);

					\draw[->,color=green!50!black,line width=2pt] (0,-1) -- (7,-2);
					\draw[->,color=green!50!black,line width=2pt] (7,-1) -- (0,-2);
					\draw[snake=brace] (7.1,-1) -- (7.1,-2) node[right,midway,font=\large] {$t=1$};
					
					\draw[->,color=green!50!black,line width=2pt] (0,-2) -- (7,-3);
					\draw[->,color=green!50!black,line width=2pt] (7,-2) -- (0,-3);
					\draw[snake=brace] (7.1,-2) -- (7.1,-3) node[right,midway,font=\large] {$t=2$};
					
					\draw[->,color=green!50!black,line width=2pt] (0,-3) -- (7,-4);
					\draw[->,color=green!50!black,line width=2pt] (7,-3) -- (0,-4);
					\draw[snake=brace] (7.1,-3) -- (7.1,-4) node[right,midway,font=\large] {$t=3$};
					
					\node[color=green!50!black,font=\huge] (dots) at (3.5,-6) {$\vdots$};

					\draw[->,color=green!50!black,line width=2pt] (0,-8) -- (7,-9);
					\draw[->,color=green!50!black,line width=2pt] (7,-8) -- (0,-9);
					\draw[snake=brace] (7.1,-8) -- (7.1,-9) node[right,midway,font=\large] {$t=N_{PC}$};

					\def\mywall{(0.01,-9.2) rectangle (6.99,-9.6)}; 
					\fill[color=brown!60!red] \mywall;
					\pattern[pattern color=black,pattern=none] \mywall;
					\node[font=\large] (end) at (3.5,-10) {End of Public Communication};
					
					\node[ ] (npc) at (3.5,-11) {$N_{PC}>1$};
					\node[ ] (fve) at (3.5,-12.5) {$\vect{F}=(F_1,F_2,\ldots,F_{N_{PC}})$}; 
				\end{scope}

			\end{tikzpicture}
		\end{adjustbox}
		\caption[Modes of interaction for two-party SKA]{Three levels (modes) of interaction for two-party SKA. 
			Note that one-way SKA is an special case of the  general noninteractive SKA. 
		}
		\label{fig:int_modes}
	\end{figure}
	
	\paragraph*{\textbf{Two-party SKA}} 
	\citeauthor{Ahlswede1993} showed that both two-party SK and PK capacities can be achieved with one-way SKA \cite[Preposition 1 and Theorem 3]{Ahlswede1993}. That is, 
	\begin{equation}\label{eq:SK_PK_NI_Gen_2P}
		(\text{when~} m=2) \quad \begin{array}{c}
			C_{SK}^{\rightarrow} = C_{SK}^{NI} = C_{SK} \\
			C_{PK}^{\rightarrow} = C_{PK}^{NI} = C_{PK}
		\end{array}.
	\end{equation}
	A single-letter characterization of two-party one-way WSK capacity was derived in \cite{Ahlswede1993}, where the corresponding one-way capacity achieving SKA protocol is showed to also achieve the general WSK capacity if the Markov condition $X_1 - X_2 -Z$ holds \cite[Theorem 1 and its Corollary]{Ahlswede1993}.  That is, 
	\begin{equation}\label{eq:WSK_NI_Gen_2P}
		(\text{when~} m=2 \text{~and~} X_1 - X_2 -Z) \quad 
		C_{WSK}^{\rightarrow} = C_{WSK}^{NI} = C_{WSK} .
	\end{equation}
	An example is given in \cite[Section V, Proof of Theorem 7]{Gohari2010} for which the one-way WSK capacity is strictly less than the WSK capacity which can be achieved by a noninteractive SKA where both Alice and Bob each send one public message to each other. 
	See also Example~\ref{exm:two-way} which is similar to the example given in \cite{Gohari2010}.  
	This result, proves that in general there is a non-zero gap between the one-way and general WSK capacities, i.e.,  
	\begin{equation}\label{eq:gap_2P_treepin}
		(\text{when~} m=2) \quad  C_{WSK} - C_{WSK}^{\rightarrow} > 0.
	\end{equation}
	See the source model of Fig.1 and the last part of the proof for Theorem 7 in \cite{Gohari2010} for the proof. 
	In other words, one-way SKA is not sufficient to achieve the two-party WSK capacity.

	\paragraph*{\textbf{Multiterminal SKA}} 
	Extending the statements of \eqref{eq:SK_PK_NI_Gen_2P}, \citeauthor{Csiszar2004a} showed that for multiterminal SKA, the SK and PK capacities can be achieved noninteractively \cite[Theorems 1 and 2]{Csiszar2004a}. The best known general lower bound for  multiterminal WSK capacity is the interactive lower bound of \cite{Gohari2010}. 
	For special cases of Tree-PIN model, WSK capacity can be achieved noninteractively \cite{Poostindouz2019,Vippathalla2021}.  
	In this paper, we gave %
	an interactive SKA protocol (with $N_{PC} = 2$) that achieves the WSK capacity of Tree-PIN sources with independent leakages.   
	However, it remains unknown if %
	interaction is %
	required for achieving the WSK capacity in general.

	To investigate if there is a non-zero gap 
	between the general multiterminal WSK capacity and the noninteractive WSK capacity, 
	it is sufficient 
	to know expressions for both capacities 
	at least for a special class of multiterminal source models. 
	For Tree-PIN, we proved %
	an expression for WSK capacity, 
	but the noninteractive WSK capacity of Tree-PIN is not known.  
	In the following, we use a specific example of a Tree-PIN source model (see Figure \ref{fig:Tree-PIN-m-3}) 
	to show that there is a non-zero gap between the WSK capacity and 
	the highest key rate of
	known noninteractive SKA methods. 
	We prove a lower bound on the  noninteractive WSK capacity of this example source model 
	which is strictly less than the WSK capacity.  
	However, we leave the problem of tightening (or closing) 
	this gap for future work.

	\begin{figure}[t]
		\centering
		\begin{adjustbox}{ width =0.6\textwidth}
			
			\begin{tikzpicture}[font=\bfseries\Huge]
				\tikzstyle{place}=[circle,draw=black!90,fill=blue!10,very thick,inner sep=3pt, minimum size=42pt,line width=2pt,font=\bfseries\Huge]
				\tikzstyle{eve}=[circle,draw=black!90,fill=red!10,very thick,inner sep=3pt, minimum size=42pt,line width=2pt]

				\node (n1) at ( 0+2,0)  [place] {1};
				\node (n2) at ( 10,0) [place] {2};
				\node (n3) at ( 20-2,0) [place] {3};
				
				\node (e) at ( 10,-5) [eve] {E};

				\draw[blue!40!red,line width=7pt,round cap-round cap,shorten <=1mm,shorten >=1mm] (n1.east) node [above right] {$V_{12}$} -- (n2.west) node [above left] {$V_{21}$} ;
				
				\draw[yellow!40!red,line width=7pt,round cap-round cap,shorten <=1mm,shorten >=1mm] (n3.west) node [above left] {$V_{32}$}  -- (n2.east) node [above right] {$V_{23}$} ;
				
				\draw[snake=snake, line before snake=10mm, line after snake=10mm,segment aspect=5, segment amplitude=5pt,blue!40!red,line width=3pt,round cap-round cap,shorten <=1mm,shorten >=1mm] (n2.south west) -- (e.north west) node [pos=0.8,left] {$Z_{12}$} ;
				
				\draw[snake=snake, line before snake=10mm, line after snake=10mm,segment aspect=5, segment amplitude=5pt,yellow!40!red,line width=3pt,round cap-round cap,shorten <=1mm,shorten >=1mm] (n2.south east) -- (e.north east) node [pos=0.8,right] {$Z_{23}$} ;

			\end{tikzpicture}
		\end{adjustbox}
		\caption[The Tree-PIN model of Example~\ref{exm:NI-lessthan-Gen}.]{The Tree-PIN model of Example~\ref{exm:NI-lessthan-Gen}. Here $X_1=V_{12}$, $X_2=(V_{21},V_{23})$, $X_3=V_{32}$, and Eve's wiretapped side information is $Z=(Z_{12},Z_{23})$. 
		}
		\label{fig:Tree-PIN-m-3}
	\end{figure}

	\begin{example}\label{exm:NI-lessthan-Gen}
		Consider the wiretapped Tree-PIN source model of Figure \ref{fig:Tree-PIN-m-3}. In this setting, %
		$\mc M=\{1,2,3\}$, $X_1=V_{12}$, $X_2=(V_{21},V_{23})$, $X_3=V_{32}$, and Eve's wiretapped side information is $Z=(Z_{12},Z_{23})$, and  the Markov relations $V_{12} - V_{21} - Z_{12}$ and $V_{32} - V_{23} - Z_{23}$  hold. 
		Further, assume $I(V_{21};Z_{12}) , I(V_{23};Z_{23}) > 0$. 
		When $\mc A=\mc M$, the WSK capacity of this model is given by Theorem \ref{thm:Tree-PIN} as 
		$$C_{WSK} =\min\{ I(V_{12};V_{21}|Z_{12}) , I(V_{23};V_{32}|Z_{23}) \}.$$
		We prove the following lower bound on the noninteractive WSK capacity of this model  
		\begin{equation}\label{eq:NI_WSK_lower}
			C_{WSK}^{NI}  \geq r_L^{NI} := H(X_2|Z) - 
			\max\{ H(X_2|X_1) , H(X_2|X_3) \} ,
		\end{equation}
		which is less that the general WSK capacity, i.e., 
		\begin{equation}\label{eq:NI_WSK_VS_WSK}
			C_{WSK}  - r_L^{NI} > 0 .
		\end{equation}
		\begin{IEEEproof}[Proof of Inequalities \eqref{eq:NI_WSK_lower} and \eqref{eq:NI_WSK_VS_WSK}]
			We first calculate the noninteractive lower bound $r_L^{NI}$ of \eqref{eq:NI_WSK_lower}, 
			by considering 
			Protocol \ref{prot:NIa} ($\vect{\Pi_{E5}^a}$). 
			The key rate of this protocol %
			immediately follows from 
			the Slepian-Wolf source coding Theorem \cite{Slepian1973a} and 
			the generalized Leftover Hash Lemma of \cite{Hayashi2016}. 
			
			The %
			noninteractive Protocol \ref{prot:NIa} ($\vect{\Pi_{E5}^a}$), 
			is in the style of one-way SKA and the SKA protocol of \cite{Gohari2010} in which 
			some terminals participate in public discussion and some don't (are silent.) 
			Protocol \ref{prot:NIa} 
			works as follows. Terminal 2, sends a public message such that  
			terminal 1 and terminal 3 can recover $X_2^n$. Using the common randomness $X_2^n$ all
			terminals extract their copies of the final key  by using  universal hashing. 

			\begin{algorithm}[t]
				\caption{First Noninteractive SKA for Tree-PIN of Example 5 ($\vect{\Pi_{E5}^a}$)}
				\label{prot:NIa}
				\DontPrintSemicolon
				\SetKwInput{PorKnw}{Public Knowledge}
				\SetKwInput{PorAsm}{Assumption}
				\SetKwInput{PorPar}{Parameter}
				\PorKnw{$P_{Z X_{\mc M}}$ and a family $\mc H$ of universal hash functions $h_s: \mc X_2^n \to \mc K$ where $s\in\mc S$.}
				\KwIn{Observations ($n-$IID samples) $X_{1}^n, X_{2}^n, X_{3}^n$}
				\KwOut{Copies of the final key $K_1, K_2, K_3$}
				\SetKwFor{MyFor}{for}{}{end~for}
				\SetKwIF{If}{ElseIf}{Else}{if}{then}{else if}{else}{end~if}
				
				\BlankLine\BlankLine%
				
				\tcp*[h]{Information Reconciliation}	\\
				
				Terminal 2 sends public message $F_2$
				
				All terminals recover $X_2^n$
				
				\BlankLine\BlankLine%
				
				\tcp*[h]{Privacy Amplification}	\\
				
				All terminals agree on a random seed $s\in\mc S$ using the public channel
				
				All terminals extract their keys from $X_2^n$ by $K_j = h_s(X_2^n) ~\forall j\in\{1,2,3\}$

			\end{algorithm}
			
			The asymptotic key rate of this protocol can be calculated using 
			Lemma 8 of \cite{Hayashi2016} as 
			\begin{IEEEeqnarray*}{rCl}
				r_K(\vect{\Pi_{E5}^a})   &\stackrel{\text{(a)}}{=}& H(X_2|Z) - 
				\min_{F_2} \lim_{n \rightarrow \infty} \frac{1}{n}\log\supp(F_2)   \\
				&\stackrel{\text{(b)}}{=}& H(X_2|Z) - \max\{ H(X_2|X_1) , H(X_2|X_3) \},
			\end{IEEEeqnarray*}
			where (a) follows from the fact that the common randomness which is used for 
			group key extraction is RV $X_2$ and 
			(b) is due to the Slepian-Wolf source coding Theorem \cite{Slepian1973a}. 
			
			Thus, the noninteractive lower bound is then given by 
			\begin{equation*}
				r_L^{NI} %
				= H(X_2|Z) - 
				\max\{ H(X_2|X_1) , H(X_2|X_3) \}. 
			\end{equation*}

			Next, we prove inequality \eqref{eq:NI_WSK_VS_WSK}. %
			Assume that $C_{WSK} = I(V_{12};V_{21}|Z_{12})$. 
			Then, %
			\begin{IEEEeqnarray*}{rCl}
				r_L^{NI} &=& H(X_2|Z) - 
				\max\{ H(X_2|X_1) , H(X_2|X_3) \} \\
				&\leq& 
				H(X_2|Z) - 
				H(X_2|X_1) \\
				&=& 
				H(V_{21}|Z_{12}) + H(V_{23}|Z_{23}) 
				- H(V_{21}|V_{12}) - H(V_{23}) \\
				&=& I(V_{12};V_{21}|Z_{12}) - I(V_{23}|Z_{23}) \\
				&<& C_{WSK} ,
			\end{IEEEeqnarray*}
			where the last inequity holds since $I(V_{23};Z_{23}) > 0$. 
			Using the same line of argument we can show that $r_L^{NI} < C_{WSK} $  if 
			the WSK capacity is $C_{WSK} = I(V_{23};V_{32}|Z_{23})$. 		   
		\end{IEEEproof}
		
		\begin{remark}
			Finally, we point out that, to our knowledge, 
			Protocol \ref{prot:NIa} ($\vect{\Pi_{E5}^a}$) gives the highest 
			known noninteractive key rate for this example. In fact in the following, 
			we show that  
			the alternative noninteractive approach of SKA by omniscience also leads to the same lower bound. 
			
			Consider the noninteractive Protocol \ref{prot:NIb} ($\vect{\Pi_{E5}^b}$), 
			which 
			is in the style of SKA by omniscience, similar to the  
			SKA protocol of \cite{Csiszar2004a}. 
			Protocol \ref{prot:NIb} 
			works as follows. 
			Terminal 2, sends a public message such that  
			terminal 1 and terminal 3 can recover $X_2^n$. 
			Then, terminal 1 (and 3), send public messages $F_1$ (and $F_3$), such that
			other terminals can recover $X_1^n$ (and $X_3^n$). 
			Using the common randomness $X_{\mc M}^n$ all
			terminals extract their copies of the final key  by using  universal hashing. 
			Let $\vect F = (F_1, F_2, F_3)$ denote the overall public communication of this protocol. 
			
			The asymptotic key rate of this protocol also can be calculated using 
			Lemma 8 of \cite{Hayashi2016} as 
			\begin{IEEEeqnarray*}{rCl}
				r_K(\vect{\Pi_{E5}^b})   &\stackrel{\text{(a)}}{=}& H(X_{\mc M}|Z) - 
				\min_{\vect F} \lim_{n \rightarrow \infty} \frac{1}{n}\log\supp(\vect F)   \\
				&\stackrel{\text{(b)}}{=}& 
				H(X_{\mc M}|Z) - \max\{ H(X_2|X_1) , H(X_2|X_3) \} - H(X_1|X_2) - H(X_3|X_2),  
			\end{IEEEeqnarray*}
			where (a) follows from the fact that the common randomness which is used for 
			group key extraction is RV $X_{\mc M} = (X_1, X_2, X_3)$ and 
			(b) is due to the Slepian-Wolf source coding Theorem \cite{Slepian1973a}.

			\begin{algorithm}[t]
				\caption{Second Noninteractive SKA for Tree-PIN of Example 5 ($\vect{\Pi_{E5}^b}$)}
				\label{prot:NIb}
				\DontPrintSemicolon
				\SetKwInput{PorKnw}{Public Knowledge}
				\SetKwInput{PorAsm}{Assumption}
				\SetKwInput{PorPar}{Parameter}
				\PorKnw{$P_{Z X_{\mc M}}$ and a family $\mc H$ of universal hash functions $h_s: \mc X_{\mc M}^n \to \mc K$ where $s\in\mc S$.}
				\KwIn{Observations ($n-$IID samples) $X_{1}^n, X_{2}^n, X_{3}^n$}
				\KwOut{Copies of the final key $K_1, K_2, K_3$}
				\SetKwFor{MyFor}{for}{}{end~for}
				\SetKwIF{If}{ElseIf}{Else}{if}{then}{else if}{else}{end~if}
				
				\BlankLine\BlankLine%
				
				\tcp*[h]{Information Reconciliation}	\\
				
				Terminal 1 sends public message $F_1$
				
				Terminal 2 sends public message $F_2$
				
				Terminal 3 sends public message $F_3$
				
				Terminals 1 and 3 recover $X_2^n$
				
				Terminals 1 and 2, use $F_3$ and $X_2^n$ to recover $X_3$ 
				
				Terminals 3 and 2, use $F_1$ and $X_2^n$ to recover $X_1$ 
				
				\BlankLine\BlankLine%
				
				\tcp*[h]{Privacy Amplification}	\\
				
				All terminals agree on a random seed $s\in\mc S$ using the public channel
				
				All terminals extract their keys from $X_{\mc M}^n$ by $K_j = h_s(X_{\mc M}^n) ~\forall j\in\{1,2,3\}$

			\end{algorithm}
			
			Noting that $H(X_1|X_2 Z) = H(X_1|X_2)$ and $H(X_3|X_2 X_1 Z) = H(X_3|X_2)$, implies that both SKA protocols have the same asymptotic key rate,  $r_K(\vect{\Pi_{E5}^a})   =  r_K(\vect{\Pi_{E5}^b})$.

		\end{remark}

	\end{example}

	In summary, the above example, 
	suggests that  known 
	noninteractive SKA approaches %
	cannot 
	achieve the 
	general WSK capacity. %

	\section{Conclusion} \label{sec:concl_treepin}

	We considered the wiretapped PIN and wiretapped Tree-PIN models. For wiretapped Tree-PIN we proved the WSK capacity and proposed an efficient capacity achieving SKA protocol. The protocol
	has two rounds and 
	uses any  capacity achieving two-party SKA as a subroutine so terminals can obtain pairwise keys. By  extending the two-party capacity achieving protocols of \cite{Hayashi2016} and \cite{Sharifian2020} to the case of Tree-PIN, we derived new finite-length lower bounds on the maximum achievable key length. We also proved a  finite-length upper bound for the general wiretapped Tree-PIN, and another lower bound for the special case of Tree-PIN studied in \cite{Poostindouz2019}. Finally, for wiretapped PIN, we proved a lower and an upper bound for  WSK capacity. The bounds are tight when $\mc A=\mc M$ or $|\mc A| =2$.
	We extended the Tree-PIN model to two other general cases and proved corresponding 
	WSK capacities. 
	Finally, we investigated the problem of noninteractive key agreement in 
	an example of wiretapped Tree-PIN model, and 
	our analysis suggests that the noninteractive approach for SKA is 
	not sufficient for achieving the general WSK capacity.

	\section*{Acknowledgment} %
	This research is in part supported by Natural Sciences and Engineering Research Council of Canada, Discovery Grant program.

	\bibliographystyle{IEEEtranN}
	{%
		\small
		\bibliography{treepin}
	}

	\appendices
	
	\counterwithin{theorem}{section}

	{\color{black}
		\section{Statistical Distance}\label{sec:SD}
		Let $X$ and $Y$ be two random variables (RVs) defined over the same finite alphabet $\mc W$. 
		The statistical (variation) distance between $X$ and $Y$ has the following equivalent definitions 
		\begin{align*}
			\sd(X;Y) & = \frac{1}{2} \sum_{w\in \mc W} \left\vert P_X(w) - P_Y(w) \right\vert    \\
			& = \max\limits_{\mc{T}\subseteq\mc{W}}  \sum_{w\in \mc T}  P_X(w) - P_Y(w) \\
			& = \sum_{w\in {\mc{T}^*}}  P_X(w) - P_Y(w),                                
		\end{align*}
		where ${\mc{T}^*} = \{w\in\mc{W}~\vert~ P_X(w) \geq P_Y(w) \}$. See Definition 11.1 and Lemma 11.1 of \cite{Mitzenmacher2005}.

		Lemma \ref{thm:SD_Single_VS_Jojnt}, and its corollaries  
		that follow next are direct consequences of the  
		definition of statistical distance. %
		We present their proofs here for completeness. 
		
		\begin{lemma}\label{thm:SD_Single_VS_Jojnt}
			For two random variables $X$ and $X'$ over the same alphabet $\mc{X}$ and two random variables $Y$ and $Y'$ over the same alphabet $\mc{Y}$ we have $\sd(X,X')\leq \sd(XY,X'Y').$
		\end{lemma}

		\begin{IEEEproof}
			Let $(X,Y)\sim P_{XY}$ and $(X',Y')\sim Q_{X'Y'}$. Then
			\begin{align*}
				\sd(XY,X'Y') & = \max\limits_{\mc{T}\subseteq\mc{X}\times\mc{Y}}  \sum_{(x,y)\in \mc T}  P_{XY}(x,y) - Q_{X'Y'}(x,y)               \\
				& \geq \sum_{(x,y)\in \mc{V}}  P_{XY}(x,y) - Q_{X'Y'}(x,y)                                                            \\
				& = \sum\limits_{x\in\overline{\mc{X}}} \sum\limits_{y\in\mc{Y}} P_{XY}(x,y) - \sum\limits_{y\in\mc{Y}} Q_{X'Y'}(x,y) \\
				& = \sum\limits_{x\in\overline{\mc{X}}}  P_X(x) - Q_{X'}(x) =\sd(X,X'), %                                           
			\end{align*} 
			where $\mc{V}=\overline{\mc{X}}\times\mc{Y}$, with $\overline{\mc{X}}=\{x\in\mc{X}~\vert~P_X(x)\geq Q_{X'}(x) \}$.
		\end{IEEEproof}
		\begin{corollary}\label{thm:SD_cor_1}
			If $\sd(XY,X'Y')\leq\eps$ then $\sd(X,X') \leq\eps$ and $\sd(Y,Y') \leq\eps$ .
		\end{corollary}
		\begin{corollary}\label{thm:SD_cor_2}
			(i) For  any three RVs $X$, $X'$, and $Y$ we have $\sd(X,X')\leq \sd(X Y,X' Y)$. \\
			(ii) If $Y$ is independent from $X$ and $X'$, then $\sd(X,X') = \sd(X Y,X' Y)$.
		\end{corollary}
		\begin{corollary}\label{thm:SD_cor_3}
			Suppose $X$ and $X'$ are correlated RVs independent from correlated RVs $Y$ and $Y'$. Then, $\sd(X Y,X' Y') \leq \sd(X,X') + \sd(Y,Y')$.
		\end{corollary}
		\begin{IEEEproof}
			Using the triangle inequality we have
			\begin{IEEEeqnarray*}{rCl}
				\sd(X Y,X' Y') &\leq& \sd(X Y,X' Y) +\sd(X' Y,X' Y') \\
				&=& \sd(X,X') + \sd(Y,Y'),
			\end{IEEEeqnarray*}
			where the equality is due to preposition (ii) of Corollary~\ref{thm:SD_cor_2}.
		\end{IEEEproof}

	}%

	\section{Proof of Upper Bound Lemma~\ref{lemma:Upper-bound-Tree}}\label{sec:app:Tree-PIN-AnotM}

	In this section, we prove Lemma~\ref{lemma:Upper-bound-Tree}. We prove that for a Tree-PIN specified by the graph $G=(\mc M,\mc E)$ and probability distribution $P_{ZX_{\mc M}},$ we have
	\begin{align*}
		C_{WSK}^{\mc A}(P_{X_\mc{M} Z}) \leq  \min_{\substack{i,j\in\mc M \\ \mathrm{~s.t.~}e_{ij}\in\mc E_{\mc A}}} I(V_{ij};V_{ji}|Z_{ij}), 
	\end{align*} where $G_{\mc A} = (\mc M_{\mc A}, \mc E_{\mc A})$ is the subtree of $G$ with the least number of edges that connects all nodes of $\mc A$.

	\begin{IEEEproof}[Proof of Lemma~\ref{lemma:Upper-bound-Tree}]
		Recall that (due to Lemma~\ref{thm:PKisUpp}, see also \cite[Theorem~4]{Csiszar2004a})
		\begin{align*}
			C_{WSK}^{\mc A}(P_{X_\mc{M} Z}) \leq C_{PK}^{\mc A|\{m+1\}}(P_{X_\mc{M} Z}), 
		\end{align*}
		where $C_{PK}$ denotes the PK capacity of the associated PIN model given by $\mc M'=[m+1]$ and $G'=(\mc M', \mc E')$ with a dummy node $m+1$ representing the adversary (i.e., $X_{m+1}=Z$). From Theorem~\ref{thm:PK-Cap-CN04} we know
		\begin{align*}
			C_{PK}^{\mc A|\{m+1\}}(P_{X_\mc{M} Z}) = H(X_{\mc M}|Z)-R_{CO}(X_\mc{A}|Z), 
		\end{align*} where $R_{CO}(X_\mc{A}|Z)$
		denotes the solution to the Linear Programming (LP) %
		problem of Figure~\ref{fig:LP-Tree-app}, defined over real numbers \cite{Csiszar2004a}.

		\begin{figure}[ht]
			\centering
			\eqbox{
				\begin{IEEEeqnarraybox}{s.t.s.s}
					\text{Minimize:}  &&  $\sum\limits_{j\in\mc M} R_j$ &\\
					\text{Subject to:}&&  $\sum\limits_{j\in\mc B} R_j \geq H(X_{\mc B}|X_{\mc B^c},Z),~\quad\forall \mc B\subsetneq \mc M, ~\mc A \nsubseteq  \mc B$ &(a)\\%\text{~and}\\
					&& $R_j \in\RR^+, ~\quad\forall j\in\mc M.$ &(b)
				\end{IEEEeqnarraybox}
			}
			\caption{The LP problem of finding $R_{CO}(X_\mc{A}|Z)$.}
			\label{fig:LP-Tree-app}
		\end{figure}

		We prove that
		\begin{align}\label{eq:Rco-Tree-app}
			R_{CO}(X_\mc{A}|Z) = H(X_{\mc{M}}|Z) - \min_{\substack{i,j\in\mc M \\ \mathrm{~s.t.~}e_{ij}\in\mc E_{\mc A}}}  I(V_{ij}; V_{ji}|Z_{ij}).
		\end{align}

		The proof is by first, proving the following lower bound \eqref{eq:Rco-lower-bound-Tree-app} and then presenting a rate assignment that achieves the equality, hence proving Equation~\eqref{eq:Rco-Tree-app}.
		\begin{align}\label{eq:Rco-lower-bound-Tree-app}
			R_{CO}(X_\mc{A}|Z) \geq H(X_{\mc{M}}|Z) - \min_{\substack{i,j\in\mc M \\ \mathrm{~s.t.~}e_{ij}\in\mc E_{\mc A}}} I(V_{ij}; V_{ji}|Z_{ij}).
		\end{align}
		
		\begin{IEEEproof}[Proof of Inequality~\eqref{eq:Rco-lower-bound-Tree-app}]
			The terminals in $\mc M$ form a Tree-PIN $G=(\mc M,\mc E)$. By cutting (removing) an arbitrary edge $e_{i'j'}\in\mc E$ that connects nodes $i'$ and $j'$, we will have two trees $G_{\langle\mc B\rangle}=(\mc B,\mc{E}_{\mc{B}})$ and $G_{\langle\mc B^c\rangle}=(\mc B^c,\mc{E}_{\mc{B}^c})$, such that $\mc P=\{\mc B, \mc B^c \}$ is a partition of $\mc M$, and nodes $i'$ and $j'$ each belong to one part of the partition -- %
			and $\mc{E}_{\mc{B}^c} \cup \mc{E}_{\mc{B}} = \mc E\setminus\{e_{i'j'} \}$.
			Consider the constraints of the LP problem in Figure~\ref{fig:LP-Tree-app} written two times for subsets  $\mc B$ and $\mc B^c$ individually, and note that  $\mc A \nsubseteq  \mc B $ and $\mc A \nsubseteq \mc B^c $. We will have, %
			\begin{align} \label{eq:Sum_R_B}
				\sum\limits_{j\in\mc B} R_j   & \geq H(X_{\mc B}|X_{\mc B^c},Z), \\  \label{eq:Sum_R_Bc}
				\sum\limits_{j\in\mc B^c} R_j & \geq H(X_{\mc B^c}|X_{\mc B},Z). 
			\end{align}
			From Slepian-Wolf source coding theorem we know that inequality \eqref{eq:Sum_R_B}, implies that if a decoder has access to side information $X_{\mc B^c}$ and $Z$, %
			then by receiving the public messages broadcasted by terminals in $\mc B$, the decoder can reliably recover $X_{\mc B}.$ Also, recall that $X_{\mc B}=\bigcup_{i\in\mc B} V_{ij}$.
			Due to the 
			mutual independence of $\{(V_{ij},V_{ji}, Z_{ij})\}$'s, 
			we get $H(X_{\mc M}|Z)=\sum_{i,j} H(V_{ij},V_{ji}|Z_{ij})$, and thus we can translate inequalities \eqref{eq:Sum_R_B} and \eqref{eq:Sum_R_Bc} to
			\begin{align*}
				\sum\limits_{j\in\mc B} R_j   & \geq \sum\limits_{\substack{i<j \\ \text{~s.t.~} e_{ij}\in\mc{E}_{\mc{B}}  }} H(V_{ij},V_{ji}|Z_{ij})+H(V_{i'j'}|V_{j'i'},Z_{ij}), \\
				\sum\limits_{j\in\mc B^c} R_j & \geq \sum\limits_{\substack{i<j \\ \text{~s.t.~} e_{ij}\in\mc{E}_{\mc{B}^c}  }} H(V_{ij},V_{ji}|Z_{ij})+H(V_{j'i'}|V_{i'j'},Z_{ij}).
			\end{align*}
			
			By adding these two inequalities, we arrive at
			\begin{align*}
				\sum\limits_{j\in\mc M} R_j & \geq \sum\limits_{\substack{i<j                                    \\ \text{~s.t.~} e_{ij}\in\mc{E}_{\mc{B}}  }} H(V_{ij},V_{ji}|Z_{ij}) + \sum\limits_{\substack{i<j \\ \text{~s.t.~} e_{ij}\in\mc{E}_{\mc{B}^c}  }} H(V_{ij},V_{ji}|Z_{ij}) \\
				& \qquad + H(V_{i'j'}|V_{j'i'},Z_{ij}) + H(V_{j'i'}|V_{i'j'},Z_{ij}) \\
				& = \sum\limits_{\substack{i<j                                       \\ \text{~s.t.~} e_{ij}\in\mc{E}  }} H(V_{ij},V_{ji}|Z_{ij}) - I(V_{i'j'};V_{j'i'}|Z_{ij})  \\
				& = H(X_{\mc M}|Z) - I(V_{i'j'};V_{j'i'}|Z_{ij}),                    
			\end{align*}
			where $e_{i'j'}$ denotes the edge that connects the two trees $G_{\langle\mc B\rangle}$ and  $G_{\langle\mc B^c \rangle}$. We also used the facts that $\mc{E}_{\mc{B}^c} \cup \mc{E}_{\mc{B}} = \mc E\setminus\{e_{i'j'} \}$ and that the sets $\{X_j|~\forall j\in\mc M\}$ and $\{V_{jk}|~ j<k , e_{jk}\in\mc E \}$ are indeed equivalent.
			The above inequality holds for any pair  $i'$ and $j'$ of terminals with $e_{i'j'}\in\mc E$ and their induced partition  $\{\mc B,\mc B^c\}$, where $\mc A \nsubseteq  \mc B $ and $\mc A \nsubseteq \mc B^c $. Thus, %
			\begin{align*}
				R_{CO}(X_\mc{A}|Z) & \geq \max\limits_{\substack{i,j\in\mc M        \\ \mathrm{~s.t.~}e_{ij}\in\mc E_{\mc A}}}  \{ H(X_{\mc M}|Z) - I(V_{ij};V_{ji}|Z_{ij}) \}, \\
				& = H(X_{\mc M}|Z) - \min_{\substack{i,j\in\mc M \\ \mathrm{~s.t.~}e_{ij}\in\mc E_{\mc A}}}  I(V_{ij};V_{ji}|Z_{ij}),
			\end{align*}
			which proves the Inequality~\eqref{eq:Rco-lower-bound-Tree-app}.
		\end{IEEEproof}
		
		To complete the proof of Equation~\eqref{eq:Rco-Tree-app}, we prove that there exists a rate assignment protocol that achieves the bound in~\eqref{eq:Rco-lower-bound-Tree-app}.

		\begin{figure}[t]
			\centering
			\eqbox{
				\centering
				\begin{IEEEeqnarraybox}{s.s}
					\text{Let $(i^*,j^*)$ s.t.} & $I(V_{i^*j^*};V_{j^*i^*}|Z_{i^*j^*})$ 
					\\
					& \qquad 
					$ = \min\limits_{\substack{i,j\in\mc M \\ \mathrm{~s.t.~}e_{ij}\in\mc E_{\mc A}}}  I(V_{ij};V_{ji}|Z_{ij})$, \text{and}\\
					\text{for any $j\in\mc M$ let}  & $R_j=\sum\limits_{i\in\Gamma(j)} \widetilde{R}^{(j)}_{i}.$ \\
					\midrule\\
					\text{To minimize $\sum\limits_{j\in \mc M} R_j$} & \\ \\
					\text{assign~}&  $\widetilde{R}^{(j^*)}_{i^*}=H(V_{j^*i^*}|V_{j^*i^*},Z_{i^*j^*}),$ \\
					~& $\widetilde{R}^{(i^*)}_{j^*}=H(V_{i^*j^*}|V_{i^*j^*},Z_{i^*j^*}),$ and\\
					\text{$\forall ~e_{ij}\neq  e_{i^*j^*}$,}~~~& $\text{with~} d(i,i^*) < d(j,i^*),$ \\%\in\mc E\setminus \{e_{i^*j^*} \}\qquad $}& \\
				\text{assign~}&$\widetilde{R}^{(j)}_{i}= H(V_{ji}|V_{ij},Z_{ij}),$ \text{ and~}\\
				~& $\widetilde{R}^{(i)}_{j}= H(V_{ij}|Z_{ij}).$%
			\end{IEEEeqnarraybox}
		}
		\caption{The rate assignment that achieves $R_{CO}(X_\mc{A}|Z)$.}
		\label{fig:Rate-assignment-2}
	\end{figure}

	\begin{IEEEproof}[Proof of Equation~\eqref{eq:Rco-Tree-app}]
		First, let $(i^*, j^*)$ be defined as follows, $$I(V_{i^*j^*};V_{j^*i^*}|Z_{i^*j^*}) = \min\limits_{\substack{i,j\in\mc M \\ \mathrm{~s.t.~}e_{ij}\in\mc E_{\mc A}}}  I(V_{ij};V_{ji}|Z_{ij}).$$
		Then for each terminal $j\in\mc M$ we let the communication rate $R_j$ be chosen %
		according to the rate assignment in Figure~\ref{fig:Rate-assignment-2}, where $R_j=\sum_{i\in\Gamma(j)} \widetilde{R}^{(j)}_{i}$ and the rate assignment protocol assigns values to all $\widetilde{R}^{(j)}_{i}$ components.

		This rate assignment satisfies the following equations, %
		\begin{align}\label{eq:couple-rates}
			\widetilde{R}^{(i)}_{j} + \widetilde{R}^{(j)}_{i} & = H(V_{ij},V_{ji}|Z_{ij}), ~\quad\forall i,j\in\mc M \text{~s.t.~} e_{ij}\in\mc E\setminus\{e_{i^*j^*}\}, \\ \label{eq:couple-rates-2}
			\widetilde{R}^{(i^*)}_{j^*}                       & = H(V_{i^*j^*}|V_{j^*i^*},Z_{i^*j^*}),                                                                    \\ \label{eq:couple-rates-3}
			\widetilde{R}^{(j^*)}_{i^*}                       & = H(V_{j^*i^*}|V_{i^*j^*},Z_{i^*j^*}),                                                                    
		\end{align}
		which leads to the following sum rate:
		\begin{align} \nonumber
			\sum\limits_{j\in\mc M} R_j & = \sum\limits_{j\in\mc M} \sum\limits_{i\in\Gamma(j)} \widetilde{R}^{(j)}_{i}= \sum\limits_{\substack{i<j \\ \text{~s.t.~} e_{ij}\in\mc{E}  }} \widetilde{R}^{(j)}_{i} + \widetilde{R}^{(i)}_{j} \\ \nonumber
			& = \sum\limits_{\substack{i<j}} H(V_{ij},V_{ji}|Z_{ij})  - I(V_{i^*j^*};V_{j^*i^*}|Z_{i^*j^*})             \\ \nonumber
			& = H(X_{\mc M}|Z) - I(V_{i^*j^*};V_{j^*i^*}|Z_{i^*j^*})                                                    \\
			& = H(X_{\mc M}|Z) -\min\limits_{\substack{i,j\in\mc M                                                      \\ \mathrm{~s.t.~}e_{ij}\in\mc E_{\mc A}}} I(V_{ij};V_{ji}|Z_{ij}). \label{eq:sum-rate-acheives-app}
		\end{align}
		
		Thus, the rate assignment indeed achieves the lower-bound of Inequality~\eqref{eq:Rco-lower-bound-Tree-app}.
		We, however, need to show that this rate assignment satisfies %
		the constraints of the LP problem described in Figure~\ref{fig:LP-Tree-app}.
		
		First, note that condition (b) in
		the LP in Figure~\ref{fig:LP-Tree-app} is satisfied as all assigned rates are non-negative.
		The constraints (a) in the LP can be rewritten for %
		an arbitrary subset of terminals (nodes) $\mc B \subsetneq \mc M, \mc A\nsubseteq \mc B$ as
		\begin{align}\label{eq:Ineq-rate-assignment-app}
			\sum\limits_{j\in\mc B} R_j & \geq \sum\limits_{\substack{i\in\mc B, j\in\mc B  }} H(V_{ij},V_{ji}|Z_{ij}) + \sum\limits_{\substack{i\in\mc B, j\notin\mc B  }} H(V_{ij}|V_{ji},Z_{ij}). 
		\end{align}

		We show in the following that the rate assignment of Figure~\ref{fig:Rate-assignment-2}, satisfies the inequality~\eqref{eq:Ineq-rate-assignment-app} for any arbitrary subset $\mc B \subsetneq \mc M, \mc A\nsubseteq \mc B$. For a given subset $\mc B$ let $\mc E_{\mc B}$ be the set of all edges contained in $\mc B$ (i.e., $\mc E_{\mc B}= \{e_{ij}|~ e_{ij}\in\mc E, \text{~and~} i\in\mc B, \text{~and~} j\in\mc B \}$). Then, depending on a given subset $\mc B$ there are two different cases: $I)~ e_{i^*j^*} \notin\mc{E}_{\mc{B}},$ and $II)~ e_{i^*j^*} \in\mc{E}_{\mc{B}}$. The proof is given for all the cases.

		\paragraph*{Case I) $e_{i^*j^*} \notin\mc{E}_{\mc{B}}$ -- }
		
		The left hand side of the inequality~\eqref{eq:Ineq-rate-assignment-app}, can be written as,
		\begin{align*}
			\sum\limits_{j\in\mc B} R_j
			& = \sum\limits_{j\in\mc B} \sum\limits_{i\in\Gamma(j)} \widetilde{R}^{(j)}_{i} \\
			& = \sum\limits_{j\in\mc B} \left( \sum\limits_{\substack{i\in\Gamma(j)         \\ i\in\mc B}}  \widetilde{R}^{(j)}_{i}  +
			\sum\limits_{\substack{i\in\Gamma(j) \\ i\notin\mc B}}  \widetilde{R}^{(j)}_{i} \right)\\
			& \stackrel{(a)}{\geq}   \sum\limits_{\substack{i<j                             \\ \text{~s.t.~} e_{ij}\in\mc{E}_{\mc{B}}  }} \widetilde{R}^{(j)}_{i} + \widetilde{R}^{(i)}_{j} +  \sum\limits_{\substack{i<j \\ \text{~s.t.~} i\in\mc B, j\notin\mc B  }} H(V_{ij}|V_{ji},Z_{ij}) \\
			& \stackrel{(b)}{=} \sum\limits_{\substack{i<j                                  \\ \text{~s.t.~} e_{ij}\in\mc{E}_{\mc{B}}  }} H(V_{ij},V_{ji}|Z_{ij}) + \sum\limits_{\substack{i<j \\ \text{~s.t.~} i\in\mc B, j\notin\mc B  }} H(V_{ij}|V_{ji},Z_{ij}),
		\end{align*}
		where, in the (a) we used the fact that $H(V_{ij}|Z_{ij})\geq H(V_{ij}|V_{ji},Z_{ij}),$
		and in (b) we used %
		Equation~\eqref{eq:couple-rates}.

		\paragraph*{Case II) $e_{i^*j^*} \in \mc{E}_{\mc{B}}$ -- }
		
		The left hand side of the inequality~\eqref{eq:Ineq-rate-assignment-app}, can be written as,
		\begin{align*}
			\sum\limits_{j\in\mc B} R_j
			& = \sum\limits_{j\in\mc B} \sum\limits_{i\in\Gamma(j)} \widetilde{R}^{(j)}_{i}                            \\
			& = \sum\limits_{j\in\mc B} \left( \sum\limits_{\substack{i\in\Gamma(j)                                    \\ i\in\mc B}}  \widetilde{R}^{(j)}_{i}  +
			\sum\limits_{\substack{i\in\Gamma(j) \\ i\notin\mc B}}  \widetilde{R}^{(j)}_{i} \right)\\
			& = \widetilde{R}^{(j^*)}_{i^*} + \widetilde{R}^{(i^*)}_{j^*} + \sum\limits_{\substack{i<j                 \\ \text{~s.t.~} e_{ij}\in\mc{E}_{\mc{B}}\setminus\{e_{i^*j^*} \}  }} \widetilde{R}^{(j)}_{i} + \widetilde{R}^{(i)}_{j} + \sum\limits_{j\in\mc B} \sum\limits_{\substack{i\in\Gamma(j) \\ i\notin\mc B}}  \widetilde{R}^{(j)}_{i} \\
			& \stackrel{(a)}{=} \widetilde{R}^{(j^*)}_{i^*} + \widetilde{R}^{(i^*)}_{j^*} + \sum\limits_{\substack{i<j \\ \text{~s.t.~} e_{ij}\in\mc{E}_{\mc{B}}\setminus\{e_{i^*j^*} \}  }} H(V_{ij},V_{ji}|Z_{ij}) + \sum\limits_{j\in\mc B} \sum\limits_{\substack{i\in\Gamma(j) \\ i\notin\mc B}} H(V_{ji}|Z_{ij}) \\
			& \stackrel{(b)}{=} \widetilde{R}^{(j^*)}_{i^*} + \widetilde{R}^{(i^*)}_{j^*} + \sum\limits_{\substack{i<j \\ \text{~s.t.~} e_{ij}\in\mc{E}_{\mc{B}}\setminus\{e_{i^*j^*} \}  }} H(V_{ij},V_{ji}|Z_{ij}) + \sum\limits_{\substack{i<j  \\j\in\mc B,  i\notin\mc B}} H(V_{ji}|V_{ij},Z_{ij}) + I(V_{ij};V_{ji}|Z_{ij})  \\
			& \stackrel{(c)}{\geq}  \sum\limits_{\substack{i<j                                                         \\ \text{~s.t.~} e_{ij}\in\mc{E}_{\mc{B}}  }} H(V_{ij},V_{ji}|Z_{ij}) +  \sum\limits_{\substack{i<j  \\j\in\mc B,  i\notin\mc B}} H(V_{ji}|V_{ij},Z_{ij}).
		\end{align*}

		In (a), we used %
		Equation~\eqref{eq:couple-rates}, and the rules of the rate assignment protocol, %
		and in (b) we used $H(V_{ji}|Z_{ij})=H(V_{ji}|V_{ij},Z_{ij})+I(V_{ij};V_{ji}|Z_{ij})$, and in (c) we observe that %
		$\mc B\subsetneq\mc M$, which means %
		there always exists at least one node $i\notin\mc B$ in $G_{\mc A}$ such that  $i\in\Gamma(j)$ for some node $j\in\mc B$. Thus, on the right-hand-side of (c) there is always a $I(V_{ij};V_{ji}|Z_{ij})$ and by definition  $I(V_{ij};V_{ji}|Z_{ij}) \geq I(V_{i^*j^*};V_{j^*i^*}|Z_{i^*j^*})$. Also, note that due to \eqref{eq:couple-rates-2} and \eqref{eq:couple-rates-3} we have $I(V_{i^*j^*};V_{j^*i^*}|Z_{i^*j^*}) +  \widetilde{R}^{(j^*)}_{i^*} + \widetilde{R}^{(i^*)}_{j^*} = H(V_{i^*j^*},V_{j^*i^*}|Z_{i^*j^*}).$ 

		With the proof of Case I  and Case II, the proof of Equation~\eqref{eq:Rco-Tree-app} is complete.
	\end{IEEEproof}
	
	Equation~\eqref{eq:Rco-Tree-app} immediately implies that
	\begin{equation*}
		C_{WSK}^{\mc A}(P_{X_\mc{M} Z}) \leq \min_{\substack{i,j\in\mc M \\ \mathrm{~s.t.~}e_{ij}\in\mc E_{\mc A}}}  I(V_{ij};V_{ji}|Z_{ij}). ~\hfill\IEEEQEDhereeqn
	\end{equation*}
\end{IEEEproof}

\section{Proof of Lower Bound Lemma \ref{lemma:Ach-Tree}}\label{sec:app:Ach}

We prove that SKA protocol~\ref{prot:SKA-Tree-PIN} 
achieves the key capacity of any
wiretapped Tree-PIN. 
The proof has three parts: (i) proof of key rate, (ii) proof of reliability, 
and (iii) proof of secrecy.

\begin{IEEEproof}[Proof of Lemma \ref{lemma:Ach-Tree}]
	We prove that for any given Tree-PIN with terminals $\mc M=[m]$ and graph $G=(\mc M,\mc E)$ and distribution $P_{ZX_{\mc M}}$, there exists an SKA protocol that achieves the upper-bound of Lemma~\ref{lemma:Upper-bound-Tree} on the wiretap secret key capacity of key agreement for $\mc A=\mc M$. We assume that each terminal $j\in\mc M$ can execute $|\Gamma(j)|$ two-party (pairwise) SKA protocols $\{\myop{\pi}_{ij}|~i\in\Gamma(j) \}$, for extracting pairwise secure keys between terminal (node) $j$ and its neighbors.

	Without loss of generality, we assume that the Tree-PIN, is labeled such that node $1$ is adjacent to node $2$ and $|\Gamma(1)|=1$. Thus, the edge $e_{12}$ will be included in all paths from node $1$ to other nodes in the tree. If the path from $i_1$ to node $i_f$, goes through the nodes $i_2,i_3,\ldots,i_{f-1}$, then we denote the path from $i_1$ to $i_f$ by $\mathrm{Path}(i_1\rightarrow i_f) = (e_{i_1 i_2},e_{i_2 i_3},\cdots, e_{i_{f-1} i_f})$.

	All terminals in $\mc M$ will participate in an SKA protocol, described in the pseudo-code~\ref{prot:SKA-Tree-PIN}. In the first phase of the protocol, each terminal $j$ obtains a shared secret key with each member of $\Gamma(j)$.
	Let   %
	$S_{ij} = \myop{\pi}_{ij}(V_{ij}^n , V_{ji}^n)$ denote the pairwise shared key for any adjacent nodes $i$ and $j$.
	Then, all terminals cut the first $\ell$ bits of their obtained keys, so that all pairwise keys have the same length. The shortened pairwise keys are $S^\prime_{ij}=S_{ij}\vert_{\ell}$. The parameter $\ell$ is a protocol parameter that has to be calculated before running the protocol, according to the known joint distribution $P_{Z X_{\mc M}}$.
	
	During the public communication phase of protocol~\ref{prot:SKA-Tree-PIN}, each node $j$ finds the unique node $j^*\in\Gamma(j)$ that is closest to node $2$. For any other node $k\in\Gamma(j)\setminus\{j^*\}$, node $j$ broadcasts $F_{jk}=S^\prime_{jj^*}\oplus S^\prime_{jk}$. Thus, the total number of broadcasts by node $j$ is $|\Gamma(j)|-1$. Note that each broadcast only uses local variables of node $j$. %

	In the last phase of the protocol, terminals $1$ and $2$ set their final shared keys to be $K_1=K_2=S^\prime_{12}$, and the rest of the terminals calculate their obtained keys $K_j$ using the public broadcasted messages (see Protocol~\ref{prot:SKA-Tree-PIN}, line~\ref{line:final_calculation}).

	\begin{IEEEproof}[\textbf{Proof of Key Rate}]
		It is known that \cite{Ahlswede1993,Maurer1993} the two-party WSK capacity of a pair of terminals $i$ and $j$ with access to  $n-$IID copies of random variables  $V_{ij}$ and $V_{ji}$ is $I(V_{ij};V_{ji}|Z_{ij})$ when $V_{ij}-V_{ji}-Z_{ij}$ holds. That is, there exists a family of $(\epsilon_n,\sigma_n)$ SKA protocols with $\lim_{n \rightarrow \infty} (\epsilon_n )= \lim_{n \rightarrow \infty} (\sigma_n)= 0$, where $\myop{length}(S_{ij}) = \floor{n\left(I(V_{ij};V_{ji}|Z_{ij}) -\Delta_n \right)}$ for some $\Delta_n(\epsilon_n +\sigma_n)$ such that $\lim_{n \rightarrow \infty} \Delta_n = 0$. To start protocol~\ref{prot:SKA-Tree-PIN}, fix  an arbitrary $\delta>0$ which is smaller that $\min_{i,j} I(V_{ij};V_{ji}|Z_{ij})$ and choose any $\ell$ such that
		$$\ell\leq n\left( \min_{i,j} I(V_{ij};V_{ji}|Z_{ij}) - \delta -\Delta_n\right).$$

		Due to the reliability of the protocol (proved next), every node $j\in \mc M$, can obtain the same key $K=S^\prime_{12}$ with length $\ell$.
		Thus, the SKA protocol~\ref{prot:SKA-Tree-PIN}, can achieve the asymptotic SK rate of
		\begin{align*}
			r_{K}(\vect{\Pi_{TP}})  = \lim\limits_{n\rightarrow\infty}\frac{1}{n}\myop{length}(S^\prime_{12}) & = \lim\limits_{n\rightarrow\infty}\frac{1}{n} \ell                                          \\
			& \leq \lim\limits_{n\rightarrow\infty} \min_{i,j} I(V_{ij};V_{ji}|Z_{ij}) - \delta -\Delta_n 
			\\
			&                                                                                             
			= \min_{i,j} I(V_{ij};V_{ji}|Z_{ij}) - \delta.
		\end{align*}
		Since, $\delta$ can take any small value, then as $\delta\rightarrow 0$, the SK rate of \ref{prot:SKA-Tree-PIN} will be arbitrary close to $C=\min_{i,j} I(V_{ij};V_{ji}|Z_{ij})$. %
	\end{IEEEproof}
	
	Next, we show that the WSK capacity achieving SKA protocol~\ref{prot:SKA-Tree-PIN} is secure and reliable for any given Tree-PIN. To prove this claim, we need to show
	\begin{itemize}
		\item \textbf{Reliability}: Showing that $\pr{K_1=K_2=\cdots=K_m=K} \rightarrow 1$ as $n\rightarrow\infty$, and
		\item \textbf{Secrecy}: Showing that $\sd\left((K,{\vect{F}} ,Z),(U,{\vect{F}},Z)\right) \rightarrow 0$ as $n\rightarrow\infty$.
	\end{itemize}

	\begin{IEEEproof}[\textbf{Proof of Reliability}]
		Let $K_j$ denote the final key calculated by terminal $j$. We show that  $K_1=K_2=\cdots=K_m=S^\prime_{12}=K,$ if all $m-1$ pairwise $(\epsilon_n,\sigma_n)-$SKs $S_{ij}$ are established. %
		For any node $j\in\mc M\setminus \{1,2\}$ there is only   one path to node $2$. This path is of the form $\mathrm{Path}(j\rightarrow 2)=(e_{jj^*}, e_{j^* i_1}, e_{i_1 i_2},e_{i_2 i_3},\cdots, e_{i_f 2})$, where node $j^*$ is the unique neighbor of $j$ which is closest to node $2$ and $i_k$'s ($i=1\ldots f$) are the labels for all the nodes (except for $j, j^*$ and $2$) that are in the path of $j$ to $2$.%
		
		In protocol~\ref{prot:SKA-Tree-PIN}, line~\ref{line:broadcast}, node $k$ broadcasts $F_{kj}=S^\prime_{i_1 k}\oplus S^\prime_{kj}$. Thus, node $j$ who has access to the key $S^\prime_{kj}$ can perfectly recover $S^\prime_{i_1 k}$ by computing $S^\prime_{kj}\oplus F_{kj}$. Also, node $i_1$ (which is connected to $i_2$ and $k$) has broadcasted $F_{i_1 k}= S^\prime_{i_2 i_1}\oplus S^\prime_{i_1 k}$. Node $j$ who has now have recovered $S^\prime_{i_1 k}$, can recover  $S^\prime_{i_2 i1}$ as well, by computing $S^\prime_{i_1 k}\oplus F_{i_1 k}$. This chain of recovering local keys will continue until $S^\prime_{12}$ is recovered by computing $K_j=S^\prime_{kj}\oplus F_{kj}\oplus F_{i_1 k} \oplus F_{i_2 i_1}\oplus F_{i_3 i_2}\oplus \cdots \oplus F_{2 i_f}$, which proves that $K_j=S^\prime_{12}$ for any $j\in\mc M$.
		
		This requires all $m-1$ pairwise $(\epsilon_n,\sigma_n)-$SKs $S_{ij}$ to be established. The error probability of each pairwise key is bounded by $\epsilon_n$, thus the error probability of establishing the global key is $(m-1)\epsilon_n=|\mc E|\epsilon_n$. Therefore, $\pr{K_1=K_2=\cdots=K_m=K}\leq 1- \epsilon'_n,$ with $\epsilon'_n=|\mc E|\epsilon_n$ where $\epsilon_n$ such that $\lim_{n \rightarrow \infty} \epsilon_n= 0$.
	\end{IEEEproof}

	\begin{IEEEproof}[\textbf{Proof of Secrecy}]
		We need to prove the secrecy of the global shared key $K$. Without loss of generality, assume that all adjacent terminal pairs $i$ and $j$ with $e_{ij}\in\mc E$ have established a binary pairwise $(\epsilon_n,\sigma_n)-$SK $S_{ij}$ with length $\ell=\floor{n(C-\delta)}$, where $C=\min_{i,j} I(V_{ij};V_{ji}|Z_{ij})$. Note that for any $e_{ij}\in\mc E$ we have $\sd((S_{ij},Q_{ij},Z),(U,Q_{ij},Z_{ij}))\leq \sigma_n$, where $U$ is the uniform distribution over $\{0,1\}^{\ell}$ and $Q_{ij}$ denotes the public communication used to generate $S_{ij}$. 
		To recall the definition of statistical distance please see Appendix \ref{sec:SD}.

		\remove{ %
			Recall that the statistical distance between RVs $X$ and $Y$ with the same alphabet $\mc W$, %
			is defined as
			\begin{align*}
				\sd(X,Y) & = \frac{1}{2} \sum_{w\in \mc W} \left\vert P_X(w) - P_Y(w) \right\vert    \\
				& = \max\limits_{\mc{T}\subseteq\mc{W}}  \sum_{w\in \mc T}  P_X(w) - P_Y(w) \\
				& = \sum_{w\in {\mc{T}^*}}  P_X(w) - P_Y(w),                                
			\end{align*}
			where ${\mc{T}^*} = \{w\in\mc{W}~\vert~ P_X(w) \geq P_Y(w) \}$.
		}%

		Let $Q$ denote the collection of all public communications required to establish all $|\mc E|=m-1$ pairwise keys $S_{ij}$, and let $F$ denote the collection of all public communications broadcasted by all terminals during the SKA protocol~\ref{prot:SKA-Tree-PIN} and $\vect{F}=(F,Q)$ be the overall public communication.
		For any given Tree-PIN $P_{ZX_{\mc M}}$ with $G=(\mc M,\mc E)$ we prove that
		\begin{align*}
			\sd((K,\vect{F},Z),(U,\vect{F},Z)) & = \sd((K,F,Q,Z),(U,F,Q,Z))                                 \\
			& \leq                                                       
			\sd((K,F,Q,Z),(U,U^{|\mc E|-1},Q,Z)) + 
			\sd((U,U^{|\mc E|-1},Q,Z),(U,F,Q,Z)) \\
			& \leq |\mc E|\sigma_n + |\mc E|\sigma_n = 2|\mc E|\sigma_n, 
		\end{align*}
		where $U^{d}$ is the uniform distribution over $\mc K ^d = \{0,1\}^{d\ell}$. 
		
		First we show that {\em ``the combination $(K,F)$ uniquely gives all pairwise keys $\{ S_{ij} \}_{i<j}$''}. Recall that any pairwise key belongs to the alphabet $\mc K=\{0,1\}^{\ell}$. Let $\vect s=\{s_{ij}\}_{i<j} \in\mc K^{|\mc E|}  $ be an instance of all pairwise keys. Note that $F=F(\vect S)$ is a set of $m-2$ linear functions 
		of the random vector $\vect S$. 
		According to Protocol~\ref{prot:SKA-Tree-PIN} each terminal $j\in\mc M$ broadcasts $|\Gamma(j)|-1$ messages. Also recall that for any tree $|\mc E|=m-1$, so, the total number of public messages is $\sum_{j\in\mc M}  |\Gamma(j)|-1 = 2|\mc E|-m=m-2$. 
		Thus, the $m-2$ elements of $F$ are not sufficient for uniquely finding all $m-1$ pairwise keys in $\vect S$. %
		However, the combination of $F$ and the final key $K$ resulted by the SKA protocol~\ref{prot:SKA-Tree-PIN} is sufficient for unique recalculation of all pairwise keys. Remember that $K=S_{12}$ and with all the public messages of terminal $2$ one can recover all pairwise keys accessible to terminal $2$ since they are all of the form $F_{2j}=S_{12}\oplus S_{2j}$ for all $j\in\Gamma(2)\setminus\{1\}$. Now with access to these pairwise keys one can recover all pairwise keys accessible to any terminal $j\in\Gamma(2)\setminus\{1\}$. This chain of calculation will continue until all pairwise keys are recovered.
		
		Since $(K,F)$ uniquely gives $\{ S_{ij} \}_{i<j},$ then $\sd((K,F,Q,Z),(U,U^{|\mc E|-1},Q,Z))\leq \sd((\{ S_{ij} \},Q,Z),(U^{|\mc E|},Q,Z))\leq  |\mc E|\sigma_n.$
		Also, %
		we have $\sd((U,U^{|\mc E|-1},Q,Z),(U,F,Q,Z))\leq  |\mc E|\sigma_n,$ %
		because,
		\begin{align*}
			& \sd((U,U^{|\mc E|-1},Q,Z),(U,F,Q,Z))                                                                                                                                                                                                       \\
			& \qquad\qquad=  \sd((F,Q,Z),(U^{|\mc E|-1},Q,Z))                                                                                                                                                                                            \\
			& \qquad\qquad\stackrel{\text{(a)}}{=} \sum_{(f,q,z)\in\mc T^*}  P_{QZ}(q,z)P_{F|QZ}(f|q,z) - P_{QZ}(q,z) P_{U^{|\mc E|-1}}(f)                                                                                                               \\
			& \qquad\qquad= \sum_{(q,z)\in\mc T^*}  P_{QZ}(q,z) \sum_{f\in\mc T^*}  P_{F|QZ}(f|q,z) - \frac{1}{|\mc K|^{|\mc E|-1 }}                                                                                                                     \\ %
			& \qquad\qquad\stackrel{\text{(b)}}{\leq} \sum_{(q,z)\in\mc T^*}  P_{QZ}(q,z) \sum_{f\in\mc T^*}  P_{F|QZ}(f|q,z) - \frac{1}{|\mc K|^{|\mc E|}}                                                                                              \\ %
			& \qquad\qquad\stackrel{\text{(c)}}{=} \sum_{(q,z)\in\mc T^*}  P_{QZ}(q,z) \sum_{\vect s\in \mc S^*(\mc T^*)} \prod_{i<j} P_{S_{ij}|Q_{ij}Z}(s_{ij}|q_{ij},z) - \frac{1}{|\mc K|^{|\mc E|}}                                                  \\ %
			& \qquad\qquad\stackrel{\text{(d)}}{\leq} \max_{\mc T\subseteq \mc Q\times\mc Z\times\mc K^{|\mc E|}}  \sum_{(q,z)\in\mc T}  P_{QZ}(q,z) \sum_{\vect s\in \mc T} \prod_{i<j} P_{S_{ij}|Q_{ij}Z}(s_{ij}|q_{ij},z) - \prod_{i<j} P_{U}(s_{ij}) \\
			& \qquad\qquad\stackrel{\text{(e)}}{=} \sd((\{ S_{ij} \},Q,Z),(U^{|\mc E|},Q,Z))                                                                                                                                                             \\
			& \qquad\qquad\stackrel{\text{(f)}}{\leq} \sum_{i<j} \sd((S_{ij},Q_{ij},Z_{ij}),(U,Q_{ij},Z_{ij}))                                                                                                                                           %\\
			%
			%& \qquad\qquad
			\leq |\mc E|\sigma_n,                                                                                                                                                                                                          
		\end{align*}
		where in (a) %
		$\mc T^*=\{ (f,q,z)| P_{QZ}(q,z)P_{F|QZ}(f|q,z) \geq P_{QZ}(q,z) P_{U^{|\mc E|-1}}(f)  \}$
		which is due to definition statistical distance   
		and in equality (c) $\mc S^*(\mc T^*)$ is defined as $\mc S^*(\mc T^*)= \{ \vect{s}~ \vert~ \vect s\in\mc K^{|\mc E|} ~\text{and}~ F(\vect s)=f, ~\forall f\in \mc T^*  \}$.  %
		Inequality (b) is due to the fact that for any $(f,q,z) \in \mc T^*$ we have $P_{F|QZ}(f|q,z) \geq P_{U^{|\mc E|-1}}(f)$.  Relations (d) and (e) are due to the definition of the statistical distance. 
		Inequality (f) follows from Corollary \ref{thm:SD_cor_3}. 
		
		Hence, the final key $K$ obtained from the SKA protocol~\ref{prot:SKA-Tree-PIN} is an $(|\mc E|\epsilon_n,2|\mc E|\sigma_n)-$SK where $\lim_{n \rightarrow \infty} (\epsilon_n )= \lim_{n \rightarrow \infty} (\sigma_n)= 0$ and the security proof is complete.
	\end{IEEEproof}
	
	With the reliability, security, and key rate proofs, the proof of Lemma~\ref{lemma:Ach-Tree} is complete.~\IEEEQEDhere
\end{IEEEproof}

\section{Proof of Theorem \ref{thm:SOA-lower} and Proposition \ref{thm:semi-lower}}\label{sec:app:FL-lower}

\begin{IEEEproof}
	We first recall the SKA protocol that attains the WSK capacity of Tree-PIN. 
	Terminals in $G_{\mc A}$ --the smallest sub-tree that connects terminals in $\mc A$ -- will generate pairwise keys. Note that in this step, terminals will generate pairwise $(\eps',\sigma')-$SKs, where $\eps'=\frac{\eps}{|\mc E_{\mc A}|}$ and  $\sigma'=\frac{\sigma}{2|\mc E_{\mc A}|}$. Next, all terminals will announce the length of their pairwise keys, and then all pairwise keys will be cut to the minimum length so every pairwise key has the same length. After this, middle nodes (terminals) will broadcast appropriate XOR public messages according to the SKA protocol described earlier. According to the proof of Lemma \ref{lemma:Ach-Tree}, the final extracted key is an $(\eps,\sigma)-$SK.

	If the pairwise keys are generated by the interactive protocol of Hayashi et. al \cite{Hayashi2016}, Theorem 15, terminals $i$ and $j$ can obtain a pairwise key of length $$\ell_{ij} = nI(V_{ij};V_{ji}|Z_{ij})- \sqrt{n \Delta_{ij} }Q^{-1}(\eps'+\sigma') - \frac{11}{2}\log n +  \mc O(1).$$

	If the  pairwise keys are generated by the OW-SKA Protocol of \cite{Sharifian2020}, then
	terminals $i$ and $j$ can obtain a pairwise key of length 
	$$\ell_{ij} = nI(V_{ij};V_{ji}|Z_{ij})-  Q^{-1}(\epsilon') \sqrt{n\Delta'_{ij}}   
	-Q^{-1}(\sigma') \sqrt{n\Delta''_{ij}} -\log n +  \mc O(1),$$
	or
	$$ \ell_{ij} = nI(V_{ij};V_{ji}|Z_{ij})
	-\sqrt{2n}\log({|\mc X|+3})(\sqrt{\log \frac{1}{\eps'}} + \sqrt{\log \frac{1}{\sigma'}}) -\log n +  \mc O(1).
	$$
	To understand the difference between these two achievability approximations and their applications, see \cite{Sharifian2020}.
	
	For the special case when $V_{ij}=V_{ji}$, there is no need for information reconciliation, %
	and thus we can use the key extraction bound of \cite{Hayashi2019}. Thus, for this case, terminals $i$ and $j$ can obtain a pairwise key of length $$\ell_{ij} = nH(V_{ij}|Z_{ij})-   \sqrt{n \Delta''_{ij}} Q^{-1}(\sigma') -\frac{1}{2}\log n +  \mc O(1).
	$$
	
	By utilizing either SKA approaches, the length of the final key agreed by all terminals in $G_{\mc A}$ is
	$$\ell = \min_{\substack{i,j\in\mc M \\ \mathrm{~s.t.~}e_{ij}\in\mc E_{\mc A}}} \ell_{ij},$$
	and hence the proof is complete.
	It's easy to see that with either of these approaches, the SKA protocol~\ref{prot:SKA-Tree-PIN} attains the capacity of Theorem~\ref{thm:Tree-PIN}.
\end{IEEEproof}

\end{document}